\documentclass[]{aa} 

\usepackage{graphicx}
\usepackage{txfonts}
\usepackage{xcolor}
\usepackage{comment}
\usepackage{hyperref} % added by James McK for links
\newcommand{\bl}[1]{\mbox{\boldmath$ #1 $}}

\begin{document}

%   \title{Primordial dust rings and the first generation of planetesimals in gravitoviscous protoplanetary disks}

  \title{Dust growth and pebble formation in the initial stages of protoplanetary disk evolution}

%   \subtitle{VI. ???}

   \author{ Eduard I. Vorobyov\inst{1,2},
    Igor Kulikov\inst{3}, Vardan G. Elbakyan\inst{2,4},  James McKevitt\inst{1}, and Manuel G\"{u}del\inst{1}
          }
   \institute{ University of Vienna, Department of Astrophysics, T\"urkenschanzstrasse 17, 1180, Vienna, Austria; 
   \email{eduard.vorobiev@univie.ac.at} 
\and
Research Institute of Physics, Southern Federal University, Stachki Ave. 194, Rostov-on-Don 344090, Russia  
\and
Institute of Computational Mathematics and Mathematical Geophysics SB RAS, Lavrentieva ave., 6, Novosibirsk, 630090 Russia
\and
Fakultät für Physik, Universität Duisburg-Essen, Lotharstraße 1, D-47057 Duisburg, Germany
}

\date{}

%   \date{Received September 15, 2020; accepted November 16, 2020}
   
   \titlerunning{Dust growth and planetesimal formation}
   \authorrunning{Vorobyov et al.}

  \abstract
  {}
  % aims heading (mandatory)
   {The initial stages of planet formation may start concurrently with the formation of a gas-dust protoplanetary disk. This makes the study of the earliest stages of protoplanetary disk formation crucially important. Here we focus on dust growth and pebble formation in a protoplanetary disk that is still accreting from a parental cloud core.  }
   {We have developed an original three-dimensional numerical hydrodynamics code, which computes the collapse of rotating clouds and disk formation on nested meshes using a novel hybrid Coarray Fortran-OpenMP approach for distributed and shared memory parallelization. Dust dynamics and growth are also included in the simulations.}
   {We found that the dust growth from $\sim 1~\mu$m to 1--10~mm  already occurs in the initial few thousand years of disk evolution but the Stokes number hardly exceeds 0.1  because of higher disk densities and temperatures compared to the minimum mass Solar nebular. The ratio of the dust-to-gas vertical scale heights remains rather modest, 0.2--0.5, which may be explained by the perturbing action of spiral arms that develop in the disk soon after its formation. The dust-to-gas mass ratio in the disk midplane is highly nonhomogeneous throughout the disk extent and is in general enhanced by a factor of several compared to the fiducial 1:100 value.  Low $\mathrm{St}$ hinders strong dust accumulation in the spiral arms compared to the rest of the disk and the nonsteady nature of the spirals is also an obstacle.
   The spatial distribution of pebbles in the disk midplane exhibits a highly nonhomogeneous and patchy character.  The total mass of pebbles in the disk increases with time and reaches a few tens of Earth masses after a few tens of thousand years of disk evolution.}
   {We found that protoplanetary disks with an age $\le 20$~kyr can possess notable amounts of pebbles and feature dust-to-gas density enhancements in the disk midplane. Hence, these young disks can already be ripe for the planet formation process to start. 
   Multidimensional numerical models of disk formation that consider the coevolution of gas and dust including dust growth are important to improve our understanding of planet formation.}


   \maketitle

\section{Introduction}
Protoplanetary disks form during the collapse of slowly rotating prestellar cloud cores. These objects are mainly composed of molecular hydrogen and helium with a small admixture of interstellar sub-micron-sized dust grains. During the initial stages of evolution, the disk is surrounded by a massive gas-dust envelope, which makes it difficult to directly observe the processes that take place in the disk vicinity. Yet, this stage may be crucial for the subsequent processes of planet formation, because here the initial stages of dust growth from sub-$\mu$m grains to millimeter-sized aggregates should occur. The accelerated formation of protoplanets is indirectly supported by the detection of ring- and gap-like structures in young protoplanetary disks, such as  GY~91 \citep{2018Sheehan}, IRS~63 \citep{2020Segura-Cox}, and HL~Tau \citep{2015ALMABrogan}, all of which are no older than 0.5~Myr. These structures can presumably be formed by the gravitational torques from yet unseen planets with a mass equal to or greater than Neptune \citep{2012ZhuNelson, 2015Dipierro, 2018DongLi}.  

Furthermore, recent estimates of the dust mass in young Class 0 and I disks and also in more evolved T~Tauri disks revealed that the dust mass systematically decreases with the age of the system \citep{2020Tobin,2020Tychoniec}, in agreement with earlier predictions \citep{2011GreavesRice}. In particular, the available mass of dust in Class 0 and I disks is comparable to that contained in exoplanets around main-sequence stars, while the dust mass in T~Tauri disks is insufficient to explain the bulk mass distribution of exoplanets. This again accentuates the importance of the earliest stages of disk formation in the planet formation paradigm.

This observational evidence motivated recent studies of dust dynamics and growth in numerical hydrodynamics models that simulate  the transition from a collapsing prestellar core to the star and disk formation stage self-consistently. \citet{2018ZhaoCaselli} suggest that dust growth
and the removal of the small dust grains from the Mathis, Rumpl, and Nordsiek (MRN) dust size
distribution changes the magnetic resistivity
significantly and promotes disk formation in a magnetized environment.
Magnetohydrodynamics simulations of disk formation by \citet{2020Tsukamoto} demonstrate that protoplanetary disks form when dust physics is included in the calculation of resistivity regardless of the dust size distribution.
\citet{2017Bate} and \citet{2023Koga} found that dust grains with sizes greater than 10~$\mu$m can dynamically decouple from gas already during the cloud core collapse. Interestingly, dust grains with sizes $\gg 1~\mu$m  have indeed been inferred observationally in protostellar clouds \citep{2019Galametz}. The physical mechanism to explain their existence has not been fully understood. Using smoothed-particle hydrodynamics simulations, \citet{2022Bate} found that dust growth in collapsing clouds occurs only in the inner few hundreds of astronomical units, and in particular in the first hydrostatic core where dust can quickly grow in excess of 100~$\mu$m. On the other hand, large grains can be entrained with the disk wind into the protostellar envelope as was recently found by \citet{2021Tsukamoto} and \citet{2023Koga}.

Efficient dust growth in the disk formation phase has been reported in the thin-disk hydrodynamics simulations of \citet{2018VorobyovAkimkin}, who demonstrated that, depending on the value of the turbulent $\alpha$ parameter and the threshold velocity of dust fragmentation, dust can grow from sub-$\mu$m grains to millimeter- or even centimeter-sized aggregates already in the embedded phase of disk evolution. 
Fast growth agrees with analytic expectation, according to which  the dust-size doubling time is inverse proportional proportional to the dust-to-gas ratio and the local Keplerian angular velocity \citep{2016Birnstiel}. 
Using three-dimensional (3D) magnetohydrodynamics simulations, \citet{2020Lebreuilly} and \citet{2022Koga} found efficient settling and the drift of grains with sizes greater than a few tens of microns in the first core and in the newly formed disk, producing dust-to-gas ratios a factor of several greater than the initial value.
\citet{2023Marchand} employed  a precomputed dust growth model and show that grain sizes can reach $\ge 100$~$\mu$m in an inner protoplanetary disk 1~kyr after its formation. \citet{2023Tsukamoto} found that dust growth can influence the disk evolution by changing the coupling between gas and the magnetic field, making the gas density profiles steeper and the disk mass smaller.

%The dynamics and growth of dust in class~0 and I 
%More text on mumerical simulations of dust growth in the early phases of disk formation ...works of Tsukamoto, Bate, Mayer ...Latter and Riols

The fast dust growth solves only part of the planet formation problem. Grown dust must avoid a catastrophic inward radial drift caused by the mismatch between the angular velocities of gas and dust in a Keplerian disk \citep{Weidenschilling1977}. This can be achieved if the disk pressure has local maxima or strongly varying spatial gradients \citep[e.g.,][]{Drazkowska2016,Flock2015}. Water ice lines can also help to reduce the inward radial drift \citep{2016Pinilla,Molyarova2021,2022LauDrazkowska}. Spiral arms and gaseous clumps in gravitationally unstable disks can also collect dust particles if they grow sufficiently large to be characterized by the Stokes number $\sim 0.1-1.0$ \citep[e.g.,][]{2004RiceLodato,2019BaehrKlahr,2021DengMayer,2022Baehr}, though the efficiency of this process may be weakened by the transitory nature of the spiral arms and tidal destruction of gaseous clumps in young protoplanetary disks \citep{2018VorobyovAkimkin,2019VorobyovElbakyan}.

In our previous works, we have studied dust dynamics and growth in young protoplanetary disks using two-dimensional thin-disk approximation \citep{2018VorobyovAkimkin,2019VorobyovElbakyan,2020Elbakyan, Molyarova2021,Vorobyov2023a}. These studies have brought important insights in the spatial distribution of pebbles,  efficiency of dust accumulation, composition of ice mantles, and dust growth efficiency in the disk midplane. 
Here, we present our first results of dust dynamics and growth obtained in fully three-dimensional hydrodynamic simulations of a gas-dust disk in the initial stages of its formation. We developed an original code based on numerical techniques that are different from our previous thin-disk model, which allows us to test our previous conclusions and obtain new insights into the three-dimensional distribution of dust and gas in young protoplanetary disks.

The paper is organized as follows. In Sect.~\ref{sect:model} we describe the basic equations entering our model of a gas-dust disk. We also provide a brief explanation of the original code and the novel Coarray Fortran methodology for distributed memory parallelization. In Sect.~\ref{sect:results} we present our main results. The caveats and future developments are briefly discussed in Sect.~\ref{sect:caveats}. The main conclusions are summarized in Sect.~\ref{sect:conclude}. In Appendices~\ref{Appendix:coarrays}-\ref{Appendix:tests} the details of the numerical solution and an extensive suite of test problems are provided.

\section{Model description}
\label{sect:model}
In this section, we present the basic equations and the solution methods, which constitute the foundation of the Formation and Evolution Of Stars and Disks on nested grids (ngFEOSAD) code. ngFEOSAD computes the co-evolution of gas and dust during the gravitational collapse of a rotating prestellar core including the initial stages of protoplanetary disk formation using a fully three-dimensional approach. We employed Coarray Fortran for the distributed memory parallelism, which is much simpler in coding than the usual message-passing interface (MPI), while being only slightly inferior to the MPI in terms of the speed. This allows focusing on the solution of scientific problems rather than on the intricacies of parallel programming. Thanks to using fast-Fourier transforms to solve the gravitational potential on nested meshes, ngFEOSAD is fast and environmentally friendly, requiring only a few modern computer nodes to run the models.
In particular, we used four nodes with three tasks per node and 16 cores per task, amounting to 192 cores in total on AMD EPYC 7713 processors. For a typical run-time of 15 days, this corresponds roughly to 70,000 core hours.
We adopt a barotropic equation of state, which includes the initial isothermal collapse stage followed by the star and disk formation stages (see Eq.~\ref{eosvol}). We note that the forming protostar is not fully resolved in our simulations with limited spatial resolution ($\ge 0.14$~au in the central regions). Yet we do not introduce the sink cell as the results may be sensitive to the details of the sink cell implementation \citep{2014Machida, 2019VorobyovSkliarevskii, 2020Hennebelle}, which requires a careful consideration. Our approach is valid for the very early stages of evolution, when the central object may still be a first hydrostatic core ($\sim 5$~au) or in the early stages of collapse toward a protostar.

\subsection{Gas component}
The dynamics of gas is followed by solving the equations of continuity and momentum
\begin{equation}
\label{eq:cont}
\frac{{\partial \rho_{\rm g} }}{{\partial t}}   + \nabla  \cdot 
\left( \rho_{\rm g} {\bl v} \right)  = 0,  
\end{equation}
\begin{equation}
\label{eq:mom}
\frac{\partial}{\partial t} \left( \rho_{\rm g} {\bl v} \right) +  \nabla \cdot \left( \rho_{\rm
g} {\bl v} \otimes {\bl v} + \mathbb{I} P \right)   =    - \rho_{\rm g} \, \nabla \Phi 
 - \rho_{\rm d,gr} \, {\bl f},
\end{equation}
where $\rho_{\rm g}$ is the volume gas density, $\bl v$ the gas velocity, $\Phi$ the gravitational potential, $P$ the gas pressure, $\mathbb{I}$ the unit tensor, and $\bl f$ the friction force per unit dust mass between gas and dust. The gas pressure is related to the gas density via the following barotropic equation
\begin{equation}
P_k = c_{\rm s,0}^2 \, \rho_{\rm g}^{\gamma_k} \prod_{i=1}^{k-1}\rho_{{\rm c},i}^{\gamma_i-\gamma_{i+1}},
\,\,\, \mathrm{for} \,\,\, \rho_{{\rm c},k-1} \le \rho_{\rm g} < \rho_{{\rm c}, k} \, ,
\label{eosvol}
\end{equation}
where $c_{\rm s,0}=\sqrt{{\cal R} T_{\rm in}/\mu}$ is the initial isothermal sound speed, $T_{\rm in}$ the initial gas temperature, $\cal R$ the universal gas constant, and $\mu=2.33$ the mean molecular weight.  The values of the indices $i$ and $k$ distinguish the four individual components of the piecewise form shown in Table~\ref{table:1}. The first component  corresponds to the initial isothermal stage of cloud core collapse, while the other three components account for the postcollapse nonisothermal evolution with various ratios of the specific heat $\gamma_{i(k)}$ \citep{2000Masunaga,2015MachidaNakamura}.
We note that when $k=1$ the product term is unity by definition, and the pressure reduces to $P_1=c_{\rm s}^2 \rho_{\rm g}^{\gamma_1}$. We also formally set $\rho_{{\rm c},0}$ to zero. The gas pressure $P$ is set equal to $P_{k}$, where the index $k$ is defined by the condition that $\rho_{\rm g}$ falls in between the corresponding critical densities, $\rho_{{\rm c},k-1} \le \rho_{\rm g} < \rho_{{\rm c}, k}$.

\begin{table}
\center
\caption{Parameters of the barotropic relation.}
\label{table:1}
\renewcommand{\arraystretch}{1.5}
\begin{tabular*}{\columnwidth}{ @{\extracolsep{\fill}} c c c }
\hline \hline
$k$ & $\gamma_{i(k)}$ & $n_{{\rm c},{i(k)}}$  \\
\hspace{1cm} & & (cm$^{-3}$)   \\ [0.5ex]
\hline \\ [-2.0ex]
1 & 1.00 & $5{\times}10^{10}$  \\
2 & 1.6667 & $2.5{\times}10^{12}$  \\
3 & 1.4 & $1.5\times 10^{15}$ \\
4 & 1.1 & ... \\ [1.0ex]
\hline
\end{tabular*}
\center{ \textbf{Notes.} The coefficients $\gamma_i$ and $n_{{\rm c},i}$ were taken from \citet{2015MachidaNakamura}. The number density $n_{{\rm c},i}$ is related to the volume density $\rho_{{\rm c},i}$ as $\rho_{{\rm c},i}=\mu m_{\rm H} n_{{\rm c},i}$, where $m_{\rm H}$ is the mass of hydrogen atom.}
\end{table}

\subsection{Dust component}
For the dust component we use the model originally developed in \citet{2018VorobyovAkimkin} with modifications to account for the back-reaction on gas in \citet{2018Stoyanovskaya}.  In this model, dust is considered as a pressureless fluid  \citep[see][for justification]{Vorobyov2022} with two 
populations: small dust that is dynamically linked to gas and grown dust that can dynamically decouple from gas. In particular, small dust are grains with a size between $a_{\rm min}=5\times 10^{-3} \ \mu \rm m$  and $a_{*} = 1 \ \mu \rm m$ and grown dust are aggregates ranging in size from $a_{*}$ to $a_{\rm max}$. The latter is the maximum value, which is calculated using a simple dust growth model and can be variable in space and time. The value of $a_\ast$ that separates small dust from grown dust is chosen so as to make small dust dynamically coupled to gas. In the gas-dust disk simulations, dynamics of small dust is often considered in the terminal velocity approximation \citep[e.g.,][]{2014LaibePrice,2017LinYoudin}. However, as was shown in \citet[][see Appendix D]{Vorobyov2022}, the drift timescales of $\mu$m-sizes dust in a protoplanetary disk is comparable to its typical age. Hence the full coupling approximation is justified on short evolutionary timescales considered in this work. We also note that the small dust-to-gas mass ratio never exceeds 0.01 in our model and is often much lower. In particular, the corresponding local values are in the $10^{-3}$-$10^{-4}$ limits in the bulk of the disk and approach $10^{-2}$ from below only in the disk outermost regions and in the disk upper atmosphere  (see also Sect.~\ref{Sect:pebbles}). We note, however, that the settling instability may develop at the dust-to-gas mass ratios as low as $10^{-2}$ \citep{2018SquireHopkins} and neglecting the dynamics of small dust and its backreaction on gas may not be justified in this case. 
Initially, all dust in a collapsing prestellar cloud is in the form of small dust, which  can grow and turn into grown dust as the collapse proceeds, and the disk forms and evolves. It is assumed that dust in both populations is distributed over size according to a simple power law: 
\begin{equation}
N(a) = C \cdot a^{ - {\rm p}},
\label{eq:dustdistlaw}
\end{equation} where $C$ is a normalization constant and ${\rm p} = 3.5$.  We note that the power index $\rm p$ is kept constant in this study, but may also vary if the form of this variation can be inferred from a more sophisticated approach that involves solving for the Smoluchowski equation for multiple dust bins \citep[e.g.,][]{2019Drazkowska}.

The dynamics of both dust populations is followed by solving the equations of continuity and momentum 
\begin{equation}
\label{eq:cont_dust_small}
\frac{{\partial \rho_{\rm d,sm} }}{{\partial t}}   + \nabla  \cdot 
\left( \rho_{\rm d,sm} {\bl v} \right)  = - S(a_{\rm max}),  
\end{equation}
\begin{equation}
\label{eq:cont_dust_grown}
\frac{{\partial \rho_{\rm d,gr} }}{{\partial t}}   + \nabla  \cdot 
\left( \rho_{\rm d,gr} \, {\bl u} \right)  = S(a_{\rm max}),  
\end{equation}
\begin{equation}
\label{eq:mom-dust}
\frac{\partial}{\partial t} \left( \rho_{\rm d,gr}\, {\bl u} \right) +  \nabla \cdot \left( \rho_{\rm
d,gr} \, {\bl u} \otimes {\bl u}  \right)   =     - \rho_{\rm d,gr} \, \nabla \Phi + \rho_{\rm d,gr} \, {\bl f} + S(a_{\rm max}) \, {\bl v},
\end{equation}
where $\rho_{\rm d,sm}$ and $\rho_{\rm d,gr}$ are the volume densities of small and grown dust, respectively, $\bl u$ the velocity of grown dust, and $S(a_{\rm max})$ the conversion rate between small and grown dust populations. We note that the presence of the $S(a_{\rm max}) \bl{v}$ term in the last equation and the absence of the mirrored term for the small dust (recall that we assume full coupling for small dust and do not solve the corresponding dynamics equation) formally means that the dust momentum is not conserved. This should be a reasonable approximation for as long as the mass of small dust component is much smaller than that of grown dust, which is generally fulfilled as soon as the disk forms and dust grows (see Figs.~\ref{fig:pebble-time} and \ref{fig:pebble-time_lb}). 

The drag force per unit dust mass $\bl f$ can be written as
\begin{equation}
    {{\bl f}} = \dfrac{1} {2 m_{\rm d}} C_{\rm D} \, \sigma \rho_{\rm g} ({{\bl v}} - {{\bl u}}) |{{\bl v}} - {{\bl u}}|,
\end{equation}
where $\sigma$ is the dust grain cross section, $m_{\rm d}$ the mass of a dust grain, and $C_{\rm D}$ the dimensionless friction parameter. In this work, we consider only the Epstein drag regime, in which case the friction parameter can be written as
$C_{\rm d}=8 v_{\rm th}/ 3 |{{\bl v}} - {{\bl u}}| $ \citep{Weidenschilling1977}, where $v_{\rm th}$ is the mean thermal velocity of gas. The drag force then takes a simple form
\begin{equation}
    {\bl f} = \frac{{\bl v} - {\bl u}}{t_{\rm stop}},
    \label{eq:fric}
\end{equation}
where $t_{\rm stop}$ is the stopping time expressed as
\begin{equation}
\label{tstop}
    t_{\rm stop} = \frac{a \, \rho_{\rm s}} {\rho_{\rm g} v_{\rm th}},
\end{equation}
where  $\rho_{\rm s}= 3.0$~g~cm$^{-3}$ is the material density of dust grains. Since grown dust in our model has a spectrum of sizes from $a_\ast$ to $a_{\rm max}$, the size of dust grains $a$ in Equation~(\ref{tstop}) should be weighted over this spectrum. 
However, we use $a_{\rm max}$ when calculating the stopping time. Our choice is justified as follows. Since we use only two dust size bins, the span between $a_\ast$ to $a_{\rm max}$ may become as large as  several orders of magnitude during the disk evolution and the average size
of dust grains $\sqrt{a_\ast a_{\rm max}}$ becomes much smaller than $a_{\rm max}$. However, we are interested in the dynamics of dust grains that are the main dust mass carriers.
For the chosen slope $p=3.5$, large grains near $a_{\rm max}$ mostly determine the dust mass \citep{2016Birnstiel}.  Therefore, we use the maximum size of dust grains $a_{\rm max}$ when calculating the values of $t_{\rm stop}$.  A more consistent approach requires introducing multiple  narrower bins for grown dust  and setting the mean dust size for each bin, but this  is outside the scope of the current work.

\subsection{Dust growth model} 
The ngFEOSAD code includes a simple model for dust growth. Initially, in the prestelar cloud, all dust is in the form of small dust grains with a maximum size $a_{\rm max}=a_\ast$. As the cloud begins to collapse and especially during the disk formation stage the maximum size of dust grains is allowed to increase. We use the following equation describing the time evolution of $a_{\rm max}$ 
\begin{equation}
{\partial a_{\rm max} \over \partial t} + ({\bl u} \cdot \nabla ) a_{\rm max} = \cal{D},
\label{eq:dustA}
\end{equation}
where the rate of dust growth due to collisions and coagulation is computed using a monodisperse model \citep{2012Birnstiel}
\begin{equation}
\cal{D} = {\rho_{\rm d} \mathit{u}_{\rm rel} \over \rho_{\rm s}}.
\label{eq:dustrate}
\end{equation}
This rate includes the total volume density of dust $\rho_{\rm d}=\rho_{\rm d.sm}+\rho_{\rm d,gr}$, dust material  density $\rho_{\rm s}$, and relative velocity of grain-to-grain collisions defined as $\mathit{u}_{\rm rel} = (\mathit{u}_{\rm br}^2 + \mathit{u}_{\rm turb}^2)^{1/2}$, where $\mathit{u}_{\rm br}$ and $\mathit{u}_{\rm turb}$ account for the Brownian and turbulence-induced local motion of dust grains, respectively.  In particular, $u_{\rm turb}$ has the biggest contribution to dust growth in our model and is expressed as \citep{2007OrmelCuzzi}
\begin{equation}
        u_{\rm{turb}} = \sqrt{{3 \alpha \over \mathrm{St}+\mathrm{St}^{-1}}} c_{\rm s},
    \label{turb_vel}
\end{equation}
where $c_{\rm s}$ is the adiabatic speed of sound, $\alpha$ is the parameter describing the strength of turbulence in the disk, and $\mathrm{St}$ is the Stokes number defined as $\mathrm{St}=t_{\rm stop} \Omega_{\rm K}$. Here, $\Omega_{\rm K}$ is the Keplerian angular velocity. We emphasize here that this definition of the Stokes number crucially depends on the assumption of a Keplerian disk. However, young disks may deviate from the Keplerian pattern of rotation. 
%This is confirmed by difficulty with the calculation of $\Omega_{\rm K}$ on the fly in our models (recall that we do not use sink particles and the mass of the protostar can only be calculated approximately).} 
We tried to use the local angular velocity  as a proxy of $\Omega_{\rm K}$ but that worked poorly in disks perturbed by spiral density waves and mass infall from the envelope. In this paper, we set $\Omega_K$ equal to that of a 0.05~$M_\odot$ central object. Considering that we limit our runs to the very early stages of star and disk formation, this should introduce only small errors to $u_{\rm{turb}}$. 
We also note that the Stokes number in the infalling envelope should be normalized to the free-fall time rather than to $\Omega_{\rm K}$. This, however, requires determining the disk-to-envelope interface on the fly and is subject to future code improvements.

The value of $\alpha$ is a free parameter in our model and it describes the strength of the magnetorotational instability (MRI) in the disk. Since the MRI can only be directly studied in focused high-resolution magnetohydrodynamics simulations \citep[e.g.,][]{2013Bai}, we use a parametric approach and set $\alpha=0.001$ as a constant of space and time. This value is also consistent with observations \citep[see][]{2023Rosotti}. 
%which corresponds to a significantly weakened MRI as suggested by non-ideal MHD simulations and efficient dust settling observed in T~Tauri disks  
We note, however, that the MRI may be fully suppressed with $\alpha\le 10^{-4}$  \citep{2013Bai,2022Dullemond} and this may have important consequences for dust dynamics and growth \citep{2018VorobyovAkimkin}. On the other hand, spiral density waves that develop in young protoplanetary disks may stir up grown dust via the process known as gravitoturbulence \citep{2018RiolsLatter_b}.  If gravitoturbulence modifies the value of $\alpha$, which in this case may become anisotropic \citep{2021BaehrZhu},  the effect on dust growth and pebble formation can be substantial \citep{Vorobyov2023a}. 

%The equations describing these quantities can be found in \citet{2018VorobyovAkimkin}.
%When calculating the volume density of dust, we take into account dust settling by calculating the effective scale height of grown dust  $H_{\rm d}$ via the corresponding gas scale height $H_{\rm g}$, $\alpha_{\rm visc}$ parameter, and the Stokes number \citep{Kornet2001}. 

As the maximum size of dust grains $a_{\rm max}$ begins to increase, part of the small dust population is converted to grown dust. This process is described by the conversion rate $S(a_{\rm max})$, which can be expressed as
\begin{equation}
    S(a_{\rm max}) = -\frac{\Delta\rho_{\mathrm{d,sm}}}{\Delta t},
    \label{growth:rate}
\end{equation}
where $\Delta\rho_{\mathrm{d,sm}}=\rho_{\mathrm{d,sm}}^{n+1}- \rho_{\mathrm{d,sm}}^{n}$ is the mass of small dust (per unit volume) converted to grown dust during one hydrodynamic time step $\Delta t$. The expression for $\Delta\rho_{\mathrm{d,sm}}$ is derived in \citet{Molyarova2021} and \citet{Vorobyov2022} assuming the conservation of dust mass and continuous dust size distribution across $a_{*}$. It can be written as follows
\begin{equation}
\label{final}
    \Delta\rho_{\mathrm{d,sm}} = \rho_{\mathrm{d,sm}}^{n+1}- \rho_{\mathrm{d,sm}}^{n} =
    \frac
    {
    \rho_{\rm d,gr}^n \int_{a_{\rm min}}^{a_*} a^{3-\mathrm{p}}da - 
    \rho_{\rm d,sm}^n \int_{a_*}^{a_{\mathrm{max}}^{\rm n+1}} a^{3-\mathrm{p}}da
    }
    {
    \int_{a_{\rm min}}^{a_{\mathrm{max}}^{n+1}} a^{3-\mathrm{p}}da
    },
\end{equation}
where the indices $n$ and $n+1$ denote the current and next hydrodynamic steps of integration, respectively. We note that the  drift of grown dust with respect to small dust (recall that small dust is dynamically coupled with gas) can result in the development of a discontinuity in the distribution function $N(a)$ at the interface $a_{*}$ between the small and grown dust populations. However, this may occur only if the drift timescales are shorter than the dust growth timescales, which is usually not realized in our models \citep[see fig. 8 in][]{Vorobyov2022}.

The dust growth in our model is terminated at the so-called fragmentation barrier \citep{2012Birnstiel}. We note that another important limiting mechanism -- the drift barrier -- is considered self-consistently via the solution of the dust dynamics equations. The maximum size up to which dust particles are allowed to grow due to the fragmentation barrier is defined as
\begin{equation}
\label{afrag}
a_{\rm frag} = \frac{\rho_{\rm g}u^2_{\rm frag}}{3 \alpha \rho_{\rm s}  v_{\rm th} \Omega_{\rm K}},
\end{equation}
where $u_{\rm frag}$ is a threshold value for the relative velocity of colliding dust grains, above which grain-to-grain collisions lead to destruction rather than to growth and $\Omega_{\rm K}$ is the Keplerian angular velocity.
%\citep{2018Blum}. 
%The effects of varying $u_{\rm frag}$ are discussed in Sect.~\ref{sec:param}.
When $a_{\rm max}$ exceeds $a_{\rm frag}$, the growth rate $\mathcal{D}$ is set equal to zero and $a_{\rm max}$ is set equal to $a_{\rm frag}$. We note that if the local conditions in the disk change so that $a_{\rm frag}$ drops below the current value of $a_{\rm max}$ (e.g., when temperature increases or gas density decreases), then we also set $a_{\rm max}=a_{\rm frag}$. This effectively implies that part of the grown dust with sizes in the $[a_{\rm frag}: a_{\rm max}]$ range is shattered via collisions and the process of dust conversion reverses, namely, grown dust is now converted to small dust. A more sophisticated approach that takes into account the growth and fragmentation probabilities, such as in \citet{Akimkin2020}, will be considred in a follow-up study. We note that the full solution of the Smoluchowski equation \citep[e.g.,][]{2019Drazkowska} and is computationally expensive for three-dimensional numerical simulations.

%The term $S(a_{\rm max})$ that enters the equations for the dust component is the conversion rate between small and grown dust populations.  We assumed that the distribution of dust particles over size follows the form given by Equation~\eqref{eq:dustdistlaw} for both small and grown populations. 
 
%Furthermore, the distribution is assumed to be continuous at $a_{*}$. However, drift of grown dust with respect to gas (and small dust by our assumption) can result in the development of a discontinuity in the distribution function $N(a)$ at the interface $a_{*}$ between the two dust populations. Our scheme is constructed so as to preserve continuity at $a_{*}$ by writing the conversion rate of small to grown dust in the following form:

\subsection{Nested grid layout}
Numerical simulations that follow self-consistently the transition from a collapsing cloud core to the stage of star and disk formation have to deal with vastly changing spatial scales. This can be achieved by using SPH codes \citep[e.g.,][]{2022Bate} or specifically designed grid-based codes. These latter use either the spherical coordinate system with radial and angular refinement toward the coordinate center and disk midplane \citep[e.g.,][]{2017MeyerVorobyov,2016Hosokawa,2023OlivaKuiper}, or 
adaptive meshes \citep[][]{2006Fromang,2020Hennebelle}, or  nested Cartesian grids \citep[e.g.,][]{2014Machida,2017Tomida}. The ngFEOSAD code adopted the latter approach. In comparison to spherical coordinate systems it does not suffer from the problem of singularity at the coordinate center and along the rotation axis, and is therefore more suitable for studying the formation of binary and multiple stars, which can freely move around and through the coordinate center. The angular momentum conservation test with the known analytic solution \citep{1980Norman} demonstrate good performance, see Sect.~\ref{Sect:angmom}.

We use 12 nested grids with the linear size of the outermost $m=1$ grid equal to 0.09~pc. The number of grid cells per coordinate direction of each nested grid is $N=64$. The effective numerical resolution in the inner 4.5~au is 0.14~au and it remains at a sub-au level up to 36~au. The forming disk in our runs is usually contained within 60--70~au. The vertical extent of the disk (from the midplane) is usually resolved by 6--7 grid zones up to a radial distance of 10~au and slightly deteriorates at larger radial distances. No mirror symmetry with respect to the $z=0$ plane is imposed.
%, allowing disk warps and related phenomena.

\subsection{Numerical solution method}

To solve the equation of hydrodynamics for the gas component, we employ the Godunov method with a combination of HLLC (Harten–Lax–van Leer contact) and HLL (Harten–Lax–van Leer)  Riemann solvers \citep{Toro2019}. We note that the gravity and friction forces are considered separately after the Godunov step. We also consider the equation for the total energy $E$ in the test problems but in the actual simulations we employ the polytropic equation of state, which allows us to eliminate the energy equation. A more accurate approach requires also solving for the radiation transfer equation, which is planned as a future update to the ngFEOSAD code. 

The equations of hydrodynamics for the gas component are written in the vector form
%Запишем уравнения гидродинамики и уравнения для пыли в векторной форме:
\begin{equation}
\frac{\partial  \bl{Q} }{\partial t} + \nabla \cdot \bl{F(Q)} = 0,
\label{eq:godunov}
\end{equation}
where $\bl Q$ is the vector of conservative variables $(\rho_{\rm g}, \rho_{\rm d,sm}, \rho_{\rm g} \bl{v}, E)$ and $\bl{F(Q)}$ is the flux of conservative variables $(\rho_{\rm g} \bl{v}, \rho_{\rm d,sm} \bl{v}, \rho_{\rm g} \bl{v} \otimes \bl{v} + \mathbb{I} P, (E+P)\bl{v}$). We note that here we also consider the continuity equation for small dust because it is dynamically coupled to gas. At each level of the nested grid, we  use the classical finite-volume Godunov scheme to solve Equation~(\ref{eq:godunov}), in which we place the values of conservative variables at the center of the cells. The fluxes of conservative quantities at the interfaces between the cells are the solution to the Riemann problem. In the case of an interface between the nested grids, the solution to the Riemann problem is averaged for larger cells. For smaller cells, the averaging is not required. For the isothermal gas the HLLC scheme is replaced with the HLL scheme to avoid the singularity in the values of gas pressure between the limiting characteristics. The time step is calculated using the Courant-Friedrichs-Lewy condition with the Courant number set equal to 0.3.
To increase the order of accuracy of the numerical method on smooth solutions and reduce dissipation at discontinuities, we use a piece-wise parabolic reconstruction of physical variables with integrals along the sound characteristics as parameters of the Riemann problem.
A detailed description of the piece-wise parabolic reconstruction, integration along characteristics, the choice of variables for reconstructions, and the rationale for using sound characteristics can be found in \citet{Kulikov2016,Kulikov2019,Kulikov2022}. 

%Для уравнений для пыли используется гидродинамический решатель с искуственным значением давления $p_{AD} = 10^{-3} p$, где $p$ -- давление газа. Это необходимо для устранения численных артефактов в виде carbuncle-эффектов. 

%Шаг по времени $\tau$ выичсляется из Courant-Friedrichs-Lewy condition. Для повышения порядка точности численного метода на гладких решениях и уменьшения диссипации на разрывах мы используем кусочно-параболическую реконструкцию физических переменных \citep{Kulikov2016}. Для этого в качестве параметров задачи Римана используются не кусочно-постоянные значения, а интегралы вдоль звуковых характеристик: $\rho_L = \rho_L(-a_L \tau), \rho_R = \rho_R(a_R \tau)$ -- плотности, $u_L = u_L(-a_L \tau), u_R = u_R(a_R \tau)$ -- продольной скорости, $v_L = v_L(-a_L \tau), v_R = v_R(a_R \tau), w_L = w_L(-a_L \tau), w_R = w_R(a_R \tau)$ -- двух поперечных скоростей, $p_L = p_L(-a_L \tau), p_R = p_R(a_R \tau)$ -- давления. Подробное описание построения кусочно-параболической реконструкции, интегрирования вдоль характеристик, выбор переменных для реконструкций и обоснование использования звуковых характеристик подробно изложены в работах \citep{Kulikov2016,Kulikov2019,Kulikov2022}.

The dynamics equations of the grown dust component can be written in the vector form as follows
\begin{equation}
\frac{\partial  \bl{Q_{\rm d}} }{\partial t} + \nabla \cdot \bl{F(Q_{\rm d})} = 0,
\label{eq:godunov_dust}
\end{equation}
where $\bl Q_{\rm d}$ is the vector of conservative variables $(\rho_{\rm d,gr}, \rho_{\rm d,gr} a_{\rm max}, \rho_{\rm d,gr} \bl{u})$ and $\bl{F(Q_{\rm d})}$ is the flux of conservative variables $(\rho_{\rm d,gr} \bl{u}, \rho_{\rm d,gr} a_{\rm max} \bl{u},  \rho_{\rm d,gr} \bl{u} \otimes \bl{u}$). We note that we introduced here an additional variable, namely, the product of $\rho_{\rm d,gr}$ and $a_{\rm max}$. The reason for that is explained later in this section.
The equations for the dust component are also solved by the Godunov method, similar to that used to solve the hydrodynamic equations. The key difference from the method of solving the gas hydrodynamic equations is the use of a one-wave HLL Riemann solver (details can be found in Appendix~\ref{Appendix:scheme}) owing to the specifics of the pressureless dust fluid.

Once the Godunov steps for the gas and dust dynamics are accomplished, we update the gas and dust momenta due to the gravitational acceleration $\nabla \Phi$. The gradient of the potential is discretized using the standard eight-point stencil for the central difference scheme. 
The gravitational potential $\Phi$ is computed using the following triple sum
\begin{equation}
    \Phi(x_i,y_j,z_k) = -G \sum_{i^\prime,j^\prime,k^\prime} { M(x_{i^\prime},y_{j^\prime},z_{k^\prime}) \over \sqrt{(x_i-x_{i^\prime})^2 + (y_j-y_{j^\prime})^2 + (z_k-z_{k^\prime})^2} } ,
    \label{sum:pot}
\end{equation}
where $M(x_{i^\prime},y_{j^\prime},z_{k^\prime})$ is the mass of gas and dust in a numerical cell ($i^\prime,j^\prime,k^\prime$) with coordinates ($x_{i^\prime},y_{j^\prime},z_{k^\prime}$), and $G$ is the gravitational constant. To accelerate the summation over all numerical cells we employ the convolution theorem \citep{1987Binney}. The extension of the convolution method to nested grids can be found in \citet{2023VorobyovMcKevitt}, see also Appendix~\ref{App:gravpot}. 

\begin{figure*}
\begin{centering}
\includegraphics[width=1\textwidth]{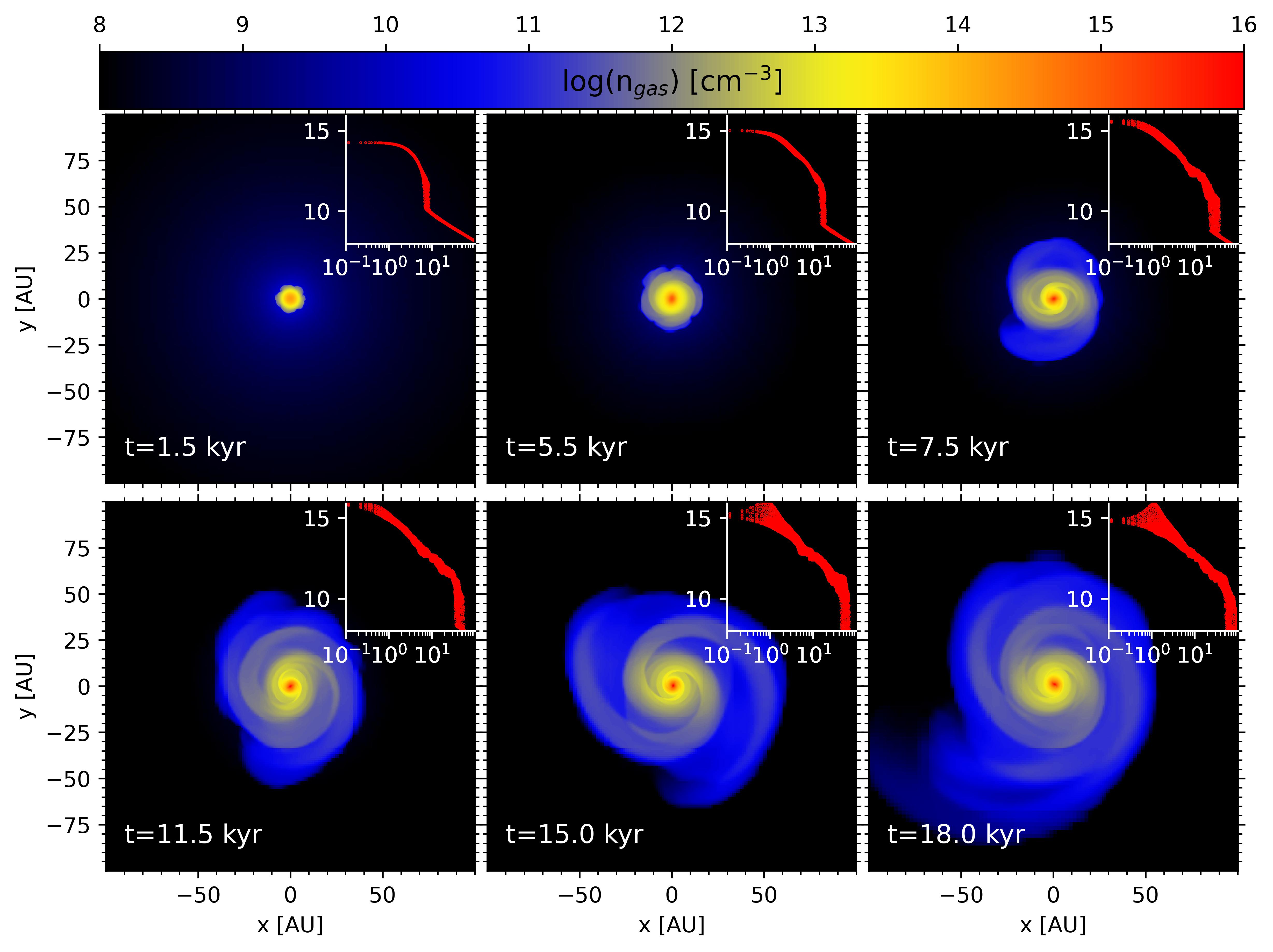}
\par\end{centering}
\caption{\label{fig:disk_evolution} Number density for different time instances in the inner 100$\times$100~AU$^2$ computational domain. The time is shown since the instance of disk formation. The inserts show the radial distribution of the gas number density for each time instance on the log scale. }
\end{figure*}

The update of the gas and dust velocities due to the friction force $\bl f$ between dust and gas is computed separately after the Godunov and gravity substeps using the following scheme \citep{2015LorenBate, 2018Stoyanovskaya}
\begin{equation}
\label{eq:ExpAll}
\bl{v}^{n+1}=\varepsilon D + E, \\ {\bl u}^{n+1}=-D+E, 
\end{equation}
\begin{equation}
\label{eq:ExpConst}
E=\displaystyle\frac{C_1}{\varepsilon + 1}, \\ 
D=\displaystyle\frac{C_2}{\varepsilon+1}\exp\left({-\displaystyle\frac{\varepsilon+1}{t_{\rm stop}}\Delta t}\right),
\end{equation}
\begin{equation}
\label{eq:ExpInit}
C_1={\bl v}^n+\varepsilon {\bl u}^n, \ \ C_2={\bl v}^n-{\bl u}^n,
\end{equation}
where, $\varepsilon=\rho_{\rm d,gr}/\rho_{\rm g}$ is the grown dust-to-gas ratio.
Equations~(\ref{eq:ExpAll})--(\ref{eq:ExpInit}) were derived assuming constant values of $t_{\rm stop}$ and $\varepsilon$ during a hydrodynamical time step $\Delta t$, in which case an analytic solution for the updated gas and dust velocities is possible.  The scheme performs well on the standard Sod shock tube and dusty wave test problems (see Appendix~\ref{Appendix:dusttests}). 
%We also tested a fully implicit scheme of \citet{Stoyanovskaya2018} and found that both schemes perform similarly.

The numerical solution of Equation~(\ref{eq:dustA}) for the time evolution of the maximum dust size $a_{\rm max}$ requires certain manipulations with its original form. In particular, the solution is split into two steps. First, the growth rate $\cal D$ is set to zero and the input from dust advection is computed. Then, the second term on the left-hand side is set to zero and the input from dust growth is considered. However, the second term on the left-hand side is not a divergence of a flux and, hence, cannot be easily incorporated into the Godunov conservative numerical scheme.  

Nevertheless, Equation~(\ref{eq:dustA}) can still be recast in the conservative form with the help of the continuity equation~(\ref{eq:cont_dust_grown}). The resulting equation reads as follows
%Setting ${\cal D}=0$ and $S(a_{\rm max})=0$ and multiplying Equation~(\ref{eq:dustA})  by $\rho_{\rm d,gr}$ and Equation~(\ref{eq:cont_dust_grown}) by $a_{\rm max}$,  we obtain after adding the resulting equations 
\begin{equation}
    {\partial \left( \rho_{\rm d,gr} \, a_{\rm max} \right) \over \partial t} + \nabla \cdot \left( \rho_{\rm d,gr} \, a_{\rm max} u \right) = \rho_{\rm d,gr} \, {\cal D} + a_{\rm max} \, S(a_{\rm max}).
    \label{eq:amax_recust}
\end{equation}
The left-hand side is the continuity equation for the product 
$\rho_{\rm d,gr} \, a_{\rm max}$ and we can use the same numerical method  as for the grown dust density to solve it. 
The product $\rho_{\rm d,gr} \, a_{\rm max}$ can then be updated to account for the combined source term shown on the right-hand side of Equation~(\ref{eq:amax_recust}) using the values of $\rho_{\rm d,gr}$ and $a_{\rm max}$ from the previous time step. The updated value of $a_{\rm max}$ is finally calculated using the updated value of $\rho_{\rm d, gr}$ obtained from Equation~(\ref{eq:cont_dust_grown}).  
We checked and confirmed that $a_{\rm max}$ remains unchanged when the growth term ${\cal D}$ is set to zero during the collapse and disk formation stage. 

\subsection{Hybrid parallelization approach}
ngFEOSAD is built using a combination of Coarray Fortran and OpenMP. Coarray Fortran extends modern Fortran, specifically Intel Fortran in our case, to distributed memory systems using the Single Program Multiple Data (SPMD) model, in which all multiprocessors execute the same program but on different data. This extension is realised through the addition of a datatype to arrays called a \texttt{codimension} and with a small number of additional directives. It is built on the Message Passing Interface (MPI) standard but only uses one-sided communication as opposed to pure MPI which is two-sided. This allows a multiprocessor to directly access the memory of another multiprocessor without requiring explicit cooperation from the target multiprocessor, something called Direct Memory Access (DMA). This capability simplifies the parallel programming process and can lead to more readable code because it reduces the need for complex coordination between multiprocessors. OpenMP, on the other hand, is used for the shared memory parallelism possible with multiprocessors and works within each distributed Coarray image, requiring only a further small number of additional directives. When combined, Coarray Fortran and OpenMP enable a highly optimized, parallel implementation of our key technique - the Cartesian nested grid.

In our design, each Coarray image, or separate program instance, performs calculations for a different subset of the domain, in our case different levels of the nested mesh. This setup means that computational tasks are efficiently spread across multiple multiprocessors given the balanced workload between images. It also means that memory transfers between distributed memory spaces are limited to only those necessary between nested meshes, something minimized by our mathematical methods for example in solving the gravitational potential \citep{2023VorobyovMcKevitt}, reducing one of the main computational overheads associated with distributed memory systems. Shared memory parallelism using OpenMP, providing each parallel thread with a single dedicated physical multiprocessor core, is then used to accelerate local computation.

Our models use 12 nested grids, each with a resolution of 64$^3$ corresponding to an effective minimum grid size of 0.14~au, deployed on the Vienna Scientific Cluster (VSC-5) which has two sockets per node with each socket connected to a multiprocessor with 64 physical cores. In our case, shared memory parallelism acceleration within a nested grid saturates beyond 32 cores, and so we assign two images per socket, resulting in four images per node and three nodes in total. This results in a very efficient, load-balanced and hardware-optimized domain decomposition. In such a configuration, simulations of the disk formation and evolution to the 18 kyr stage take only 12 days to perform and require only 384 cores, a very compact and still accelerated result.
Description of the step-by-step solution procedure and more details on the parallel realization using the Coarray Fortran technique can be found in Appendix~\ref{Appendix:coarrays}.

%The solution consists of three steps. First, we construct the quantity $\rho_{\rm d,gr} \, a_{\rm max}$ and solve this equation with the right-hand side set to zero using exactly the same algorithm as for the dust density. Then, we consider the input from dust growth and small-to-grown dust conversion by setting the divergence term to zero. Finally, the value of $a_{\rm max}$ is retrieved from the known value of $\rho_{\rm d,gr}$.

\subsection{Initial and boundary conditions}
\label{Sect:init}
For the initial conditions we consider a Bonnor-Ebert sphere with mass $M_{\rm cl} = 0.8~M_\odot$ and radius $R_{\rm cl}=8\times10^3$~au. The initial temperature is set equal to 10~K. The initial dust-to-gas mass ratio is 0.01. To initiate the collapse we give the cloud an initial positive gas and dust density perturbation of 30\%, namely, we increase the corresponding densities by 30\%. The ratio of rotational to gravitational energy is $\beta=0.5$\% and the ratio of thermal to gravitational energy is 70\%. Initially, most dust is in the form of small dust grains with a maximum size of $1~\mu$m.  For computational reasons, we also add a small amount of grown dust, about 10\% of the total dust mass budget, which has a maximum size of $a_{\rm max}=2$~$\mu$m. The cloud is initially submerged into an external medium with negligible density and velocity. The gas and dust dynamics is calculated for the entire computational domain, but the gas and dust velocities in the external medium are reset to zero after every time step. With this procedure, some mass and angular momentum may leak across the boundary between the Bonnor-Ebert sphere and the external medium. The magnitude of this effect does not exceed a few per cent of the total mass and angular momentum budget initially contained in the Bonnor-Ebert sphere and affects only the outer regions of the collapsing envelope.  

\begin{figure}
\begin{centering}
\includegraphics[width=1\columnwidth]{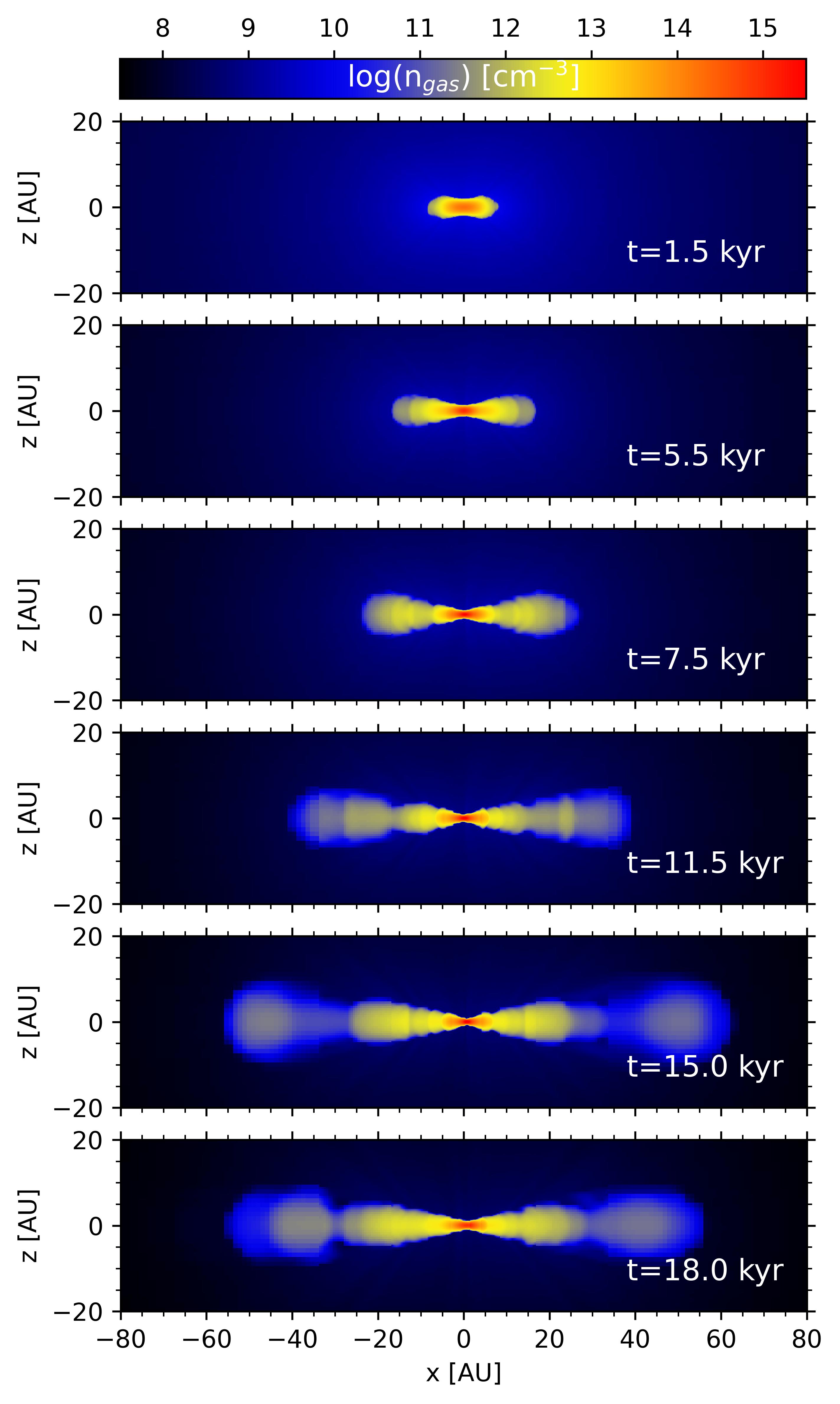}
\par\end{centering}
\caption{Vertical cuts showing the gas number density in the $x-z$ at $y=0$. The same time instances as in Figure~\ref{fig:disk_evolution} are considered.}
\label{fig:x-z}
\end{figure}

\begin{figure}
\begin{centering}
\includegraphics[width=\linewidth]{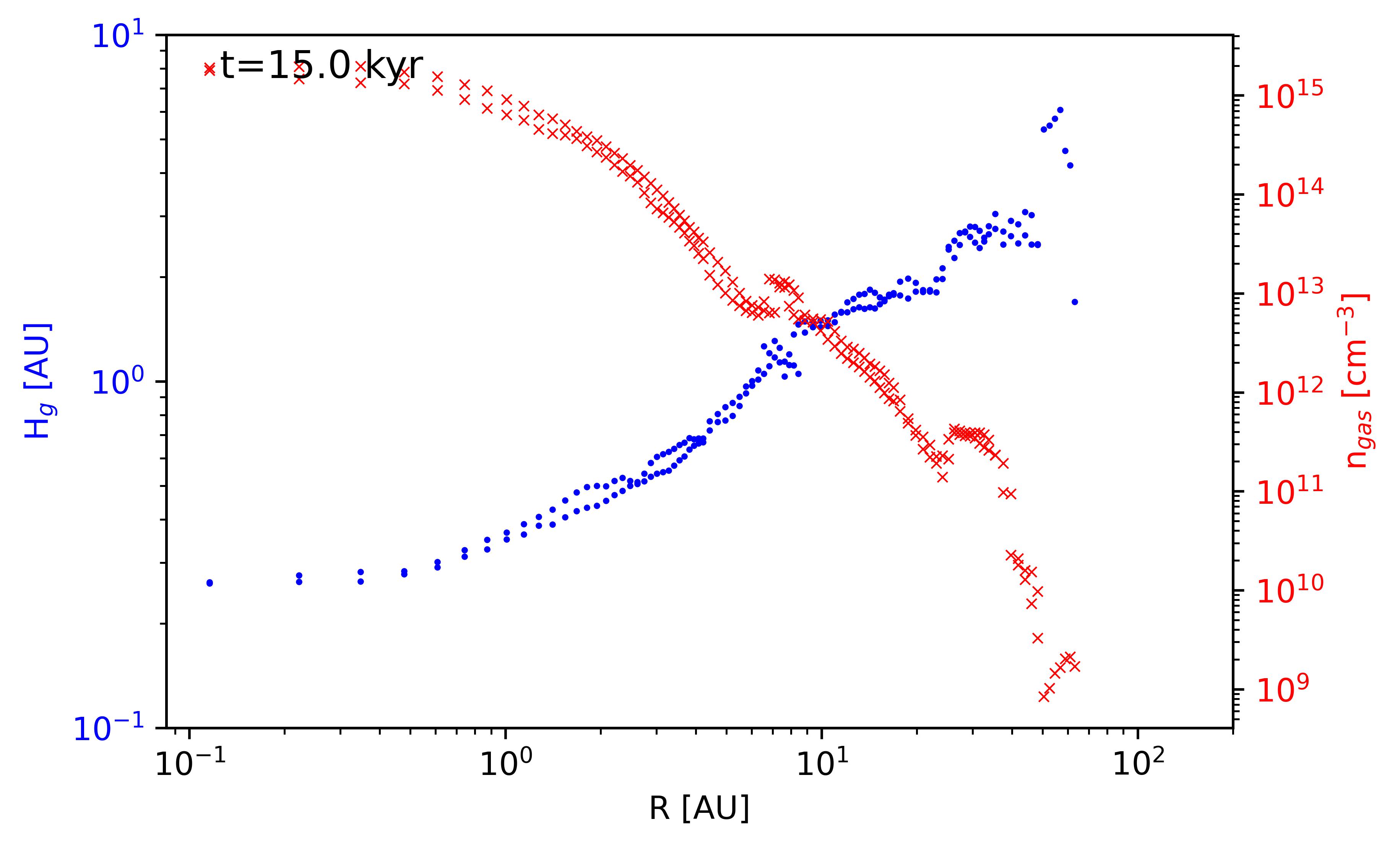} 
\par \end{centering}
\caption{Disk vertical scale height $H_{\rm g}$ (blue circles) and the gas number density in the midplane $n_{\rm gas}$ (red crosses) as a function of distance in both directions from the disk center along the $x$-coordinate.} 
\label{fig:Hg}
\end{figure}

\section{Results}
\label{sect:results}
In this section, we provide the results of our numerical simulations, focusing on the efficiency of dust growth, pebble formation, and dust settling in a young protoplanetary disk.

\subsection{Fiducial model.  Gas distribution}
Figure~\ref{fig:disk_evolution} presents the evolution of the gas number density $n_{\rm g}$ in the fiducial model in the disk midplane for 18~kyr after disk formation. A distinct two-armed spiral pattern forms after $\approx 5.0$~kyr. The calculation of the Toomre $Q$ parameter confirms that the disk is gravitationally unstable with $Q\approx 1$. The gas disk grows with time owing to accretion from the collapsing cloud and reaches a radius of $\approx 60$~au by $t=18$ kyr. 
The highest density is found in the disk center where the protostar is forming, though our numerical resolution is insufficient to follow the second Larson collapse to stellar densities.
The distribution of $n_{\rm g}$ as a function of distance from the disk center follows the $n_{\rm g} \simeq r^{-2.57}$ law. For a gravitationally unstable disk that settles in a state with $Q\sim 1$, the $n_{\rm g} \simeq r^{-3}$ law is expected. This mismatch is likely caused by the nonsteady character of the young disk and also by deviations from Keplerian rotation owing to a notable contribution from the massive disk to the stellar gravity.

%which can be expected  and $H_{\rm g}\propto r$

Figure~\ref{fig:x-z} displays the gas volume density distribution at the vertical cuts taken through the $x-z$ plane at $y=0$.  Interestingly, as the disk forms and grows in size, the vertical shape of the disk changes. At $t=1.5$ and 5.5~kyr the disk surface is rather smooth and the disk  thickness monotonically increases with distance from the disk center. However, already at $t\ge 11.5$~kyr the disk surface attains a wave-like shape. A comparison of Figures~\ref{fig:disk_evolution} and \ref{fig:x-z} suggests that the disk bulges at the position of the spiral arms and shrinks in the inter-armed regions.

To investigate how the disk thickness behaves with radial distance from the disk center, we calculated the vertical scale height of the gas disk as
\begin{equation}
    H_{\rm g} = {\Sigma_{\rm g} \over \sqrt{2\pi} \rho_{\rm g}(z=0)},
\end{equation}
where $\Sigma_{\rm g}$ is the surface density of the gas disk (the details on how we calculate this quantity are provided later in this section) and $\rho_{\rm g}(z=0)$ is the gas volume density in the disk midplane. This equation holds in the assumption of the local vertical hydrostatic equilibrium, which may not be fulfilled in a gravitationally unstable disk. Moreover, Riemann solvers may require higher vertical resolution than currently achieved in our modeling to maintain the vertical equilibrium and dust settling may exacerbate the problem further. Therefore, our calculations of $H_{\rm g}$ should be considered with caution, only as a first but still useful approximation. 

Figure~\ref{fig:Hg} presents the radial profile of $H_{\rm g}$ taken along the $x$-coordinate with $z=0$ and $y=0$ at a time instance of $t=15$~kyr. For convenience, the corresponding distribution of the gas number density $n_{\rm g}$ is also shown. The forming star occupies the inner few astronomical units and $H_{\rm g}$ is almost constant there. The value of $H_{\rm g}$   starts growing as the protostar transits to the disk at $r>2-3$~au. At $r=10$~au, the vertical scale height is $H_{\rm g}\approx 1.3$~au and at $r=50$~au $H_{\rm g}\approx 3$~au. 
Excluding the inner several au, $H_{\rm g}$ scales as $r^{0.6}$. 
The local peaks in $n_{\rm g}$ that are seen at $r\approx 8$~au and 30~au are the density enhancements in the spiral arms that cross the $y=0$ line at these locations. The values of $H_{\rm g}$ also have local maxima at the positions of the spiral arms. This analysis confirms that the disk bulges at the position of the spiral arms, at least in the polytropic approximation. This would make such spirals easier to detect observationally thanks to illumination by the stellar radiation. On the other hand, the more distant segment of the spiral arm may be hidden behind the closer segment, making the spiral structure more difficult to discern. Postpocessing with, for example, RADMC-3D is required to assess the detectability of the spirals in this case \citep[e.g.,][]{2012dullemond,2019MeyerVorobyov,2020Cadman}.

%also mention the aspect ratio ...

\begin{figure}
\begin{centering}
\includegraphics[width=\linewidth]{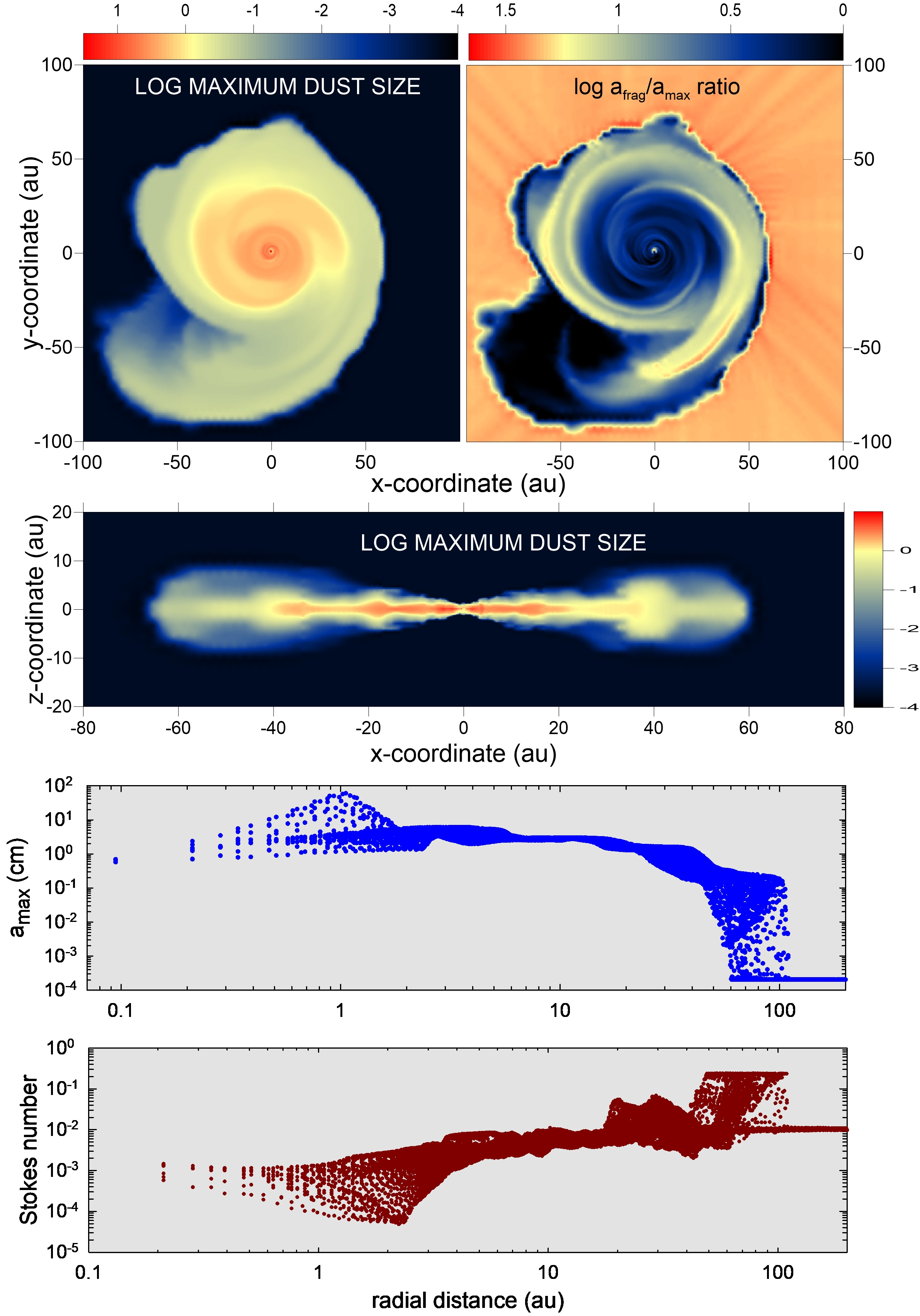} 
\par \end{centering}
\caption{Spatial and radial distribution of the maximum dust size $a_{\rm max}$ and the Stokes number at $t=18$~kyr. The top row consists of two sub-panels: the left-hand side subpanel showing $a_{\rm max}$ and right-hand side one presenting the ratio of the fragmentation barrier $a_{\rm frag}$ to the actual maximum size of dust grains $a_{\rm max}$, both in the disk midplane.
The second row is the vertical cut taken across the disk at $y=0$, third and bottom rows are $a_{\rm max}$ and $\mathrm{St}$, respectively, as a function of radial distance in the disk midplane in all directions. The scale bars for $a_{\rm max}$ are in log~cm, while the color bar for $a_{\rm frag}/a_{\rm max}$  is in log dimensionless units. 
} 
\label{fig:dustsize}
\end{figure}

\begin{figure}
\begin{centering}
\includegraphics[width=\linewidth]{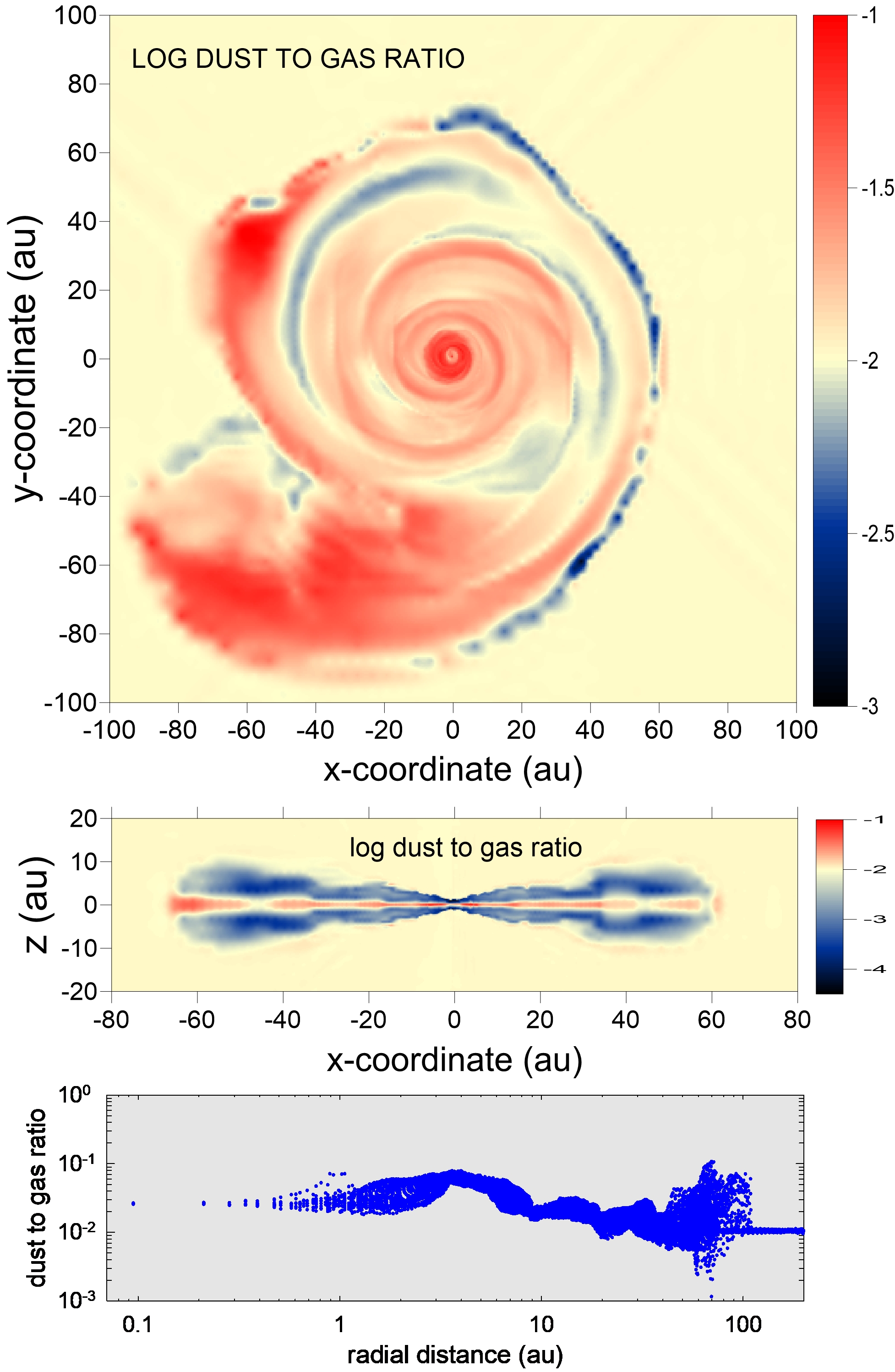} 
\par \end{centering}
\caption{Total dust-to-gas mass ratio at $t=18$~kyr. The top panel shows the two-dimensional distribution in the disk midplane, while the middle panel presents the data across the vertical cut taken at $y=0$. The bottom panel is the dust-to-gas ratio as a function of radial distance in the disk midplane in all directions. The color bars in the top and middle panels are in log units. } 
\label{fig:d2g}
\end{figure}

\subsection{Fiducial model. Dust distribution and size}

We focus now on the spatial distributions of the dust maximum size $a_{\rm max}$ and the total dust-to-gas mass ratio in the disk midplane and across the disk vertical extent. Figure~\ref{fig:dustsize} gives a detailed view of the dust maximum size at $t=18$~kyr after disk formation. 
The left-hand side subpanel in the first row and also the third row indicate that $a_{\rm max}$ in the disk midplane increases sharply from $\approx 1~\mu$m at the disk-envelope interface to a few millimeters in the disk outer regions, and then $a_{\rm max}$ grows slowly to centimeters in the inner several tens of astronomical units. The maximum values can reach decimeters at about 1~au, but we note that this region is occupied by the forming star and should be considered with care. 
The dominance of millimeter- and centimeter-sized grains imply that the spiral pattern may be detectable already at this yearly stage \citep{2020Cadman}.
We do not find any notable dust growth in the bulk of the infalling envelope because of its rarefied density and low turbulent velocity, see Eq.~(\ref{eq:dustrate}).  We acknowledge that our dust growth model is designed for  protoplanetary disks and not for infalling envelopes, for which certain modifications, for example, to the Stokes number and turbulent $\alpha$ parameter are required \citep{2023Commercon}. This is beyond the scope of the current paper. 

To estimate the efficiency of dust growth, we calculate the ratio of the fragmentation barrier $a_{\rm frag}$ to the maximum dust size $a_{\rm max}$ in the disk midplane. The right-hand side subpanel in the first row of Figure~\ref{fig:dustsize} demonstrates that dust growth has almost saturated in the inner disk where the ratio approaches unity (note that the scale bar is in log units). There is also a region of the lower spiral arm  where dust growth is saturated but this is due to low $a_{\rm frag}$ in this rarefied region. In general, however, the spiral arms are the disk regions where dust has not yet reached its upper values determined by collisional fragmentation and there is a potential for further dust growth there.

The distribution of $a_{\rm max}$ in the vertical cut taken along  $y=0$  (second row in Fig.~\ref{fig:dustsize}) is characterized by higher values closer to the disk midplane. This can be explained not only by dust settling but also by a more favorable dust growth environment in the disk midplane characterized by higher dust densities. %We note that this picture may change if a vertically variable turbulence efficiency is considered (recall that in the present work we set $\alpha$ to a spatially and temporally constant value of $10^{-3}$). 

Finally, the bottom row in Figure~\ref{fig:dustsize} presents the Stokes number as a function of radial distance from the forming protostar in the disk midplane. The Stokes number gradually increases outward. Interestingly, $\mathrm{St}$ reaches a value of 0.01 only beyond 10~au, which has important consequences for the spatial distribution of pebbles as discussed later in the text. Another important feature is that $\mathrm{St}$ never reaches values close to 1.0, as often assumed in dust dynamics simulations with a fixed dust size without growth \citep[e.g.,][]{2004RiceLodato,Boss2020}. These works considered preset isolated protoplanetary disks, more appropriate for the T Tauri phase, and found efficient dust accumulation in the spiral arms for dust grains with $\mathrm{St}\sim 1.0$ or dust size $1-10$~m. Our modeling indicates that such values of the Stokes number and dust size are not easily realizable in more realistic young protoplanetary disks formed via gravitational collapse, at least in the early 20~kyr of evolution. 

\begin{figure}
\begin{centering}
\includegraphics[width=\linewidth]{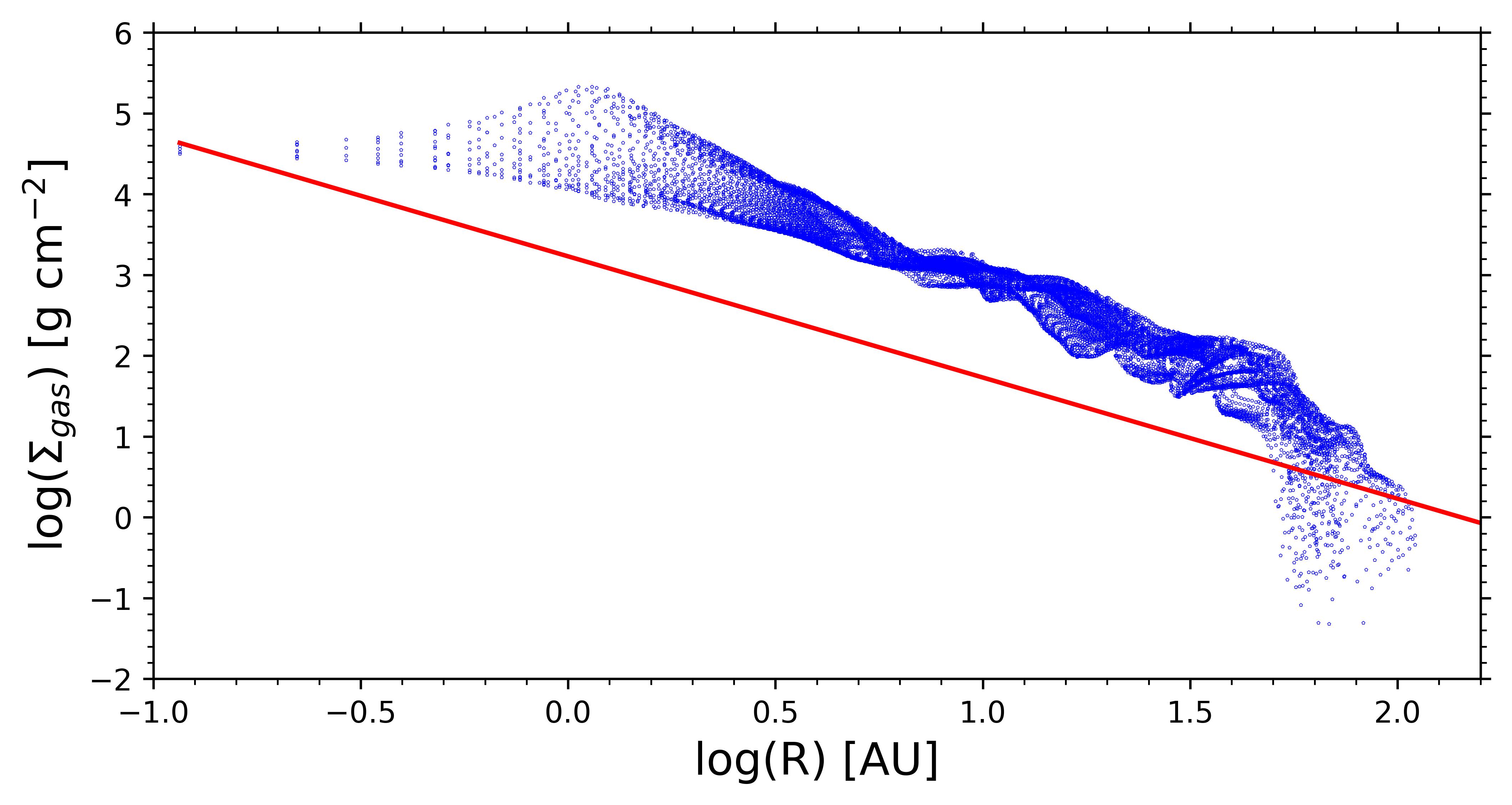} 
\includegraphics[width=\linewidth]{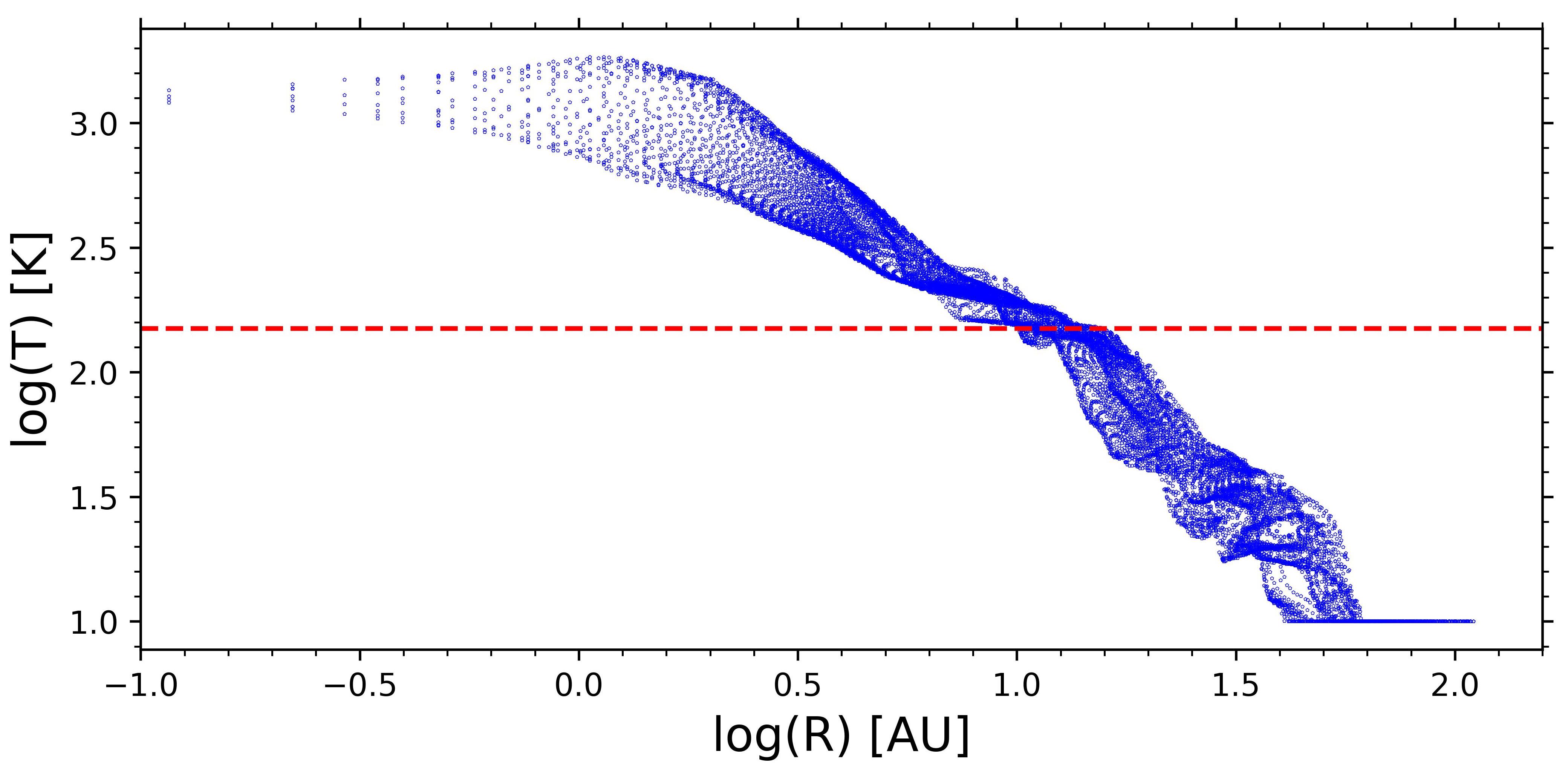} 
\par \end{centering}
\caption{Comparison of the gas surface density and temperature in our model with those of the MMSN. The blue dots in the top and bottom panels are the model gas surface densities and temperatures at t=18~kyr, respectively. The red line in the top panel shows the surface density profile for the MMSN. The dashed horizontal line in the bottom panel marks the gas temperature of 150~K, taken to be the water sublimation temperature.} 
\label{fig:MMSN}
\end{figure}

This difference can be understood if we compare the gas density and temperature of our young disk with those of the Minimum Mass Solar Nebular (MMSN). Figure~\ref{fig:MMSN} presents the radial profiles of the  gas surface density and temperature at 18~kyr. The surface density profile of the MMSN is taken from \citep{1981Hayashi} and has the following form 
\begin{equation}
    \Sigma_{\rm MMSN} = 1700 \left( {r \over \mathrm{1.0~au} } \right)^{-1.5}  [\mathrm{g~cm}^{-2}].
\end{equation}
Clearly, $\Sigma_{\rm g}$ is about an order of magnitude higher than $\Sigma_{\rm MMSN}$. The bottom panel indicates that the water snow line in our disk, taken to be at a radial position where the gas temperature equals 150~K, is beyond 10~au. This also indicates that our disk is warmer than that of the MMSN. Finally, we note that the protostellar mass at $t=18$~kyr is lower by a factor of 10 than the Sun. All these factors taken together lower the Stokes number for cm-sized dust particles below 0.1 in a young protoplanetary disk.

Figure~\ref{fig:d2g} presents the total dust-to-gas mass ratio at $t=18$~kyr. The top panel displays the corresponding values calculated in the disk midplane as 
\begin{equation}
\xi_{\rm d2g}(z=0) = {\rho_{\rm d,gr}(z=0)+\rho_{\rm d,sm}(z=0) \over \rho_{\rm g}(z=0) },
\end{equation}
The value of  $\xi_{\rm d2g}(z=0)$ throughout most of the disk midplane is elevated and can become as high as 0.1. At the same time, the spatial distribution of $\xi_{\rm d2g}(z=0)$ is highly spatially nonhomogeneous, which was also found to be the case for gravitationally unstable disks in our earlier thin-disk simulations \citep{2019VorobyovElbakyan}. The highest values of $\xi_{\rm d2g}(z=0)$ are found in between 1.0 and 10~au, and also in the outer disk parts. The latter regions, however, are characterized by quite low values of the gas and dust densities, and may be unfavorable from the point of view of planet formation. 

The middle panel in Figure~\ref{fig:d2g} presents the dust-to-gas ratio calculated across the disk vertical extent at $y=0$. The plot
reveals a strong vertical segregation in the dust-to-gas ratio, with the highest values found near the midplane and lowest in the upper/lower disk atmosphere. The disk midplane is enhanced in dust relative to the fiducial value of $\xi_{\rm d2g}=0.01$, while the disk atmosphere is strongly depleted in dust.  These variations are caused by dust growth and settling toward the disk midplane. Interestingly, there is a mild enhancement in $\xi_{\rm d2g}$   just above/below the disk surface. This enhancement is caused by grown dust falling in faster than gas in the envelope (recall that the prestellar core in our setup contained a small amount of grown dust with $a_{\rm max}=2~\mu$m, see Sect.~\ref{Sect:init}). Dust is pressureless and falls in on a free-fall time scale. Gas falls in slower because it has substantial pressure support against gravity, with the ratio of thermal to gravitational energy set equal to 70\% in the prestellar core, much higher than the corresponding ratio for rotational energy, 0.5\%.

%This grown dust component is indeed decoupled from gas because of very small gas densities in the envelope. As a result, there is a mismatch between the infall timescales of gas and grown dust. Grown dust is infalling on a classical free-fall time tff~ sqrt{1/rho}, but gas is infalling slower. This is because our initial cloud core has a ratio of thermal to gravitational energy of 70\%, much higher than that of rotational to gravitational energy. This is typical for prestellar cores. Therefore, there is substantial contribution from energy per unit volume (pressure) to the total energy budget and the gas pressure decelerates the infall of gas with respect to grown dust. } 

%which is caused by the dust drift relative to gas in the infalling envelope. The timescales considered are still too short to produce notable dust depletion in the bulk of the envelope and strong dust enhancement in the vicinity of the disk-to-envelope interface, but the effect may become more pronounced with time. 

The bottom panel presents the values of $\xi_{\rm d2g}(z=0)$ as a function of the radial distance from the disk center. The plot confirms that the dust-to-gas ratio in the disk midplane is enhanced through most of the disk extent, except for the outermost regions where its value may drop below 0.01. The peak values approaching $\xi_{\rm d2g}(z=0)=0.1$ are found at radial distances from several to ten astronomical units. This dust concentration is also revealed as a dust ring in the top panel of Figure~\ref{fig:d2g}.

\begin{figure}
\begin{centering}
\includegraphics[width=1\columnwidth]{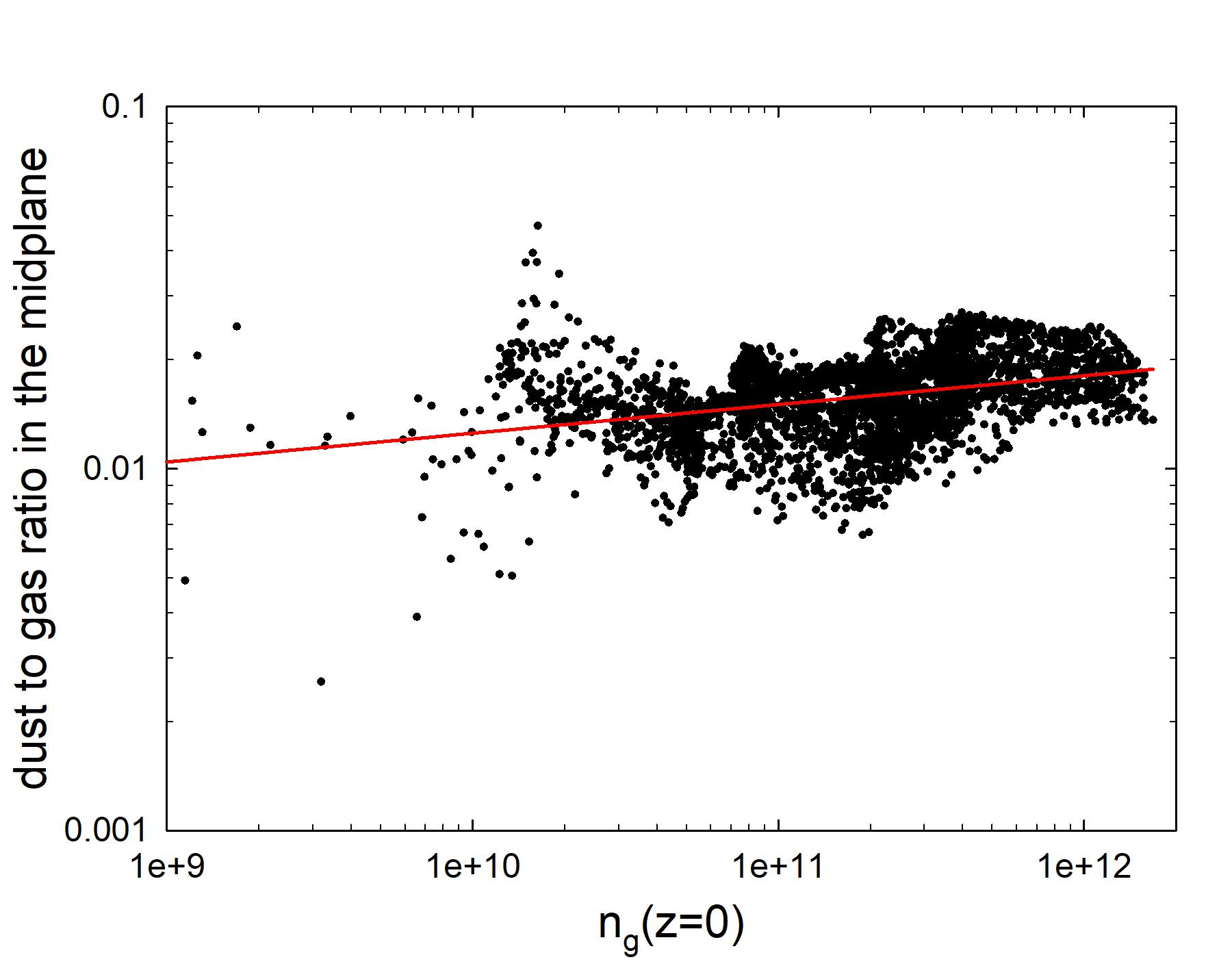}
\par\end{centering}
\caption{Total dust-to-gas mass ratio $\xi_{\rm d2g}$ as a function of the gas number density $n_{\rm g}$. All data are for the disk midplane and for a radial annulus 30--50~au capturing the most prominent spiral arm. The red line is the power-law best fit to the model data.}
\label{fig:spiral}
\end{figure}

\begin{figure}
\begin{centering}
\includegraphics[width=1\columnwidth]
{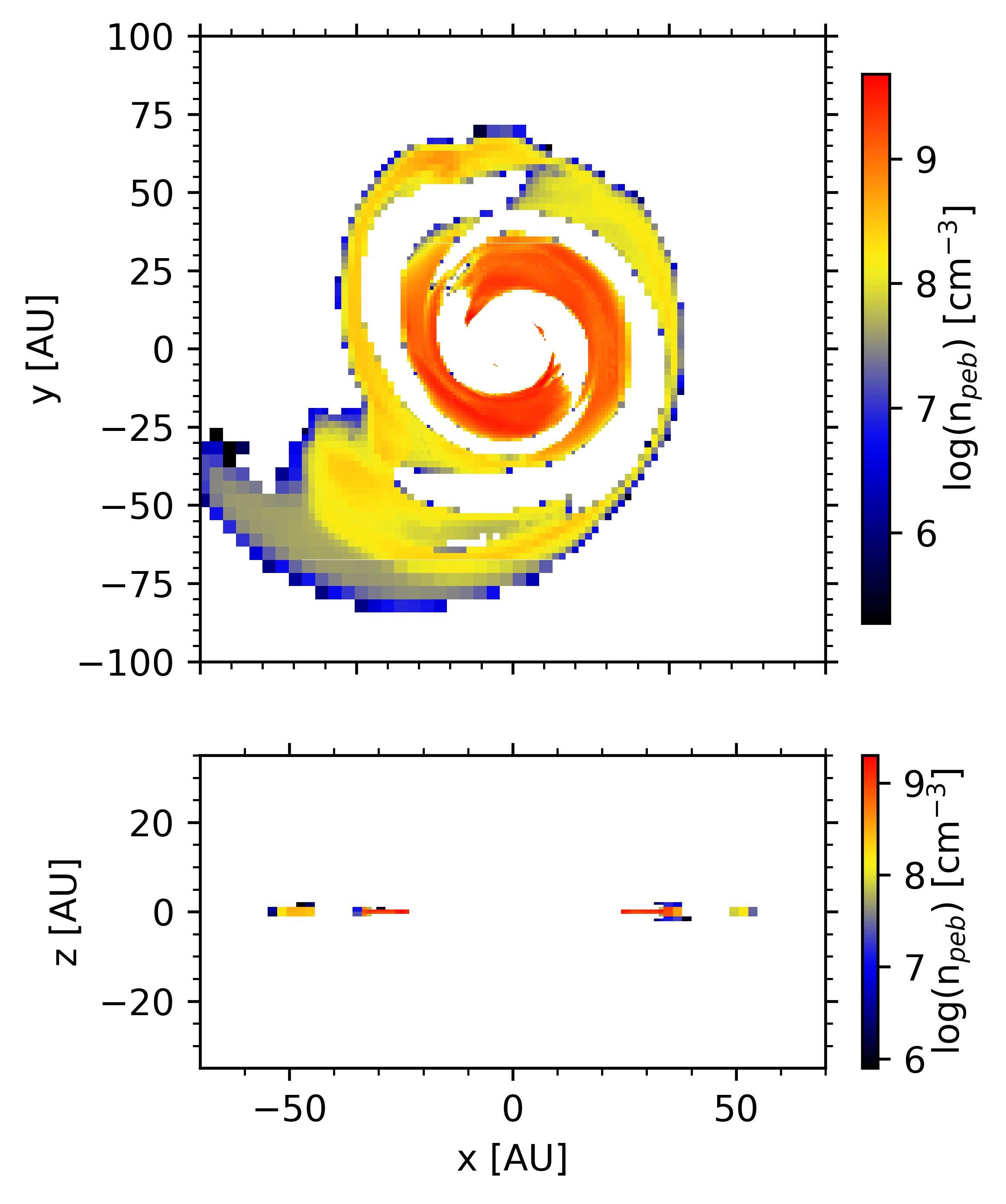}
\par\end{centering}
\caption{\label{fig:pebbles} Spatial distribution of pebbles in the disk at $t=18.1$~kyr. The top panel presents pebbles in the disk midplane, while the bottom panel -- across the disk vertical cut taken at $y=0$.}
\end{figure}

To probe if spiral arms can efficiently accumulate dust in our model, we take a disk snapshot at t=15~kyr with a well developed two-armed spiral pattern. Spiral arms are the disk regions with the enhanced gas density, and we may expect a positive correlation between $\xi_{\rm d2g}$ and $n_{\rm g}$ in the disk midplane if the accumulation of dust is present in the arms. We cut a radial region between 30 and 50~au to remove disk regions where spiral arms are less prominent. Figure~\ref{fig:spiral} displays the corresponding relation in the disk midplane and the red line plots the power-law best fit to the data of the form
\begin{equation}
    \xi_{\rm d2g} = 10^{-2.7} \, n_{\rm g}^{+0.08} [\mathrm{cm}^{-3}].
\end{equation}
The positive correlation takes place but is rather weak. The lack of strong dust concentration toward spiral arms was also reported in \citet{2018VorobyovAkimkin} and \citet{Vorobyov2022}, and hence this phenomenon is not an artifact of the thin-disk simulations as was suggested in \citet{Boss2020}. We think that the dust concentration in spiral arms is hindered by a rather low Stokes number in young disks. The nonsteady character of the spiral pattern, with the arms changing its form on timescales of a few kyr, may also work against strong dust concentrations. Our modeling emphasizes the need for realistic disk formation and dust growth models  to consider dust accumulation in spiral arms. 

\subsection{Fiducial model. Pebbles}
\label{Sect:pebbles}
It is interesting to consider the distribution and time evolution of pebbles in the disk, since they may play a crucial role in the growth of protoplanetary cores \citep{2012Lambrechts,2016Ida,2017JohansenLambrechts}. Following \citet{Vorobyov2023a},  we define pebbles as dust grains with radius greater than or equal to $a_{\rm peb}^0=0.5$~mm and $\mathrm{St}\ge \mathrm{St}_0 = 0.01$.  More specifically, we calculate the minimum pebble size as 
\begin{equation}
    a_{\rm peb}^{\rm min} = \max(a_{\rm peb}^0, a_{\rm St}^{0}),
\end{equation}
where $a_{\rm St}^0$ is the size of dust grains that have the Stokes number equal to $\mathrm{St}_0$ for the local disk conditions, $a_{\rm St}^0=a_{\rm max} \mathrm{St}_0 / \mathrm{St}$. We then search the disk locations where these conditions on the dust size and Stokes number are fulfilled. Noting that pebbles share the same size distribution with grown dust, the mass of pebbles  in each computational cell can be calculated knowing the mass of grown dust, the slope of the dust size distribution (fixed at $p=3.5$ in our work), the minimum and maximum sizes of grown dust, $a_\ast$ and $a_{\rm max}$, respectively, and the minimum size of pebbles, $a_{\rm peb}^{\rm min}$ \citep[see][for details]{Vorobyov2023a}.

Figure~\ref{fig:pebbles} shows the spatial distribution of pebbles at $t=18$~kyr. Clearly, pebbles are distributed unevenly across the disk midplane. Such a patchy spatial distribution of pebbles is also found at earlier evolution times. There are regions where pebbles are entirely absent, and most interestingly, in the inner 10~au. Dust can grow to centimeters there but the Stokes number happens to be lower than 0.01 because of the rising gas temperature and density in the innermost disk regions, see Eq.~(\ref{tstop}). This finding is in agreement with our previous work using thin-disk simulations \citep{Vorobyov2023a}, but we note that decreasing the turbulent $\alpha$ from $10^{-3}$ (as in the current work) to $10^{-4}$ may have a strong impact on the pebble distribution in the disk.   We also found that pebbles are concentrated to the midplane, which is a consequence of dust settling. 

\begin{figure}
\begin{centering}
\includegraphics[width=1\columnwidth]{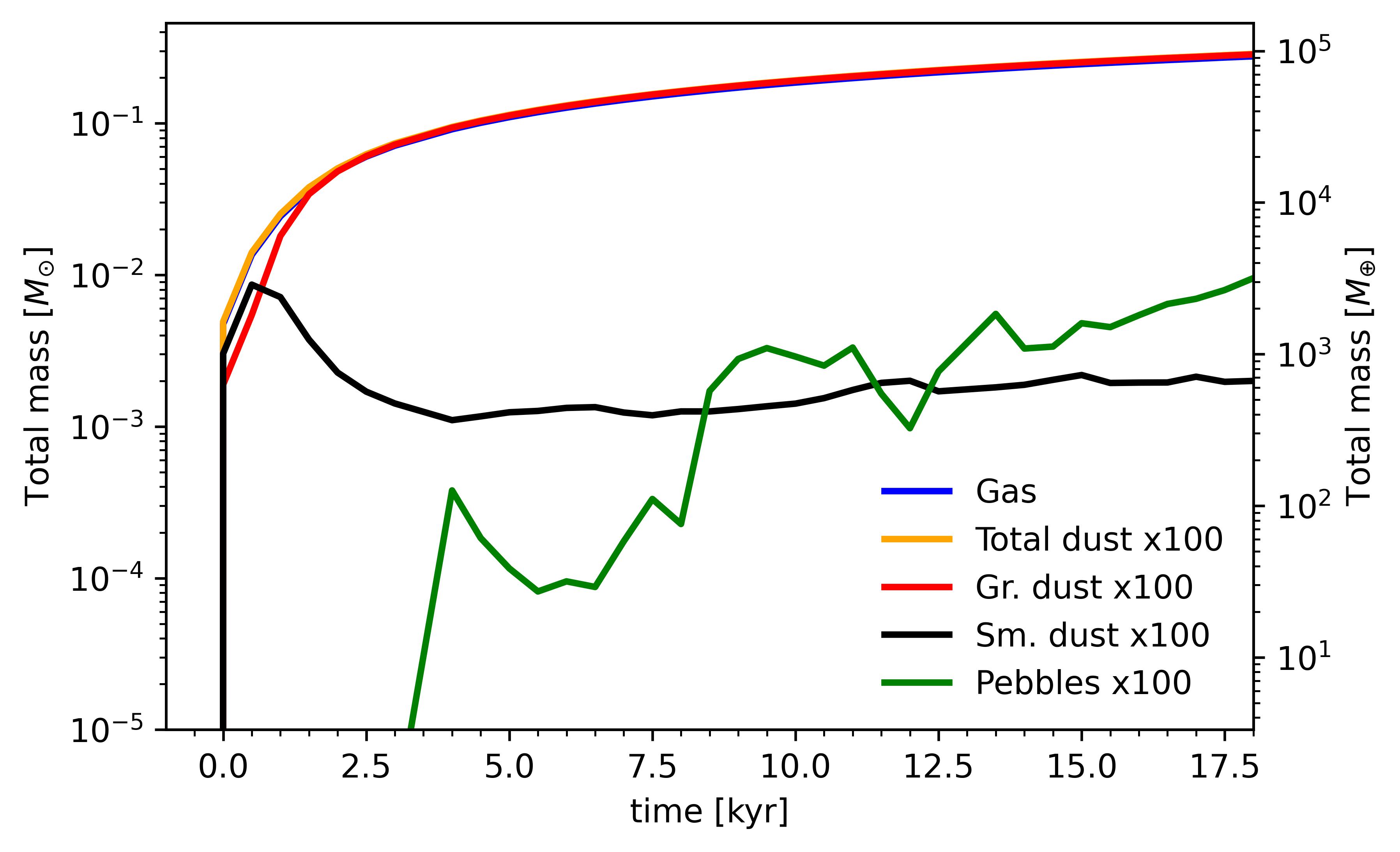}
\par\end{centering}
\caption{\label{fig:pebble-time} Time evolution of the total gas, \textbf{total} dust, \textbf{grown and small dust,} and pebble masses in the disk. Note that the dust and pebble masses are scaled up by a factor of 100.}
\end{figure}

Figure~\ref{fig:pebble-time} presents the time evolution of the total pebble mass in the disk as it forms and evolves. The total masses of gas and dust (both small and grown) in the disk are also shown for comparison. We note here that we do not distinguish between the disk and the forming protostar, so that the mass of gas and dust include both. On the other hand, pebbles are absent in the inner several astronomical units and the mass of pebbles is that contained in the disk only.
We confirm with 3D simulations that dust growth proceeds very fast, a conclusion also made in our earlier thin-disk simulations \citep{2018VorobyovAkimkin}. 
The first generation of pebbles appears after 2.5~kyr from the disk formation instance and the mass of pebbles reaches 31.7~$M_\oplus$ by the end of our simulations. It can be noted that the mass of pebbles is not growing steadily but there are episodes when its total mass decreases. This is caused by the varying local disk conditions in which pebbles find themselves as they drift through the disk. In particular, the Stokes number may decrease below the critical value of 0.01 because of the locally rising gas density and/or temperature. Also, the maximum size of dust grains may decrease if the local fragmentation barrier decreases. Overall, however, the pebble mass increases with time and appears to saturate at $\approx 31.7$~$M_\oplus$. 
 Given the uncertainties with the definition of the Stokes number in young disks, we also investigated how the pebble mass would change if $\mathrm{St}_0$ is varied by a factor of two. The corresponding mass of pebbles at $t=18$~kyr decreased/increased approximately by the same factor. 
Finally, we note that the mass of small dust is much lower than that of grown dust throughout most of the considered disk evolution period, and by the end of simulations becomes lower than that of pebbles.

To assess the efficiency of grown dust settling, we plot the radial distribution of the following ratio
\begin{equation}
    {H_{\rm d} \over H_{\rm g}} \simeq {\Sigma_{\rm d,gr} \over \Sigma_{\rm g}} \times \left( \xi_{\rm d2g}(z=0) - { \Sigma_{\rm d,sm} \over \Sigma_{\rm g} } \right)^{-1} = {\Sigma_{\rm d,gr} \over \Sigma_{\rm g}}{\rho_{\rm g} \over \rho_{\rm d,gr}}.
\end{equation}
%where $\Sigma_{\rm d}=\Sigma_{\rm d,sm}+\Sigma_{\rm d,gr}$ is the total dust surface density obtained by integrating the corresponding volume densities of dust over the vertical extent of the disk. \textbf{where $\xi_{\rm d,gr2g}(z=0)$ is the ratio of grown dust to gas at the midplane.} 
This equation is derived assuming Gaussian distributions of gas and dust volume densities in the vertical direction, which is an acceptable approximation, and assuming that small dust has the same vertical scale height as that of gas. To calculate the disk surface density, we adopted the disk tracking conditions outlined in \citet{2012Joos}. In particular, we use the following criteria to determine if a particular computational cell belongs to the disk:
 \begin{itemize}
     \item the gas rotational velocity must be faster than twice the radial infall velocity, $v_{\phi}>2v_{\rm r}$,
     \item the gas rotational velocity must be faster than twice the vertical infall velocity, $v_{\phi}>2v_{\rm z}$,
     \item gas must not be thermally supported, $\frac{\rho v_{\phi}^2}{2}>2P$,
     \item the gas number density must be higher than $10^9$~cm$^{-3}$.
 \end{itemize}
If any of the first three conditions fail, a particular grid cell is not qualified as belonging to the disk. The forth condition must always be fulfilled.

\begin{figure}
\begin{centering}
\includegraphics[width=1\columnwidth]{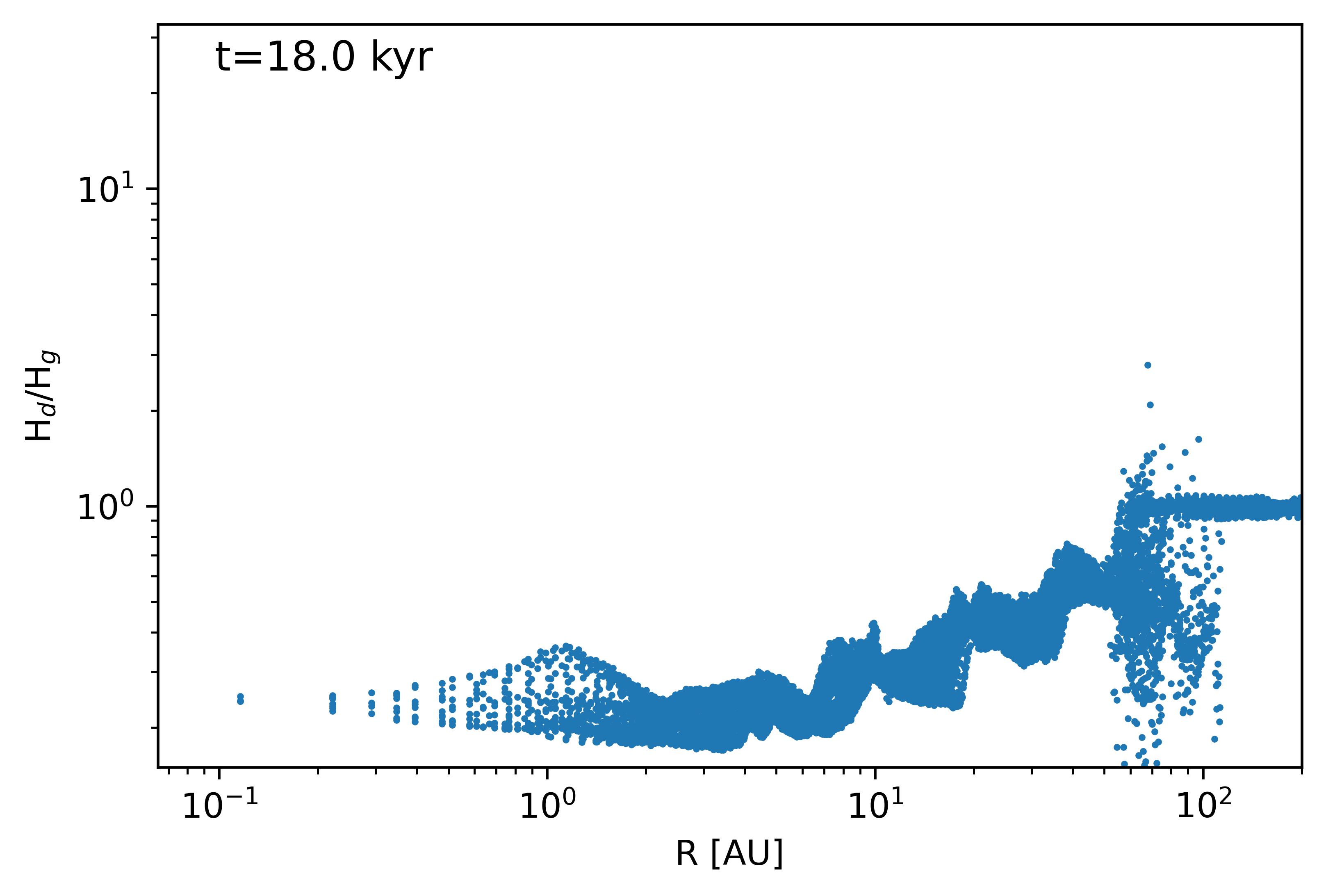}
\par\end{centering}
\caption{\label{fig:HdHg} Dust settling efficiency as a function of radial distance. Shown is the ratio of the grown dust-to-gas scale heights $H_{\rm d} / H_{\rm g}$ at $t=18.1$~kyr. }
\end{figure}

Figure~\ref{fig:HdHg} displays the ratio of the scale heights $H_{\rm d} / H_{\rm g}$ at $t=18$~kyr. Grown dust settling is present across the entire disk extent, being strongest in the inner 10~au, where $H_{\rm d} / H_{\rm g} \simeq 0.2$. At larger distances, however, dust settling weakens to $H_{\rm d} / H_{\rm g} \simeq 0.4-0.5$, most likely owing to the perturbing influence of spiral density waves, as was shown in \citet{2020Riols}. Our values (0.4--0.5) are still lower than those quoted by these authors ($\ge 0.6$), which can be attributed to different simulation techniques. We model the development of spiral arms in global disk formation simulations, while \citet{2020Riols} do that for the closed-box simulations. The regions with  $H_{\rm d} / H_{\rm g} > 1.0$  are caused by strong perturbations in the flow at the envelope to disk transition, which is a transitory effect.

\subsection{Effects of lower $\beta$}
\begin{figure}
\begin{centering}
\includegraphics[width=1\columnwidth]{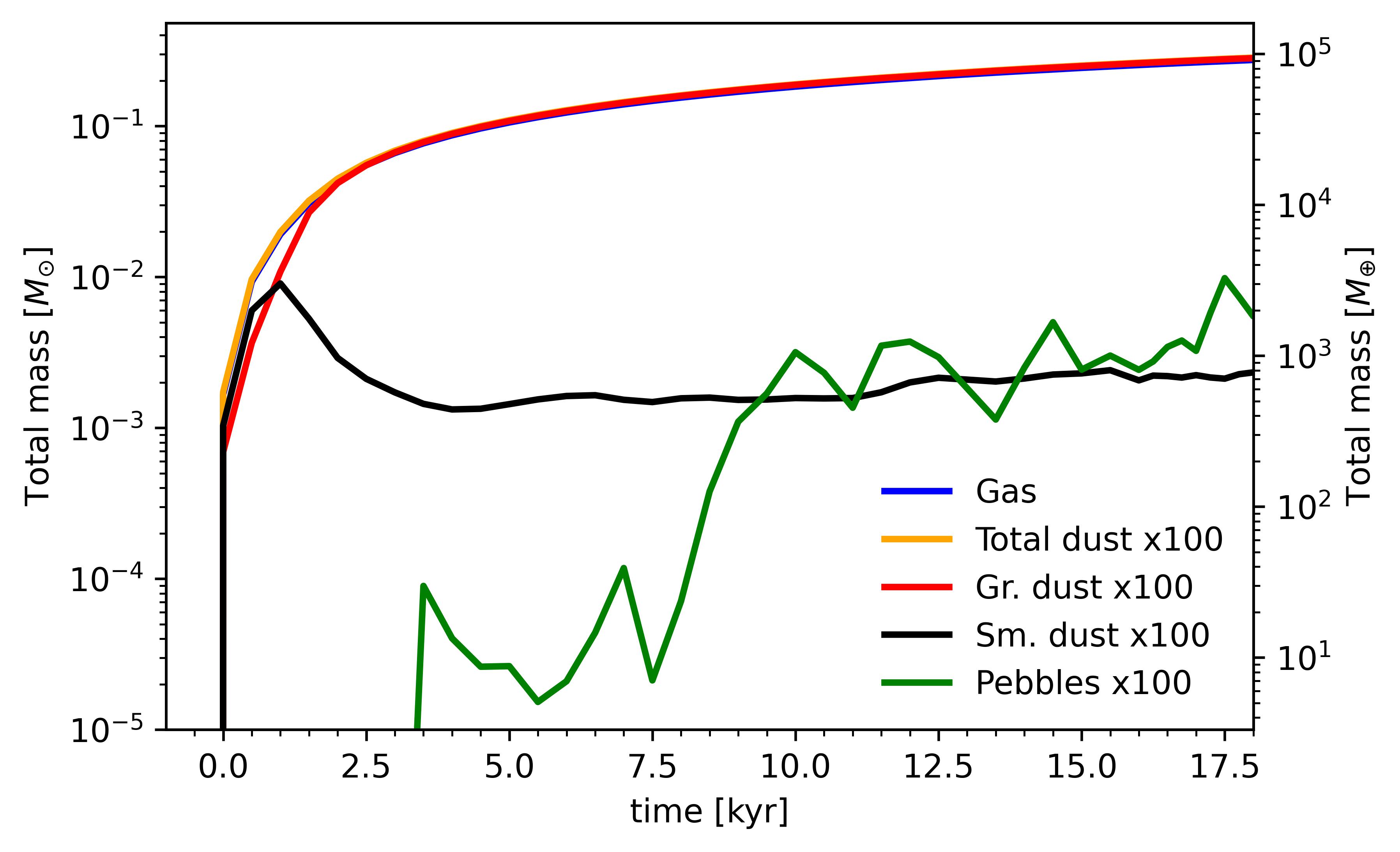}
\par\end{centering}
\caption{\label{fig:pebble-time_lb} Similar to Fig.~\ref{fig:pebble-time} but for the model with half the rotation velocity of the cloud core.}
\end{figure}

\begin{figure}
\begin{centering}
\includegraphics[width=1\columnwidth]{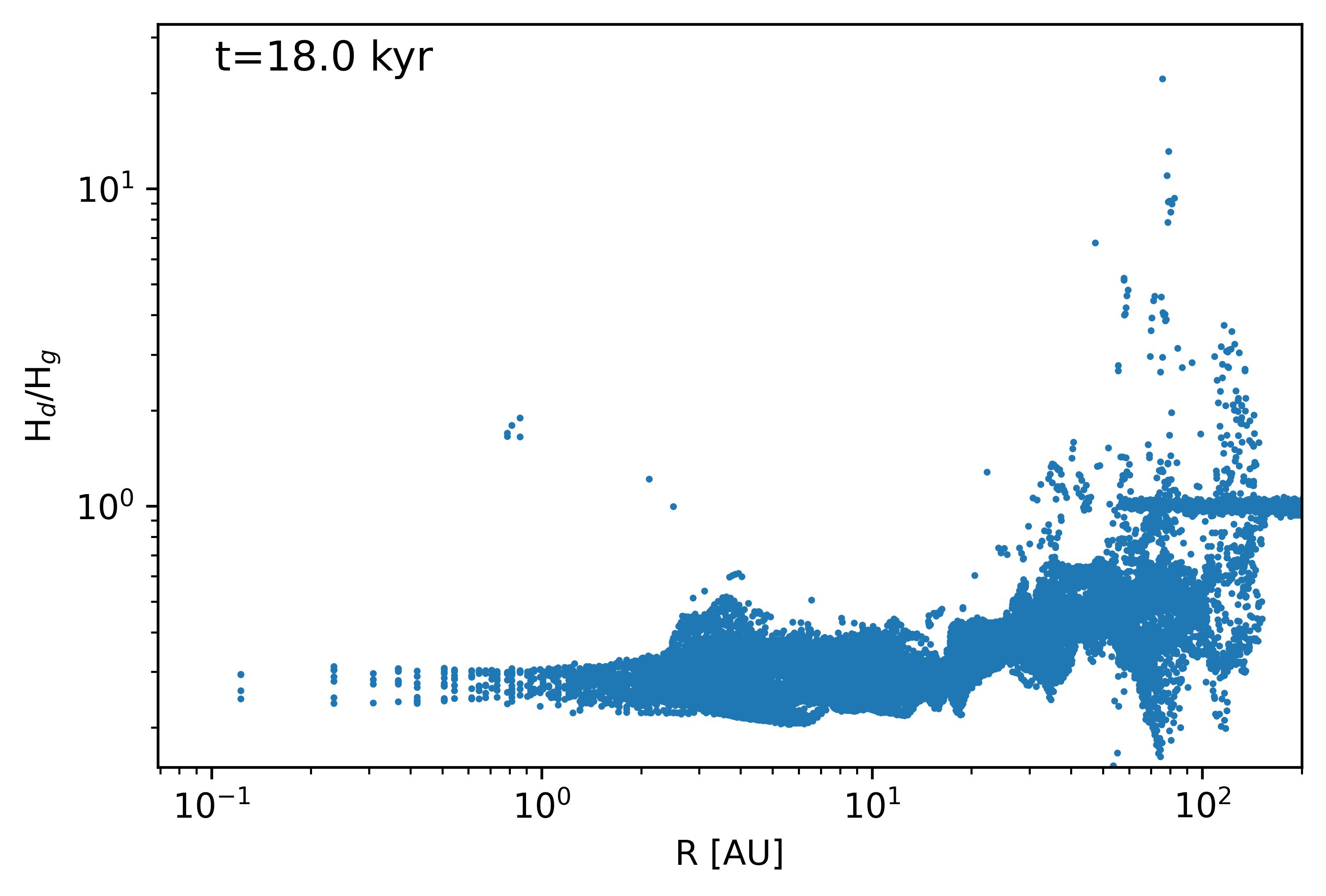}
\par\end{centering}
\caption{\label{fig:HdHg_lb} Similar to Fig.~\ref{fig:HdHg} but for the model with half the rotation velocity of the cloud core.  }
\end{figure}

We have also considered the collapse of a cloud core with half the rotation velocity as compared to the fiducial model. The disk in this model forms about 3~kyr later, but in general the disk evolution in both models follows a similar path. In Figures~\ref{fig:pebble-time_lb} and \ref{fig:HdHg_lb} we show the time evolution of the disk and pebble masses, as well as the radial distribution of the $H_{\rm d}/H_{\rm g}$ ratio. The mass of pebbles is a bit lower at the same time instance after the disk formation, t=18.0~kyr, and is equal to 18.3~$M_\oplus$. The time evolution of the total pebble mass is also nonmonotonic, which is a consequence of the spatially patchy distribution of pebbles across the disk. Dust settling in the low-rotation model is slightly more efficient, likely because the spiral density waves are weaker in this model. The difference, however, is marginal.

\section{Model caveats and future developments}
\label{sect:caveats}
This is our first step toward fully three-dimensional numerical hydrodynamics modeling of the formation of protoplanetary gas-dust disks, which builds upon our previous extensive studies of long-term gas-dust disk evolution using the thin-disk approximation \citep[e.g.,][]{2018VorobyovAkimkin,Molyarova2021,Vorobyov2023a}. 
Further developments of the ngFEOSAD code are needed to improve the realism of simulations. The barotropic approach is to be replaced with more realistic radiative transfer simulations \citep[e.g.,][]{2017FlockNelson} and magnetic fields have to be introduced to better describe the disk dynamics and flow in the envelope and the disk \citep[e.g.,][]{2020Tsukamoto,2020Lebreuilly,2023Koga}.
The simplified calculations of the thermal balance may affect the dust growth rate and the fragmentation velocity in our models, which in turn may have an impact on the spatial distribution and mass of pebbles in the disk. The lack of magnetic field may affect the dynamics of both gas and dust, with nonmagnetized disks being more massive and extended compared to their magnetized counterpart.
Dust growth is sensitive to ice mantles \citep{Okuzumi2016} and the dynamics of volatiles is to be considered together with dust evolution to accurately capture this effect \citep{Molyarova2021}.
A careful consideration to the sink cell model as a proxy of the forming star has to be given to avoid possible numerical artifacts \citep[][see also Appendix~\ref{sect:app-global}]{2014Machida,2019VorobyovSkliarevskii,2020Hennebelle}. The dust fragmentation barrier, dust settling, and dust growth depend on the $\alpha$ parameter, and  choosing a value other than $10^{-3}$ (adopted here) may have a strong impact on the dust and pebble distribution in the disk. Moreover, protoplanetary disks are unlikely to have spatially and temporally constant values of the turbulent $\alpha$ parameter and a more accurate model that considers, for example, luminosity bursts triggered by the magnetorotational instability is to be developed \citep[e.g.,][]{Bae2014,2020VorobyovKhaibrakhmanov}. 

Better dust growth models that include more dust size bins and finally transit to the solution of Smoluchowski equation are needed \citep[e.g.,][]{2019Drazkowska}, although this might be very challenging in 3D simulations from the numerical point of view. The use of two bins forced us to consider the dynamics of dust with the maximum size because they are the main dust mass carriers However, the dynamics of dust grains with smaller size is also important as they may substantially contribute to the dust opacity and, hence, to the disk thermal balance. Finally, we plan to perform higher resolution simulations with the base number of grid cells $N=128^3$ when the code is updated to use the GPUs for the gravitational potential solver. The current resolution with $N=64^3$ is insufficient to fully resolve the vertical extent of the gas disk, not to mention that of the grown dust given that dust settling is present, which may affect our conclusions on the dust settling efficiency.

\section{Conclusions}
\label{sect:conclude}
We have developed an original three-dimensional numerical hydrodynamics code on nested meshes, ngFEOSAD, which computes the coevolution of gas and dust in a protoplanetary disk, which in turn forms self-consistently via the gravitational collapse of a rotating cloud.  We employed a novel Coarray Fortran, combined with OpenMP, parallelization scheme, allowing us to efficiently use compact high-performance computing resources. The dust dynamics and growth, including pebble formation,  were studied in detail during the initial 18~kyr of disk evolution.  Our results can be summarized as follows.

The disk becomes gravitationally unstable and develops a spiral pattern already after several thousand years from its formation instance. The radial distribution of the gas volume density in the disk midplane is approximately proportional to $r^{-2.5}$, which is typical of gravitationally unstable disks. In the vertical extent, the disk bulges at the position of the spiral arms.

The dust-to-gas mass ratio in the disk midplane is highly nonhomogeneous throughout the disk extent, in agreement with our earlier thin-disk simulations \citep{2019VorobyovElbakyan} and \citet{2020Lebreuilly}. The dust-to-gas mass ratio also shows prominent vertical stratification, indicative of dust vertical settling. The ratio of the vertical scale heights of the dust and gas disks was found to be in the 0.2--0.5 range. Such a moderate dust settling can be attributed to perturbing action of the spiral arms, as was already noted by \citet{2020Riols}.

Dust grows fast in a young protoplanetary disk and reaches sizes on the order of 1--10~mm already after 15~kyr. However, the Stokes number remains rather low, $\mathrm{St}\sim 10^{-3}-10^{-1}$ throughout the disk extent, with the largest values closer to the disk outer edge. These low $\mathrm{St}$ are caused by systematically higher densities and temperatures in young disks around subsolar mass stars as compared to the corresponding characteristics in the MMSN. Low Stokes, along with a nonsteady character of the spiral pattern, hinder strong dust accumulation in the spiral arms, although in general the dust-to-gas ratio in the midplane is enhanced by a factor of several compared to the initial 1:100 value. Our modeling demonstrates the importance of coupled disk formation and dust growth calculations in realistically assessing the efficiency of spiral arms as dust traps in comparison to models of isolated disks with a fixed dust size or Stokes number \citep[e.g.,][]{2004RiceLodato,Boss2020}. 

The spatial distribution of pebbles exhibits a highly nonhomogeneous and patchy character. Interestingly, pebbles are absent in the inner 10~au because of the low $\mathrm{St}<10^{-2}$ in these regions. Pebbles are also found to be strongly concentrated to the disk midplane. The total mass of pebbles in the disk increases with time and reaches a few tens of Earth masses by 18~kyr.

%3. Efficiency of dust settling 

%4. Inefficient dust concentration to spiral arms in young disks, confirmed ...

Variations in the initial rotation rate of the collapsing cloud were found to have minor effect on our conclusions. We note that when calculating dust growth we assumed a spatially and temporally constant value of the turbulent $\alpha$ parameter.  We plan to explore the effects of this  simplification in follow-up works.

\begin{acknowledgements}  We thank the anonymous referee for constructive and useful comments and suggestions.
 This work was supported by the FWF project I4311-N27 (E. I. V., J. M.) and RFBR project 19-51-14002 (I. K. and V. E.).
Simulations were performed on the Vienna Scientific Cluster (VSC)\footnote{\href{https://vsc.ac.at/}{https://vsc.ac.at/}} and on the Narval Cluster provided by Calcul Québec\footnote{\href{https://www.calculquebec.ca/}{https://www.calculquebec.ca/}} and the Digital Research Alliance of Canada\footnote{\href{https://alliancecan.ca/}{https://alliancecan.ca/}}.
\end{acknowledgements}

\bibliographystyle{aa}
\bibliography{refs}

\begin{appendix}

\section{Solution procedure and parallel realization in Coarray Fortran}
\label{Appendix:coarrays}

\begin{figure}[ht]
	\centering
	\begin{minipage}[h]{1\linewidth}
		\center{\includegraphics[width=1\linewidth]{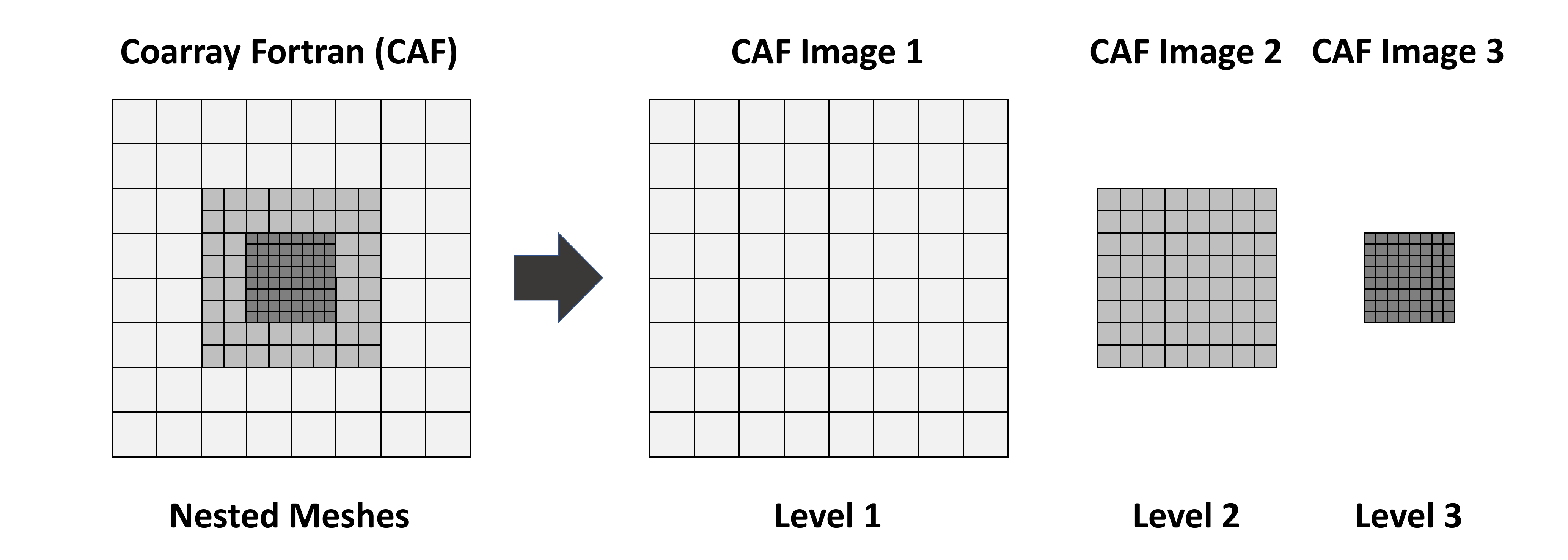}}
	\end{minipage}
	\caption{Decomposition of the computational domain. Each level of nested grids is placed on a separate CAF image. As an example, nested grids with three levels of refinement are shown. The number of coarray images is equal to the total number of nested grids.}
	\label{decomposition}
\end{figure}

In this section, we provide essential details of the solution procedure and parallelization technique employed in ngFEOSAD. For a parallel implementation of the solution procedure on nested grids, we use a geometric decomposition in which each grid is placed on a separate image of the Coarray Fortran (CAF) program  (see Figure \ref{decomposition}). Below we present the main procedures of the ngFEOSAD computational cycle.
%Для параллельной реализации будем использовать геометрическую декомпозицию расчетной области в виде размещения каждого уровня вложенного сетки на отдельном образе Coarray Fortran программы (see figure \ref{decomposition}). 

%Приведем основные процедуры главного вычислительного цикла программы
\begin{enumerate}
	\item \textbf{Projection}. This first step implements the averaging of the conserved hydrodynamic quantities (i.e., densities and momenta for both gas and dust) from a finer to a coarser grid to ensure their conservation on nested grids.  The recalculation scheme is illustrated in Figure~\ref{projection}. The conservative values are averaged  over four (or eight in the three-dimensional case) cells using conservation laws. An example of the CAF code with one-sided communications is shown, which transfers part of the array to a coarser grid. We note that the recalculation of the conservative values is carried out sequentially from a finer to a coarser grid level (not in parallel mode). While this removes the CAF (distributed memory) parallelization, it does not affect OpenMP (shared memory) parallelization. In any way, this procedure is computationally inexpensive.
 
% Процедура \textbf{Projection} реализует пересчет законов сохранения с каждой вложенной сетки на сетку более высокого уровня. Схему пересчета можно проиллюстрировать рисунком \ref{projection}. В первой части процедуры с использованием законов сохранения идет осреднение консервативных величин из четырех (или восьми в трехмерном случае) ячеек. В правой части рисунка записан CAF code of one-sided communications передачи части массива на предыдущий по номеру образ. Заметим, что пересчет законов сохранения осуществляется последовательности по образам CAF программы. Однако, с учетом малой доли времени процедуры такая организация вычислений не вызывает существенного увеличения времени общего счета.     

	\begin{figure}
		\centering
		\begin{minipage}[h]{1\linewidth}
			\center{\includegraphics[width=1\linewidth]{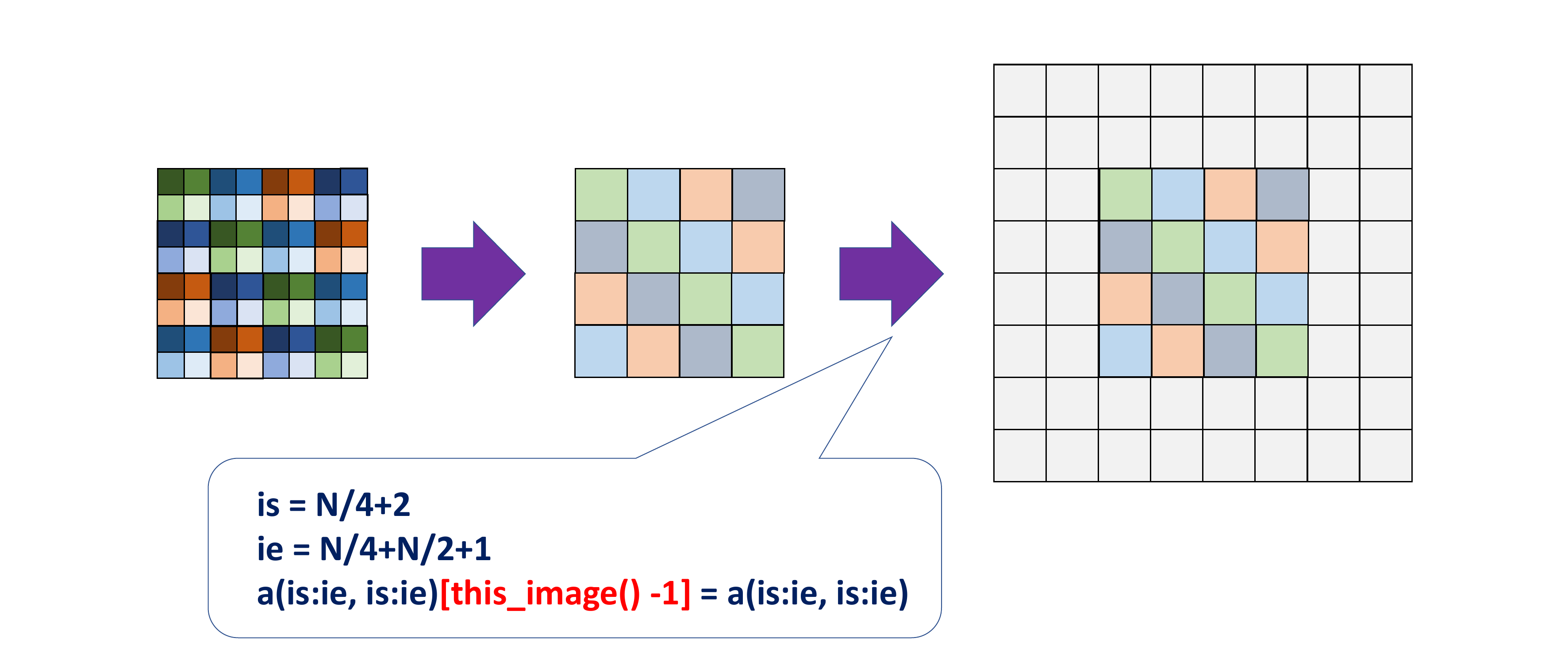}}
		\end{minipage}
		\caption{Averaging of conservative hydrodynamic quantities from a finer to a coarser grid. The colors represent different values of conserved quantities. Only inner part of the coarser grid is affected by the averaging.}
		\label{projection}
	\end{figure}  

%	\begin{figure}
%		\centering
%		\begin{minipage}[h]{0.6\linewidth}
%			\center{\includegraphics[width=1\linewidth]{./figures/godunov.pdf}}
%		\end{minipage}
%		\caption{Cells of different nesting levels in the implementation of the Godunov method. The white cells are a particular nested grid, while the colored cells represent the boundary cells on a coarser and finer levels with respect to the current grid. At the boundary with a finer grid, the fluxes of conservative values are calculated on each face of the gray cells, and then averaged over the boundary. At the boundary with a coarser grid, two (four in the three-dimensional case) gray cells participate in the solution of the Riemann problem with the same neighboring cell on the coarser grid.}
%		\label{godunov}
%	\end{figure}

\item \textbf{Boundary}. Here, the boundary conditions are implemented by adding one outer layer of cells to each level of the nested grids.
For the outer coarsest grid, the boundary conditions are determined from the physics of the problem. On all higher-level grids, the boundary values are taken from the corresponding cells on the underlying coarser grid. The gravitational potential is linearly interpolated to improve the accuracy \citep[see][]{2023VorobyovMcKevitt}. This step is implemented in a parallel mode for both CAF and OpenMP.
 
%	\item В процедуре \textbf{Boundary} реализуется постановка граничных условий. Для этого на каждом уровне вложенной сетки добавляется один внешний слой ячеек (see figure \ref{boundary}) Темно-серым обозначены фантомные граничные ячейки. Для внешней сетки граничные условия определяются из постановки задачи. На всех вложенных сетках считываются соответствующие значения из ячеек из сетки с предыдущего CAF образа. Заметим, что постановка граничных условий реализуется в параллельном режиме. 
%	\begin{figure}[ht]
%		\centering
%		\begin{minipage}[h]{0.5\linewidth}
%			\center{\includegraphics[width=1\linewidth]{./figures/boundary.pdf}}
%		\end{minipage}
%		\caption{Постановка граничных условий.}
%		\label{boundary}
%	\end{figure}

 \item \textbf{Primitive}. This step finds the physical variables (e.g., densities and velocities of both gas and dust) from conservative variables. This procedure is implemented in parallel (Carray Fortran and OpenMP) for each level of the nested grids without data exchanges between the grids.

	\item \textbf{Courant}. Here, the time step is calculated, which is the same for all nested grid levels. Since each grid level is located in its own CAF image, we use the CAF reduction procedure \textbf{call co\_max} to find the maximum wave speed over all grid levels, which is then used to determine the minimum time step applied to all nested grids.

	\item \textbf{Reconstruction}. This step computes the piece-wise parabolic reconstruction of physical variables in all cells and in all coordinate directions at each grid level. Both CAF and OpenMP parallelizations are employed.

	\item \textbf{Godunov}. Here we solve the Riemann problem for each level of the nested grids. At the boundary with a coarser grid, four cells participate in the solution of the Riemann problem with the same neighboring cell on the coarser grid. This step employs both CAF and OpenMP parallelization.

\item \textbf{Gravity}. This step accounts for the effects of disk self-gravity and the gravity of the star(s) (if formed). We note that in the present work we do not introduce a sink particles as proxy for the star but plan to do that in the future. The self-gravity of both gas and dust is considered. The detailed explanation of the applied method can be found in \citep{2023VorobyovMcKevitt}. The details of the CAF-OpenMP realization will be presented elsewhere. 

\item \textbf{Friction}. The computation cycle is completed with the calculation of the friction force between the dust and gas and its effect on the gas and dust momenta. This step employs the full power of both CAF and OpenMP parallelization.

\end{enumerate}

\section{Riemann solver for grown dust}
\label{Appendix:scheme}

To find the fluxes of conservative variables for the gas component (see Eq.~\ref{eq:godunov}), we use a combination of the HLLC and HLL Riemann solvers, the description of which can be found in \citet{Toro2019}. To find the fluxes for the pressureless grown dust (see Eq.~\ref{eq:godunov_dust}), we use a modified HLL solver. To describe the scheme for solving the Riemann problem for grown dust, we consider a configuration of the left (L) and right (R) cells, in which the values of $\rho_{\rm L}, \rho_{\rm R}$ are the dust densities, $u_{\rm L}, u_{R\rm }$ are the dust longitudinal velocity, $v_{\rm L}, v_{\rm R}, w_{\rm L}, w_{\rm R}$ are two transverse dust propagation velocities. The one-wave HLL algorithm for grown dust is then as follows.
%Для описания схемы решения задачи Римана для пыли рассмотрим конфигурацию из левой (L) и правой (R) ячеек, в которых заданы значения $\rho_L, \rho_R$ -- плотности пыли, $u_L, u_R$ -- продольной скорости пыли, $v_L, v_R, w_L, w_R$ -- двух поперечных скоростей распространения пыли. Алгоритм one-wave HLL для пыли имеет следующий вид.
\begin{enumerate}
\item We form vectors of conservative variables and their fluxes:
\begin{equation}
Q_{\rm L } = \left( \begin{array}{c}
	\rho_{\rm L} \\
	\rho_{\rm L} u_{\rm L} \\
	\rho_{\rm L} v_{\rm L} \\
	\rho_{\rm L} w_{\rm L} 
\end{array}
\right),
\quad
Q_{\rm R} = \left( \begin{array}{c}
	\rho_{\rm R} \\
	\rho_{\rm R} u_{\rm R} \\
	\rho_{\rm R} v_{\rm R} \\
	\rho_{\rm R} w_{\rm R} 
\end{array}
\right),
\end{equation}
\begin{equation}
F_{\rm L} = \left( \begin{array}{c}
	\rho_{\rm L} u_{\rm L} \\
	\rho_{\rm L} u_{\rm L} u_{\rm L}  \\
	\rho_{\rm L} v_{\rm L} u_{\rm L} \\
	\rho_{\rm L} w_{\rm L} u_{\rm L} 
\end{array}
\right),
\quad
F_{\rm R} = \left( \begin{array}{c}
	\rho_{\rm R} u_{\rm R} \\
	\rho_{\rm R} u_{\rm R} u_{\rm R} \\
	\rho_{\rm R} v_{\rm R} u_{\rm R} \\
	\rho_{\rm R} w_{\rm R} u_{\rm R} 
\end{array}
\right).
\end{equation}
\item We calculate the exact value of the dust transport rate according to the following equation: 
\begin{equation}
    S = \frac{\sqrt{\rho_{\rm L}} u_{\rm L} + \sqrt{\rho_{\rm R}} u_{\rm R}}{\sqrt{\rho_{\rm L}} + \sqrt{\rho_{\rm R}}},
\end{equation}
\item We write down a one-wave HLL scheme in the following basic form:
\begin{equation}
	F^{\rm 1WHLL} = \left \lbrace 
	\begin{array}{ll}
		F_{\rm L}, & 0 \le S, \\
		F_{\rm R}, & 0 \ge S
	\end{array}  \right. .
\end{equation}
\end{enumerate}
Similarly to the solution of the hydrodynamic equations for gas, a piece-wise parabolic representation of physical variables is used to increase the order of accuracy on smooth solutions and to reduce dissipation at discontinuities. In this case, instead of the piece-wise constant values of the physical quantities $q_{\rm L}$ and $q_{\rm R}$, where $q$ is understood as the density and velocity functions, an integral over the parabola constructed along the characteristic $S$ is considered.

\section{Numerical tests}
\label{Appendix:tests}
In this section, we provide the results of extensive testing of the main constituent parts of the ngFEOSAD code.

%The method performs well on the standard axysimmetric and non-axisymmetric test problems (oblate ellipsoid, wide-separation binary), and provides an overall second-order convergence, except near the grid interfaces where the convergence is linear. 

\subsection{Tests on the numerical scheme for gas dynamics}

Verification of the numerical method for solving hydrodynamic equations for the gas component was carried out on a set of classical one-dimensional Godunov tests, tests that consider the development of physical instabilities of the Kelvin-Helmholtz and Rayleigh-Taylor types, as well as on the Sedov test of a point explosion. Such a test suite is aimed at checking the method for correctly reproducing the characteristic gas flows and has either analytical solutions or a general estimate of the nature of the gas flow during the growth of instabilities. All presented tests in this section were performed on a single Cartesian grid with equidistant cell spacing.

\subsubsection{The Godunov tests for gas dynamics}
Here, we consider a set of Godunov tests in the form of the Sod problem with a sonic point, the Einfeldt problem with the formation of a strong rarefaction region, and the Lax problem with the formation of strong shock waves. The initial configuration for these three test problems is given in Table~\ref{ShockTubeProblem}, where $x_0$ is the initial position of the boundary between the two adjacent states (the subscripts $\rm L$ and $\rm R$ denote the left-hand side and right-hand side states, respectively). For computational experiments, 200 computational cells were used.

\begin{table*}
	\centering
	\caption{Initial configuration of the Godunov problem tests.}
	\vskip 0.4 cm
	\begin{tabular}{c|ccc|ccc|c|c}
		\hline 
		 &  $\rho_{\rm L}$ & $v_{\rm L}$ & $p_{\rm L}$ & $\rho_{\rm R}$ & $v_{\rm R}$  &  $p_{\rm R}$ & $x_{0}$ & $t$ \\ 
		\hline 
		Sod test & 1 & 0.75 & 1 & 0.125 & 0 & 0.1 & 0.3 & 0.2 \\ 
		\hline 
        Sod test for dust & 1 & 0.75 & 0 & 0.125 & 0 & 0 & 0.3 & 0.2 \\ 
		\hline 
		Einfeldt test & 1 & -2 & 0.4 & 1 & 2 & 0.4 & 0.5 & 0.15 \\ 
		\hline 
		Lax test & 1 & 0 & 1000 & 1 & 0 & 0.01 & 0.5 & 0.012 \\ 
		\hline 
	\end{tabular}  
	\label{ShockTubeProblem}
\end{table*}

\begin{figure}[ht]
	\centering
	\begin{minipage}[h]{0.48\linewidth}
		\center{\includegraphics[width=1\linewidth]{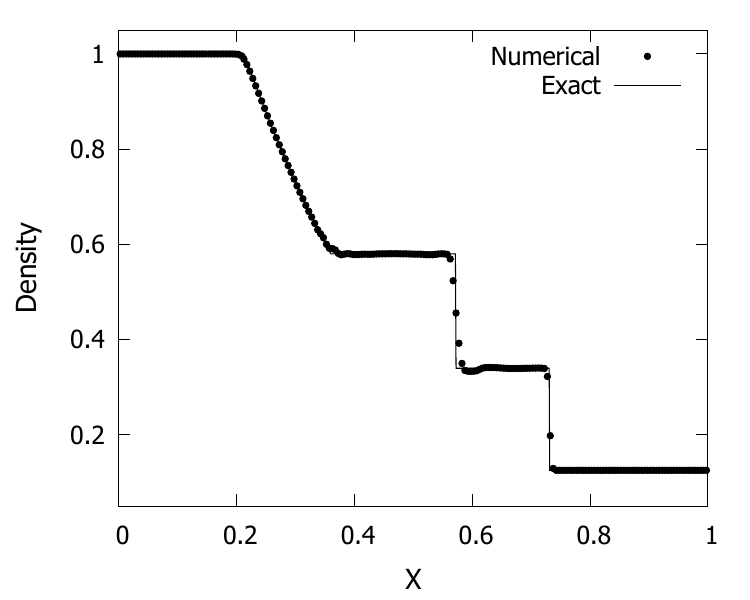}} (a)
	\end{minipage}
	\begin{minipage}[h]{0.48\linewidth}
		\center{\includegraphics[width=1\linewidth]{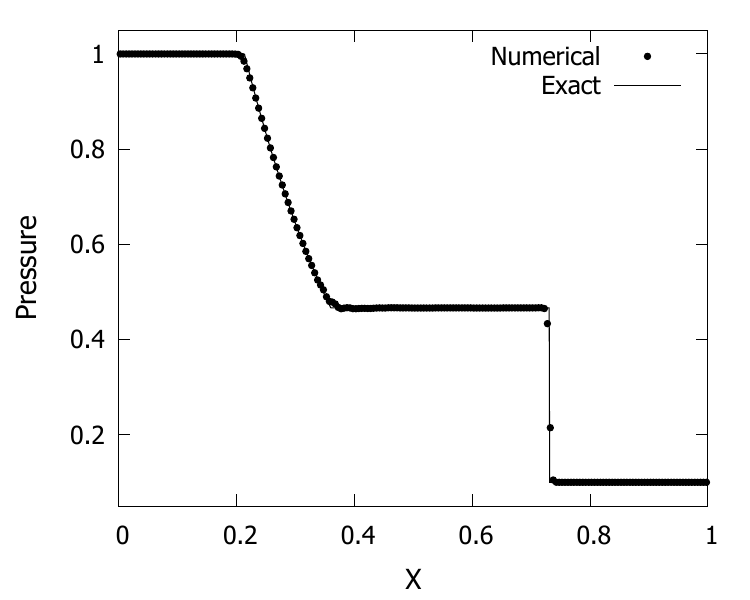}} (b)
	\end{minipage}
	\begin{minipage}[h]{0.48\linewidth}
		\center{\includegraphics[width=1\linewidth]{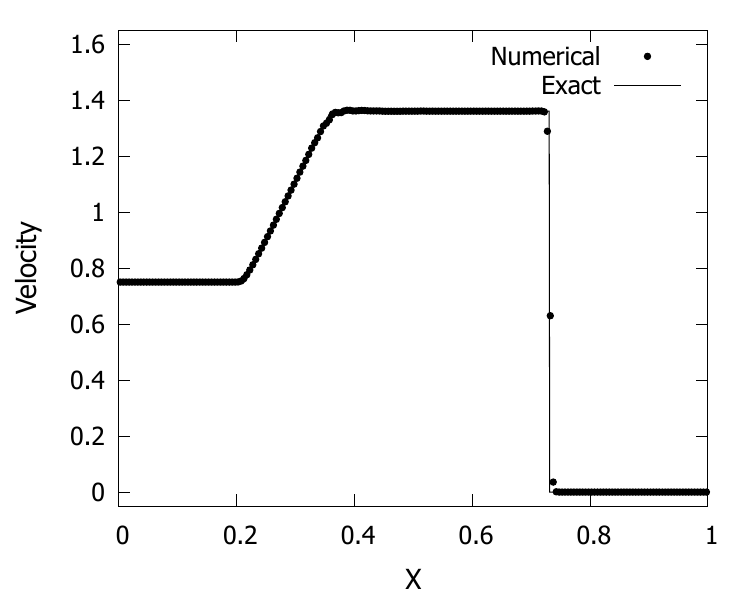}} (c)
	\end{minipage}
	\begin{minipage}[h]{0.48\linewidth}
		\center{\includegraphics[width=1\linewidth]{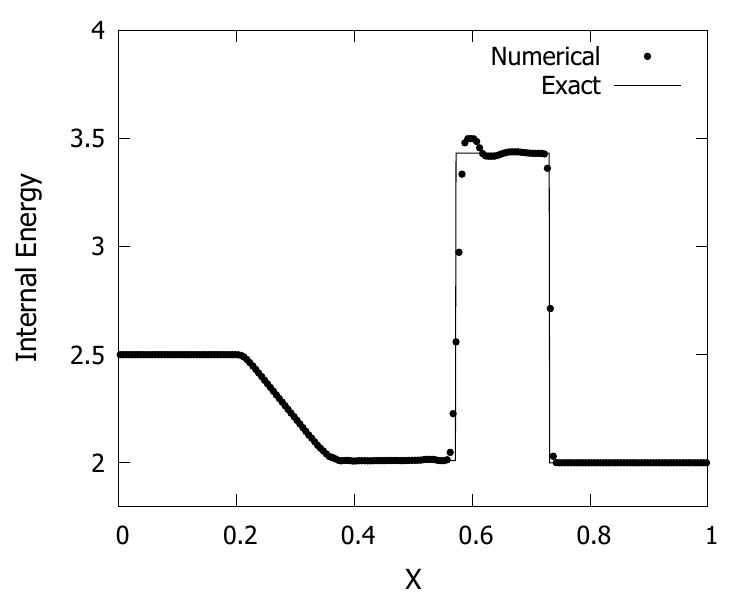}} (d)
	\end{minipage}
	\caption{Sod test problem: panel (a) displays the gas density, panel (b) -- pressure, panel(c) -- velocity, and panel (d) -- internal energy. The symbols present the numerical data, while the lines correspond to the analytical solution.}
	\label{ShockTubeSimulation1}
\end{figure}

The main goal of the Sod problem is to test the ability of the numerical method to simultaneously reproduce a weak shock wave, a contact discontinuity, and a rarefaction wave. Unlike the standard realizations of this test problem, our initial setup is intentionally complicated by the presence of a nonzero velocity at the left state, which forms initially a supersonic gas flow. Roe-type methods, when reproducing such a solution, may have a feature in the form of an entropy trace at the point of the discontinuity of the rarefaction wave. In our version of the method (see Figure~\ref{ShockTubeSimulation1}), all waves are reproduced correctly, including the absence of the entropy trace. The use of piece-wise parabolic reconstruction of physical variables makes it possible to reproduce a weak shock wave with a numerical dissipation of only two cells.

\begin{figure}[ht]
	\centering
	\begin{minipage}[h]{0.48\linewidth}
		\center{\includegraphics[width=1\linewidth]{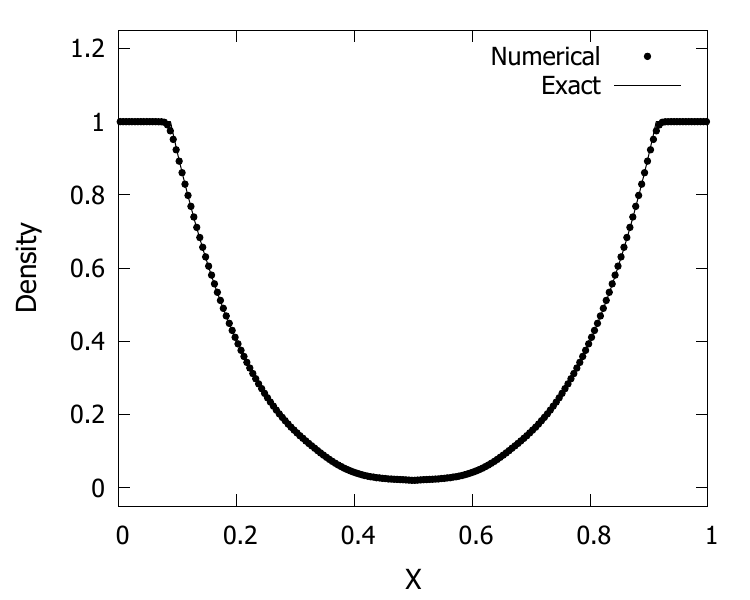}} (a)
	\end{minipage}
	\begin{minipage}[h]{0.48\linewidth}
		\center{\includegraphics[width=1\linewidth]{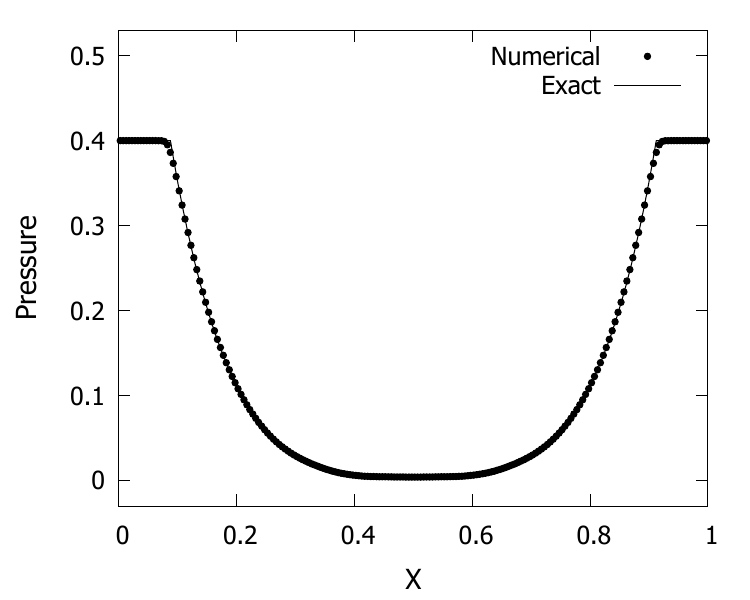}} (b)
	\end{minipage}
	\begin{minipage}[h]{0.48\linewidth}
		\center{\includegraphics[width=1\linewidth]{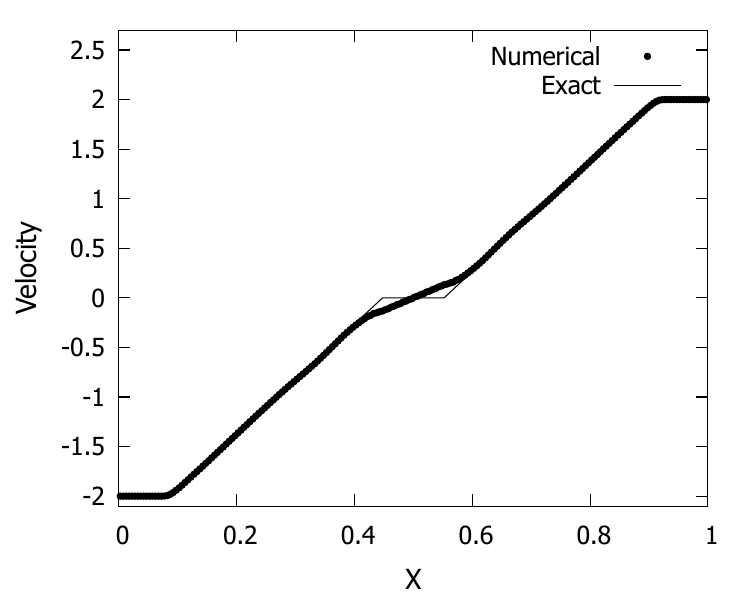}} (c)
	\end{minipage}
	\begin{minipage}[h]{0.48\linewidth}
		\center{\includegraphics[width=1\linewidth]{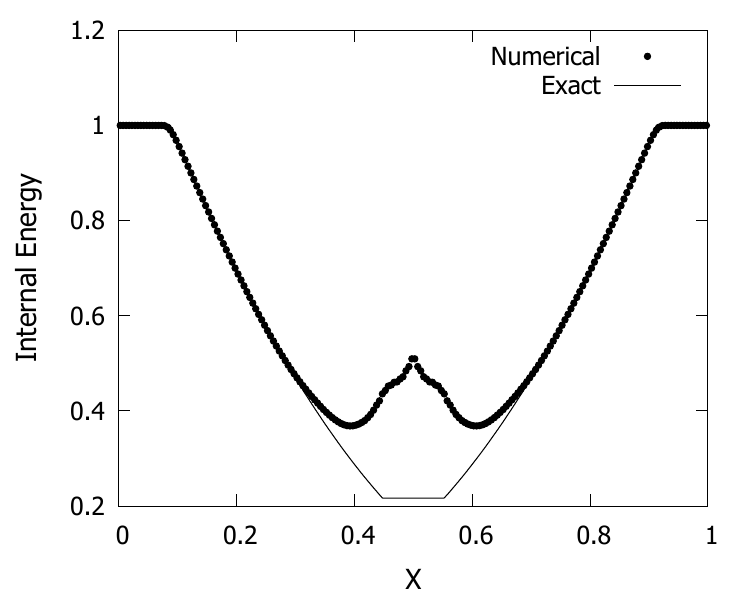}} (d)
	\end{minipage}
	\caption{Einfeldt test problem: density (a), pressure (b), velocity (c), and internal energy (d).
 The symbols present the numerical data, while the lines correspond to the analytical solution.}
	\label{ShockTubeSimulation2}
\end{figure}

The second Godunov test, the Einfeldt test, assesses the ability of a numerical method to reproduce a rarefaction region caused by gas expansion in the opposite directions. The main problem of the test is the nonphysical growth of the internal energy at the boundary between the two initial states, with a fairly good quality of reproduction of the pressure, density, and velocity distributions. We also found it to be the case in our numerical method -- the internal energy grows nonphysically in the rarefaction region (see Figure~\ref{ShockTubeSimulation2}), but such growth is limited in amplitude, rather smooth, and does not have oscillations, which speaks in favor of the chosen numerical technique.
 
\begin{figure}[ht]
	\centering
	\begin{minipage}[h]{0.48\linewidth}
		\center{\includegraphics[width=1\linewidth]{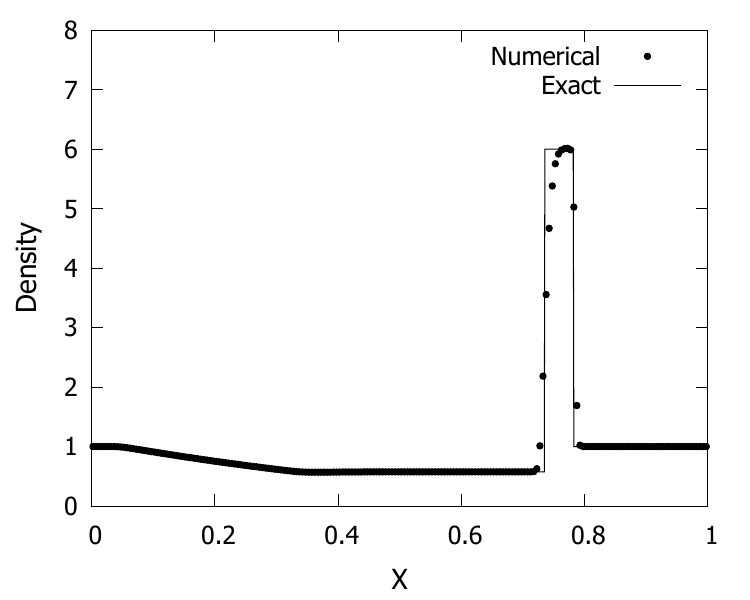}} (a)
	\end{minipage}
	\begin{minipage}[h]{0.48\linewidth}
		\center{\includegraphics[width=1\linewidth]{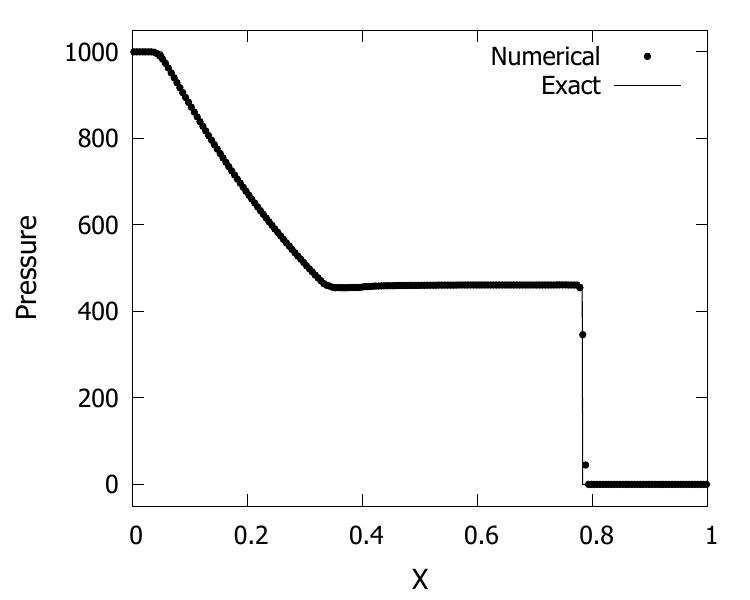}} (b)
	\end{minipage}
	\begin{minipage}[h]{0.48\linewidth}
		\center{\includegraphics[width=1\linewidth]{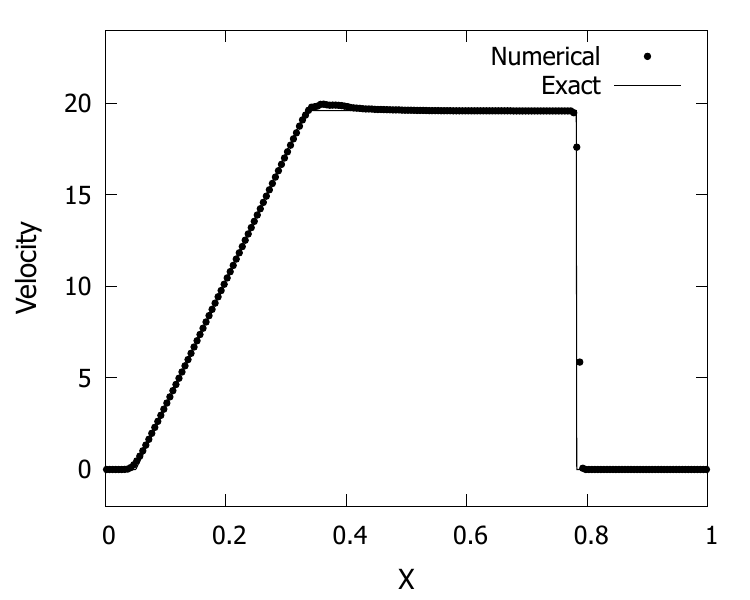}} (c)
	\end{minipage}
	\begin{minipage}[h]{0.48\linewidth}
		\center{\includegraphics[width=1\linewidth]{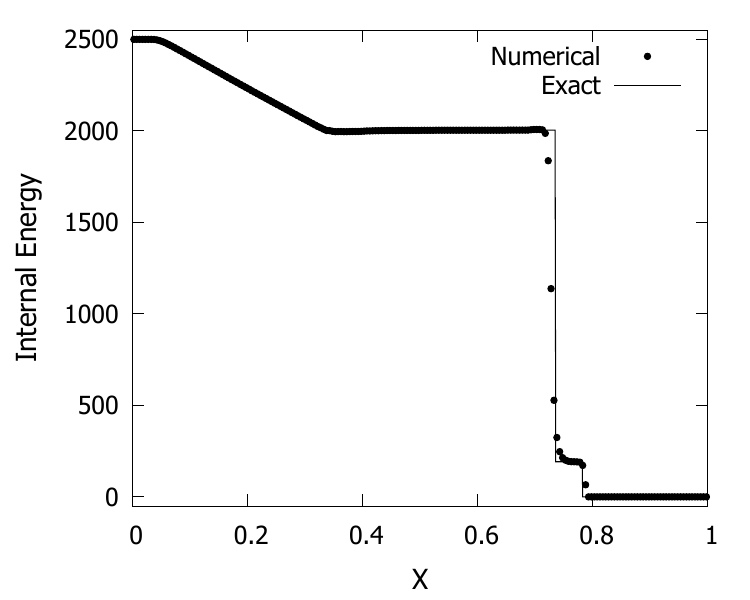}} (d)
	\end{minipage}
	\caption{Lax test problem: density (a), pressure (b), velocity (c), and internal energy (d). The symbols present the numerical data, while the lines correspond to the analytical solution.}
	\label{ShockTubeSimulation3}
\end{figure}

In the Lax test, the ability of the code to accommodate a significant pressure drop (5 orders of magnitude) is assessed. The key feature of the flow is the low-amplitude precursor wave that can be seen in the internal energy ahead of the shock wave. Our computational experiments (see Figure \ref{ShockTubeSimulation3}) indicate that such a wave is reproduced correctly without significant dissipation. We also note the absence of unphysical oscillations on the shock wave and the contact discontinuity, which may occur in high order schemes.
 
\subsubsection{The Kelvin–Helmholtz and Rayleigh–Taylor instabilities}

An important property of numerical methods is their ability not to suppress physical instabilities when they are numerically resolved. To verify this property of ngFEOSAD, we study the development of Kelvin-Helmholtz and Rayleigh-Taylor instabilities. The Rayleigh-Taylor instability develops at the boundary of two gases with different densities in the common gravitational field if the heavier component is located on top of the lighter one. The Kelvin-Helmholtz instability develops at the boundary of two gases with different densities moving relative to each other. Both types of instabilities lead to the development of nonlinear hydrodynamic turbulence.

\begin{figure}[ht]
	\centering
	\begin{minipage}[h]{0.48\linewidth}
		\center{\includegraphics[width=1\linewidth]{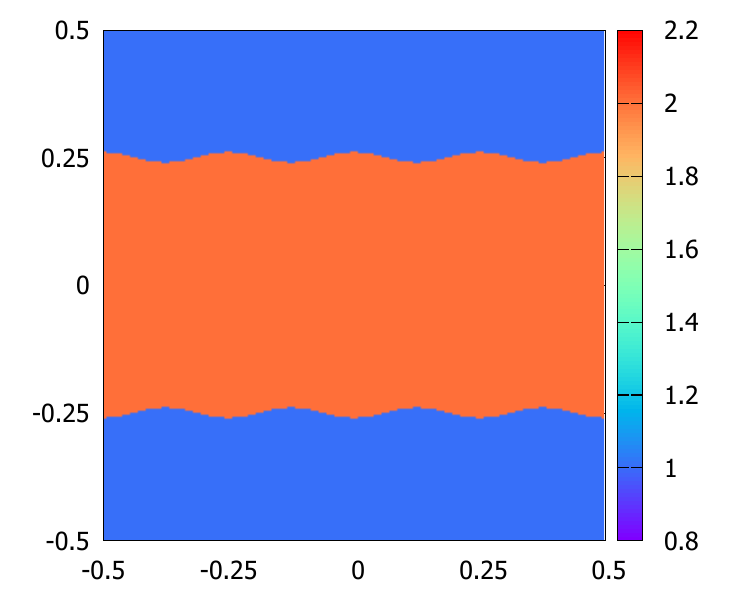}} (a)
	\end{minipage}
	\begin{minipage}[h]{0.48\linewidth}
		\center{\includegraphics[width=1\linewidth]{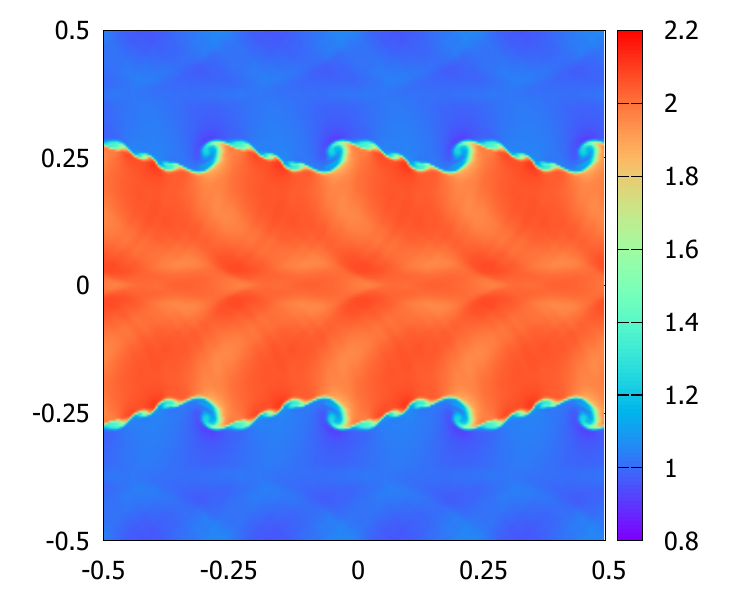}} (b)
	\end{minipage}
	\begin{minipage}[h]{0.48\linewidth}
		\center{\includegraphics[width=1\linewidth]{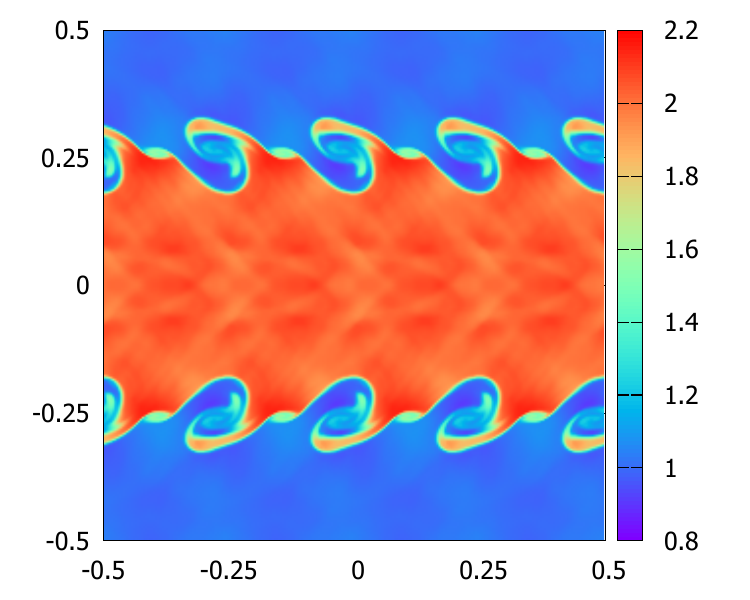}} (c)
	\end{minipage}
	\begin{minipage}[h]{0.48\linewidth}
		\center{\includegraphics[width=1\linewidth]{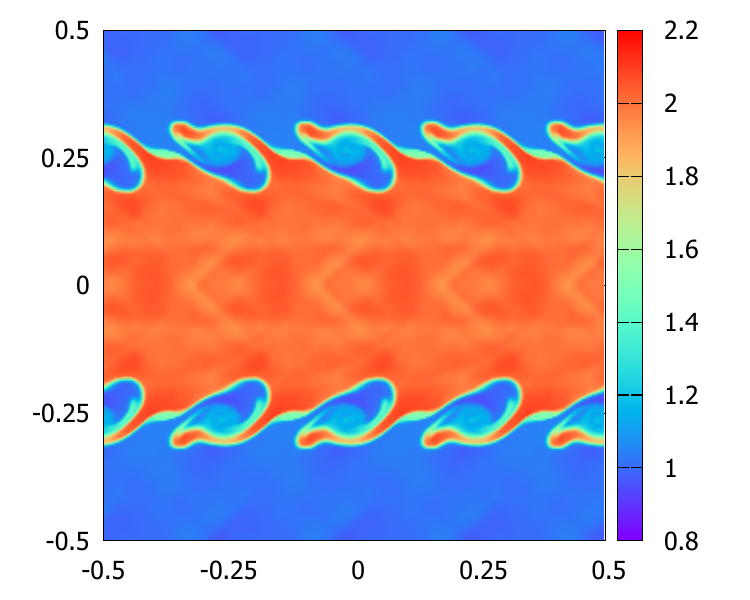}} (d)
	\end{minipage}
	\caption{Development of the Kelvin–Helmholtz instability. Panel (a) presents the initial state, panel (b) -- $t = 0.2$, panel (c) -- $t = 0.5$, and panel (d) -- $t = 0.6$. The color bar shows the gas density in dimensionless units.}
	\label{KHInstability}
\end{figure}

\textbf{Kelvin-Helmholtz instability.} We consider a square domain $[-0.5;0.5]^2$. The initial density profile and the $x$-velocity component are given as:
%Рассмотрим квадратную область $[-0.5;0.5]^2$. Начальный профиль плотности и $x$-компонента скорости задается в виде: 
\begin{equation}
	\rho_{0}(x,y) = \left \lbrace 
		\begin{array}{ll}
			1, & \vert y \vert > 0.25 + 0.01 (1 + \cos(8 \pi x)) \\
			2, & \vert y \vert \leq 0.25 + 0.01 (1 + \cos(8 \pi x)),
		\end{array}  \right. 
\end{equation}
\begin{equation}
	v_{x,0}(x,y) = \left \lbrace 
	\begin{array}{ll}
		-0.5, & \vert y \vert > 0.25 + 0.01 (1 + \cos(8 \pi x)) \\
		 0.5, & \vert y \vert \leq 0.25 + 0.01 (1 + \cos(8 \pi x)).
	\end{array}  \right. 
\end{equation}
The gas pressure is set equal to $p = 2.5$ and the adiabatic index is $\gamma = 1.4$. For the development of instability, a sinusoidal perturbation of the gas density and velocity is introduced. The development of the Kelvin-Helmholtz instability is shown in Figure~\ref{KHInstability}.
%Давление газа $p = 2.5$, показатель адиабаты $\gamma = 1.4$. Для развития неустойчивости задается синусоидальное возмущение плотности и скорости газа. Развитие неустойчивости Кельвина-Гельмгольца показано на рисунке (\ref{KHInstability}).} 
The characteristic time for the development of the Kelvin-Helmholtz instability can be estimated as:
%Характерное время развития неусточивости Кельвина-Гельмгольца оценивается уравнением:
\begin{equation}
	\tau_{\rm KH} = \frac{\lambda \left( \rho_{\rm in} + \rho_{\rm out} \right)}{v_{\rm rel} \sqrt{\rho_{\rm in} \rho_{\rm out}}},
\end{equation}
where $\lambda = 1/4$ is the inverse frequency of the sinusoidal disturbance, $v_{\rm rel} = 1$ the relative velocity of motion between gases of different densities, $\rho_{\rm in} = 2$ the density of the inner denser region, $\rho_{\rm out} = 1$ the density of the outer rarefied region. For such parameters, the characteristic development time of the Kelvin-Helmholtz instability is $\tau_{\rm KH}\approx 0.53$, which agrees well with the results of computational experiments shown in Figure~\ref{KHInstability}. Indeed, the formation of turbulent eddies occurs at the boundary of two gases at $t\approx0.5$.

\begin{figure}[ht]
	\centering
	\begin{minipage}[h]{0.24\linewidth}
		\center{\includegraphics[width=1\linewidth]{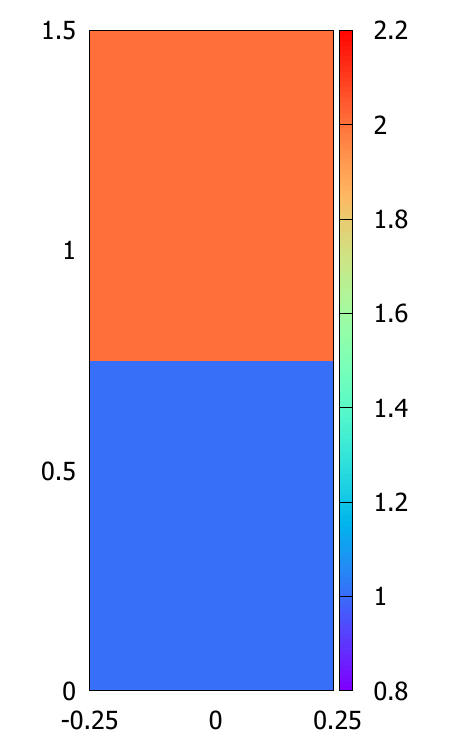}} (a)
	\end{minipage}
	\begin{minipage}[h]{0.24\linewidth}
		\center{\includegraphics[width=1\linewidth]{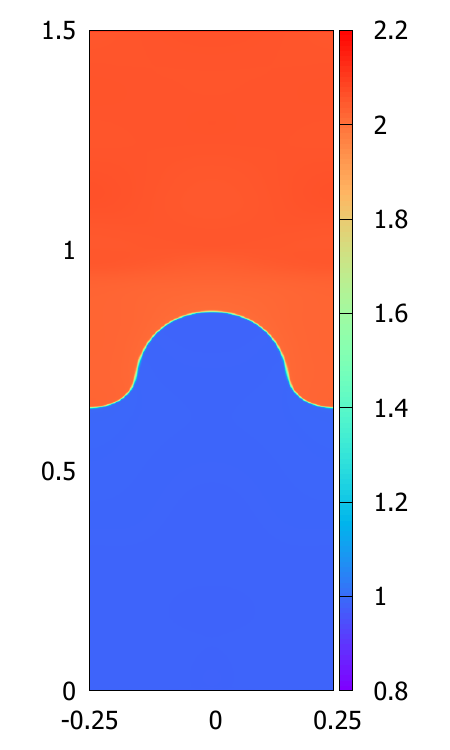}} (b)
	\end{minipage}
	\begin{minipage}[h]{0.24\linewidth}
		\center{\includegraphics[width=1\linewidth]{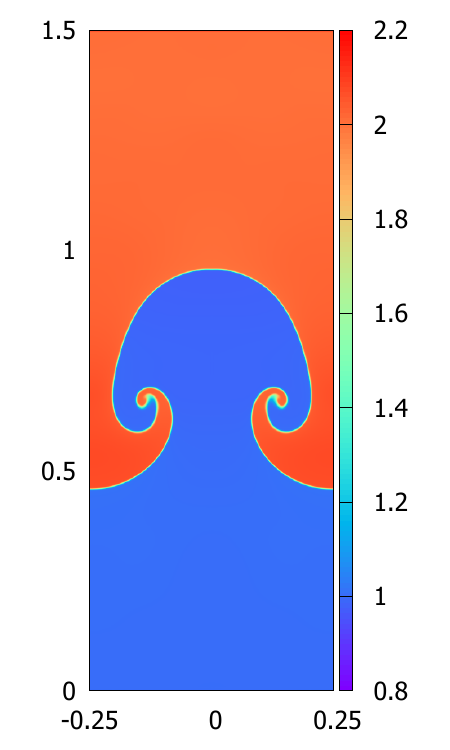}} (c)
	\end{minipage}
	\begin{minipage}[h]{0.24\linewidth}
		\center{\includegraphics[width=1\linewidth]{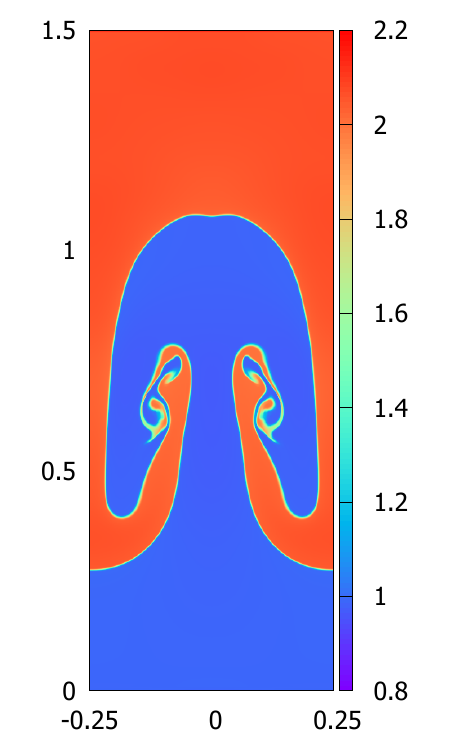}} (d)
	\end{minipage}
	\caption{Development of the Rayleigh–Taylor instability. Panel (a) presents the initial state, panel (b) -- $t = 72$, panel (c) -- $t = 10$, and panel (d) -- $t = 13$. The color bar shows the gas density in dimensionless units. }
	\label{RTInstability}
\end{figure}

\textbf{Rayleigh-Taylor instability.} We consider the area $[-0.25;0.25] \times [0;1.5]$ with the following parameters:
%Рассмотрим  область $[-0.25;0.25] \times [0;1.5]$ со следующими параметрами:
\begin{equation}
	\rho_{0}(x,y) = \left \lbrace 
		\begin{array}{ll}
			1, & y > 0.75 \\
			2, & y \leq 0.75,
		\end{array}  \right.
\end{equation}
\begin{equation}
	p = 0.15 - \rho \cdot g \cdot \vert y \vert,
\end{equation}
where $p$ is the hydrostatic equilibrium pressure, $g = 0.1$ the gravitational acceleration, $y$ the vertical coordinate, and $\gamma = 1.4$ the adiabatic exponent. The vertical component of velocity is perturbed according to the following rule:
$v_{y,0}(x,y) = A(y - 0.75)[1+\cos(2 \pi x)][1+\cos(2 \pi y)]$, where
\begin{equation}
	A(y) = \left \lbrace 
		\begin{array}{ll}
			10^{-2}, & y \le 0.01 \\
			0, & y > 0.01.
		\end{array}  \right. 
\end{equation}
The development of the Rayleigh-Taylor instability is shown in Figure~\ref{RTInstability}.
%Развитие неустойчивости Релея-Тейлора показано на рисунке 
The growth of the perturbation amplitude of the Rayleigh-Taylor instability is determined as:
%Рост амплитуды неустойчивости Релея-Тейлора определяется уравнением:
\begin{equation}
f(t) = 0.01 \exp \left( t \sqrt{A g} \right) \simeq 0.01  \exp \left( 0.25 t \right),
\end{equation}
where $A = 2/3$ is Atwood's number. For example, at time $t = 13$, the amplitude is $f(t = 13) \simeq 0.25$, which is in good agreement with the model results. Indeed, the initial position of the interface between different gas states is located at $y = 0.75$. At $t = 13$, the density perturbation propagates approximately $dy = \pm [0.25:0.3]$ from the initial position of the interface between the two states.

\subsubsection{The Sedov point explosion test}
\begin{figure}[ht]
	\centering
	\begin{minipage}[h]{0.48\linewidth}
		\center{\includegraphics[width=1\linewidth]{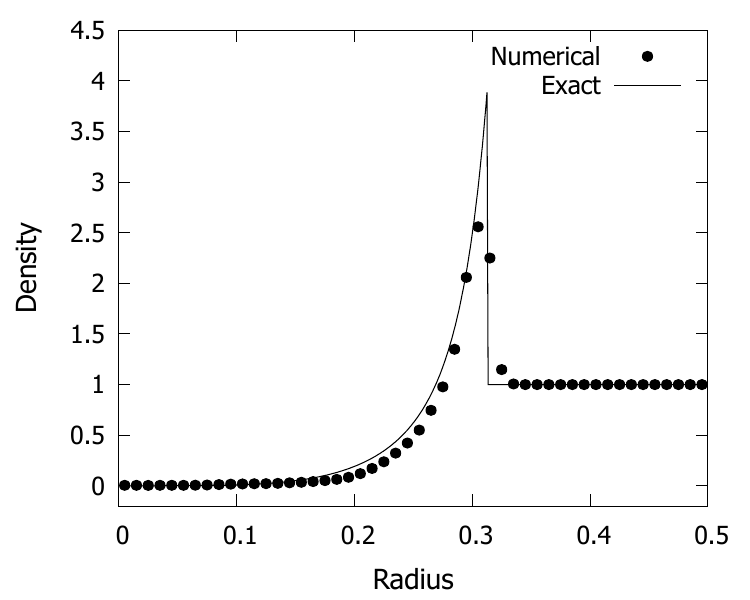}} (a)
	\end{minipage}
	\begin{minipage}[h]{0.48\linewidth}
		\center{\includegraphics[width=1\linewidth]{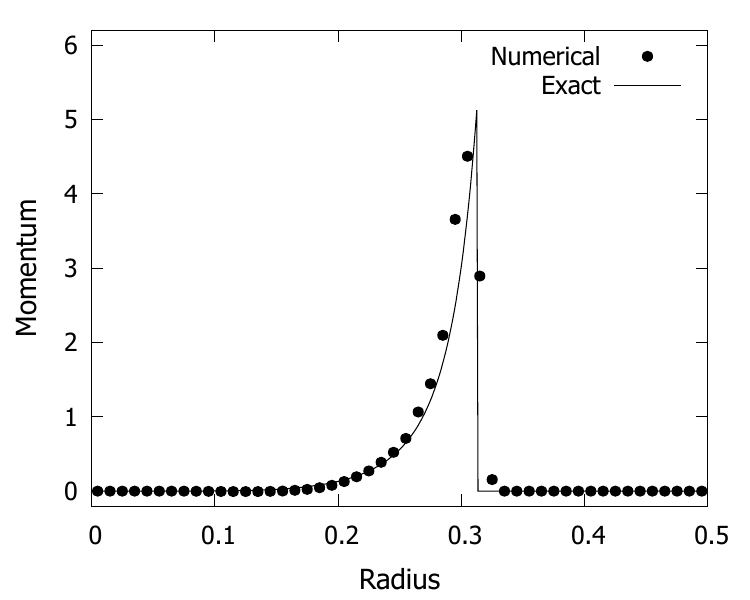}} (b)
	\end{minipage}
	\caption{Density (left) and momentum (right) in the Sedov blast wave problem. The solid lines represent the exact solution, while the symbols correspond to the model data.}
	\label{SedovSimulation}
\end{figure}

Traditionally, the Sedov test assesses the ability of a numerical method to reproduce flows with high Mach numbers, which may also occur in star formation. We set the computational domain $[-0.5;0.5]^{3}$, the adiabatic exponent $\gamma = 5/3$, the initial density in the domain $\rho_{0} = 1$, and the initial pressure $p_{0} = 10^{-5}$. At time $t = 0$, the internal energy $e_{0} = 0.6$ is released. The explosion area is limited by the radius $r_{0} = 0.01$. The computational grid with $100^3$ cells is used. The density and momentum profiles at $t = 0.05$ are shown in Figure~\ref{SedovSimulation}.

In our model test, the sound speed of the background medium is negligibly small, so the Mach number reaches $M \sim 2000$. ngFEOSAD quite well reproduces the position of the shock wave, as well as the density and momentum profiles.

\subsection{Tests on the dust dynamics scheme and friction force}
\label{Appendix:dusttests}
In this section, we tested the dust solver on the problems that consider a pure dust fluid and a mixture of gas and dust, including the friction force between the two components. 
We have considered two numerical schemes to compute the momentum exchange between the gas and dust components,  including the backreaction of dust on gas. The first method makes use of an analytic solution for the updated gas and dust velocities as described in
\citep[see, e.g.,][]{Bate2015,2018Stoyanovskaya}. The second method employs a fully implicit scheme to compute the updated gas and dust velocities as presented in \citet{2018Stoyanovskaya}. Both schemes perform well on the standard tests, such as the dusty 
wave and the Sod shock tube problems. Here, we present the results for the method that uses the analytic solution, which is also implemented in the ngFEOSAD code.

\subsubsection{Sod shock tube for a mixture of gas and dust}
This test is used to assess the ability of a numerical algorithm to accurately track the position of shock
waves and contact discontinuities when for a mixture of gas and pressureless dust.  Initial conditions involve two
discontinuous states, with a hot dense gas on
the left and cold rarified gas on the right. In particular,
we set the pressure and density of gas at $x\in[0.0.5]$ to 1.0, while at
$x\in[0.5.1.0]$ the gas pressure is 0.1 and gas density is 0.125. The velocity
of the $\gamma$=1.4 gas is initially zero everywhere. The dust component has the
same initial distribution as that of gas, but the dust pressure is set to zero.
The dust-to-gas ratio is, therefore, equal to unity everywhere.

The analytic solution for the gas and dust mixture is only known in the limit of short stopping times compared to the time of shock wave propagation. We used the SPLASH code \citep{2007Price} to generate the analytic solution. We follow the terminology of \citet{2012LaibePrice} and introduce the drag parameter $K$ and define the stopping time as $t_{\rm stop}=\rho_{\rm g}/K$, where $\rho_{\rm g}=1$ is the gas density in left half of the shock tube. 

\begin{figure}
\begin{centering}
\includegraphics[width=\linewidth]{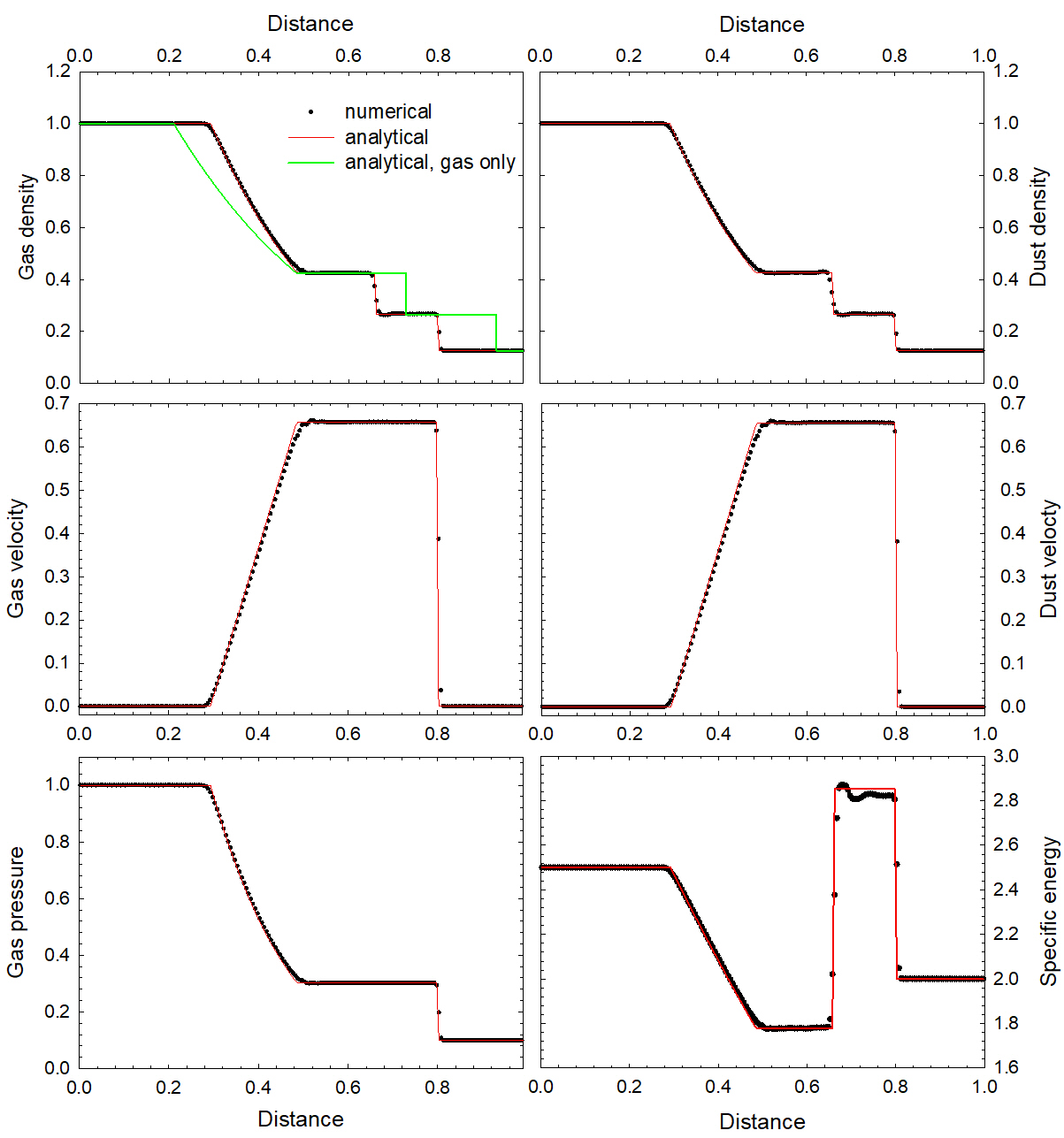} 
\par \end{centering}
\caption{Sod shock tube problem for a strongly coupled gas-dust mixture. The top and middle panels present solutions for the gas component, while the bottom rows show solutions for the dust component.
The red lines present the analytical solution and the filled circles are the numerical solution. The green line the upper left panel presents the analytic solution for the pure gas for comparison. } 
\label{fig:A1}
\end{figure}

%\begin{figure}
%\begin{centering}
%\includegraphics[width=\linewidth]{figures/figA1a.jpg} 
%\par \end{centering}
%\caption{Sod shock tube problem for a цуфлдн coupled gas-dust mixture. The filled red circles present the numerical solutions obtained using the 2D-FEOSAD code \citep{2018VorobyovAkimkin}, which employs the ZEUS integration scheme  \citep{SN1992}, while  the filled black circles correspond to ngFEOSAD, which uses the Godunov solution scheme with the Riemann solvers.} 
%\label{fig:A1a}
%\end{figure}

Figure~\ref{fig:A1} present the test results for $K=5000$, namely, for a strongly coupled gas-dust mixture. The first and second rows present the resulting gas distributions at t=0.24, while the bottom row shows those of dust. The numerical resolution is 200 grid zones and the Courant number is set to 0.2.  Clearly, the code reproduces well the position of the shock and rarefaction waves. The shocks are spread over just 2--3 grid zones, which can be expected for the 3rd order accurate parabolic interpolation scheme used in ngFEOSAD.  

%Figure~\ref{fig:A1a} also presents the results for the case of a weak coupling with $K=1$. Although there is no analytic solution, the data can be compared to similar tests in other studies. We found that our test results agree well with those obtained with grid-based and smooth particle hydrodynamics codes that use conceptually different numerical hydrodynamics schemes \citep{2012LaibePrice,2018VorobyovAkimkin}.

%\begin{equation}
%    u^{n+1}=-D+E, \ \ v^{n+1}=\varepsilon D + E,
%\end{equation}
%where
%\begin{eqnarray}
%\label{eq:ExpConst}
%E&=&\displaystyle\frac{C_1+(a_{\rm g}+\varepsilon a_{\rm d})\tau}{\varepsilon + 1}, \\  
%D&=&\displaystyle\frac{C_2}{\varepsilon+1}e^{-\displaystyle\frac{\varepsilon+1}{t_{\rm stop}}\tau}+
%\frac{(a_{\rm g}-a_{\rm d})t_{\rm stop}}{(\varepsilon+1)^2}, \\
%C_1&=&v^n+\varepsilon u^n, \ \ C_2=v^n-u^n-\displaystyle\frac{(a_{\rm g}-a_{\rm d})t_{\rm stop}}{\varepsilon+1},
%\end{eqnarray}

\subsubsection{Dusty wave}

%In particular, this test allows us to estimate the accuracy with which gas densities, gas and dust velocities are calculated during the propagating and damping of acoustic waves. 

\begin{figure}
\begin{centering}
\includegraphics[width=\linewidth]{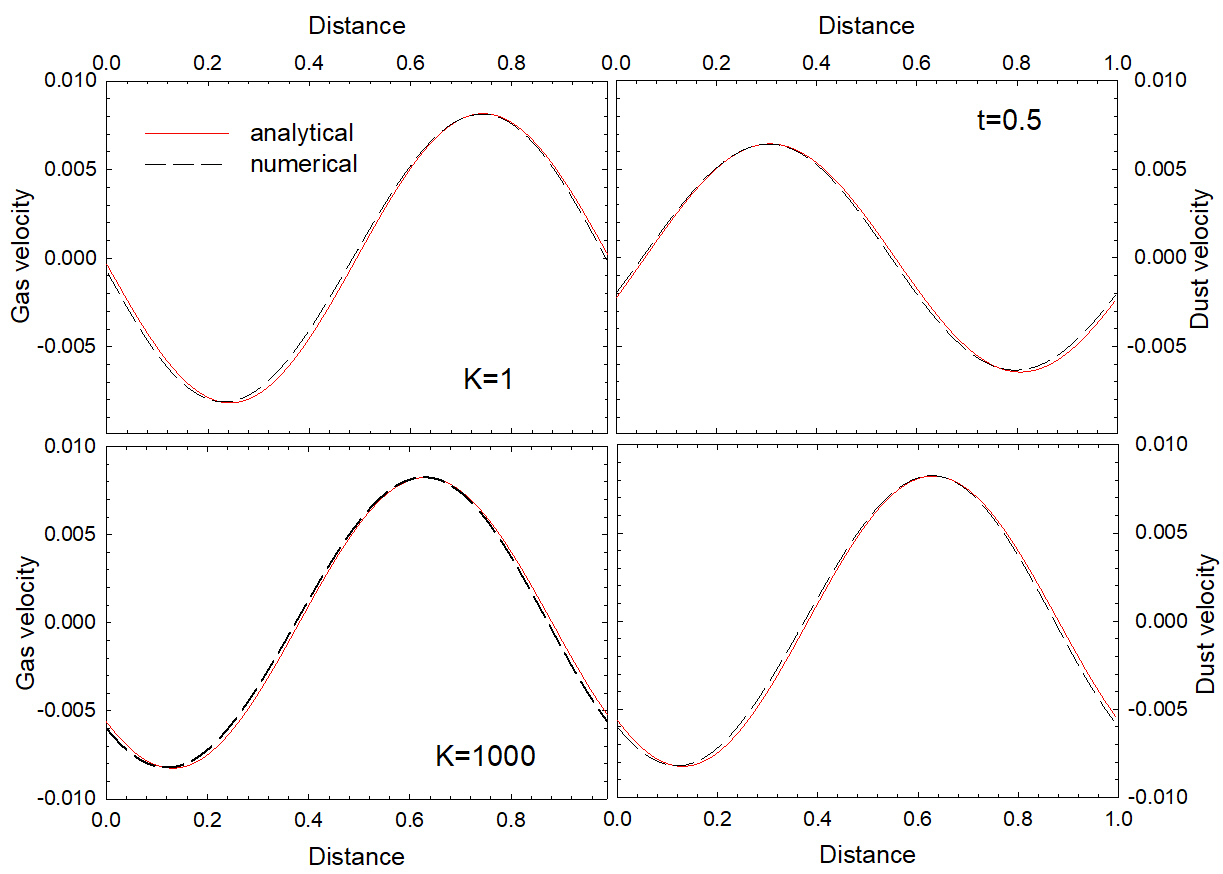} 
\par \end{centering}
\caption{Solution of the dusty wave problem for weak (top row) and strong (bottom row) coupling between gas and dust. The left and right columns present the gas and dust velocities, respectively. The black dashed curves are the
numerical solutions, while the red line shows the analytical solution.} 
\label{fig:A2}
\end{figure}
The dusty wave problem tests the propagation and damping of linear waves in the mixture of gas and dust. The advantage of this test is that it has analytical solutions 
for both weak and strong coupling between gas and dust. The initial setup consists of a sinusoidal wave  of the following
form
\begin{equation}
\label{eq:DustyWave_init1}
\rho_0=\tilde{\rho_{\rm g}}+\delta \sin(kx), \ \rho_{\rm d,0}=\tilde{\rho_{\rm d}}+\delta \sin(kx),
\end{equation}
\begin{equation}
\label{eq:DustyWave_init2}
v_0=\delta \sin(kx), \ u_0=\delta \sin(kx).
\end{equation}
We adopt isothermal gas  with $c_s$=1, $\tilde{\rho_{\rm g}}=1$, $k=1$, and $\delta=0.01$ 
throughout this section and vary $K$ and $\tilde{\rho_{\rm d}}$. All simulations were done with 
200 grid zones on unit distance  and periodic boundary 
conditions. The Courant number is set to 0.2. The analytical solution  is taken from \citet{2011LaibePrice}. 

Figure~\ref{fig:A2} presents the gas and dust  velocities at $t=0.5$ for $K=1$ and $K=1000$, namely, for the cases of weak and strong coupling. The dust-to-gas ratio is $\tilde{\rho_{\rm d}} / \tilde{\rho_{\rm g}}=1.0$. We achieve a reasonably good agreement with 
the analytical solution for both the gas and dust velocities, and also in the limits of weak and strong coupling between gas and dust. A small mismatch between the analytic and numerical solution is typical for this test \citep{2012LaibePrice,2014Loren-AguilarBate}. 
Finally, we note that both tests, the Sod shock tube and the dusty wave, took the back reaction of dust on gas into account.

\subsubsection{Sod shock tube for pure dust}

\begin{figure}[ht]
	\centering
	\begin{minipage}[h]{0.48\linewidth}
		\center{\includegraphics[width=1\linewidth]{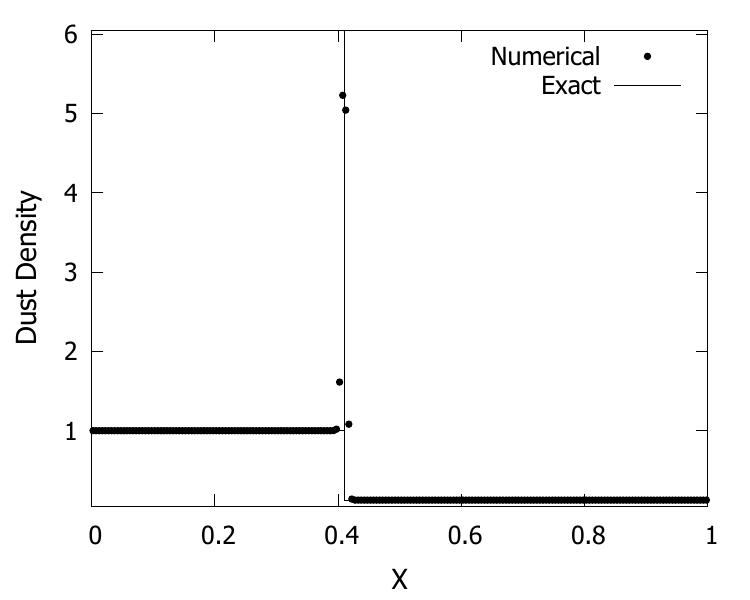}} (a)
	\end{minipage}
	\begin{minipage}[h]{0.48\linewidth}
		\center{\includegraphics[width=1\linewidth]{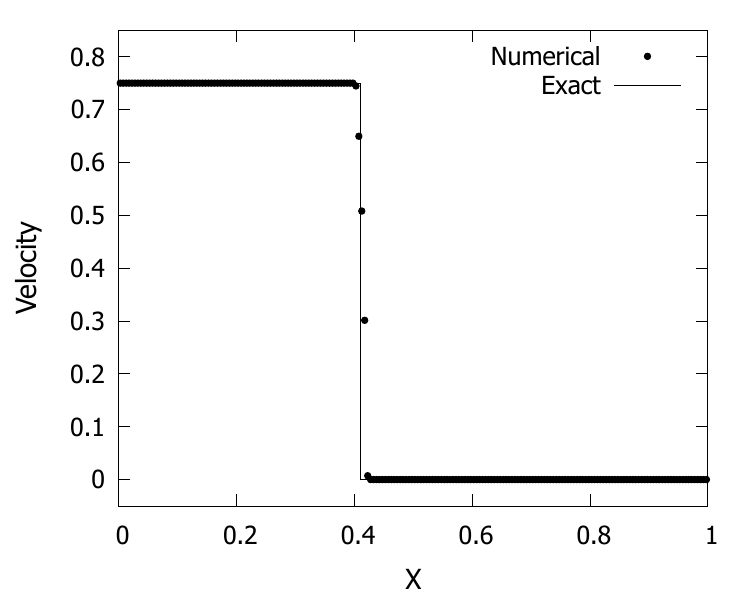}} (b)
	\end{minipage}
	\caption{Sod problem test for pure dust: density (a) and velocity (b).  The lines present the analytic solution while the symbols correspond to the numerical data.}
	\label{ShockTubeSimulationDust}
\end{figure}

Finally, we test the numerical solver on pure dust without gas. 
We consider the initial setup similar to that shown in Table~\ref{ShockTubeProblem} for the Sod test for gas, noting only that dust is pressureless.
The key in this test is the correct reproduction of the dust front propagation velocity, low dissipation of the discontinuity, as well as the reproduction of the delta function at the discontinuity front as a function of the dust density.  Figure~\ref{ShockTubeSimulationDust} demonstrates that all features of the solution are reproduced correctly and with low dissipation.

\subsection{Gravitational potential solver}
\label{App:gravpot}
Unlike many other grid-based codes that solve the Poisson equation to find the gravitational potential, ngFEOSAD employs the convolution theorem to compute the triple sum (Eq.~\ref{sum:pot}). 
A detailed explanation of the method, its extension to nested grids, and comparison with a more common method of conjugate gradients, as well as the pertinent tests can be found in \citet{2023VorobyovMcKevitt}. The solver is characterized by a global second-order convergence, except for the nested grid interfaces where the convergence is linear. The mean errors on the standard tests -- an oblate ellipsoid and a wide-separation binary -- stay below 0.05\% and 0.5\%, respectively. 

The advantage of our original method is that it can be easily parallelized on nested grids and does not require finding the boundary potential on the coarsest grid via the multipole expansion. The method can be significantly accelerated using the graphic processing units, although in the current work we do not utilize this feature.

\subsection{Global collapse problems} 
\label{sect:app-global}
In this section, we consider the test problems that are most relevant to modeling the gravitational collapse of prestellar cores. These problems are inclusive in the sense that they test the code performance on nested grids for both the gravitational potential and hydrodynamics solvers. 

\subsubsection{Collapse of a cold nonrotating gaseous sphere}
\begin{figure}
\begin{centering}
\includegraphics[width=\linewidth]{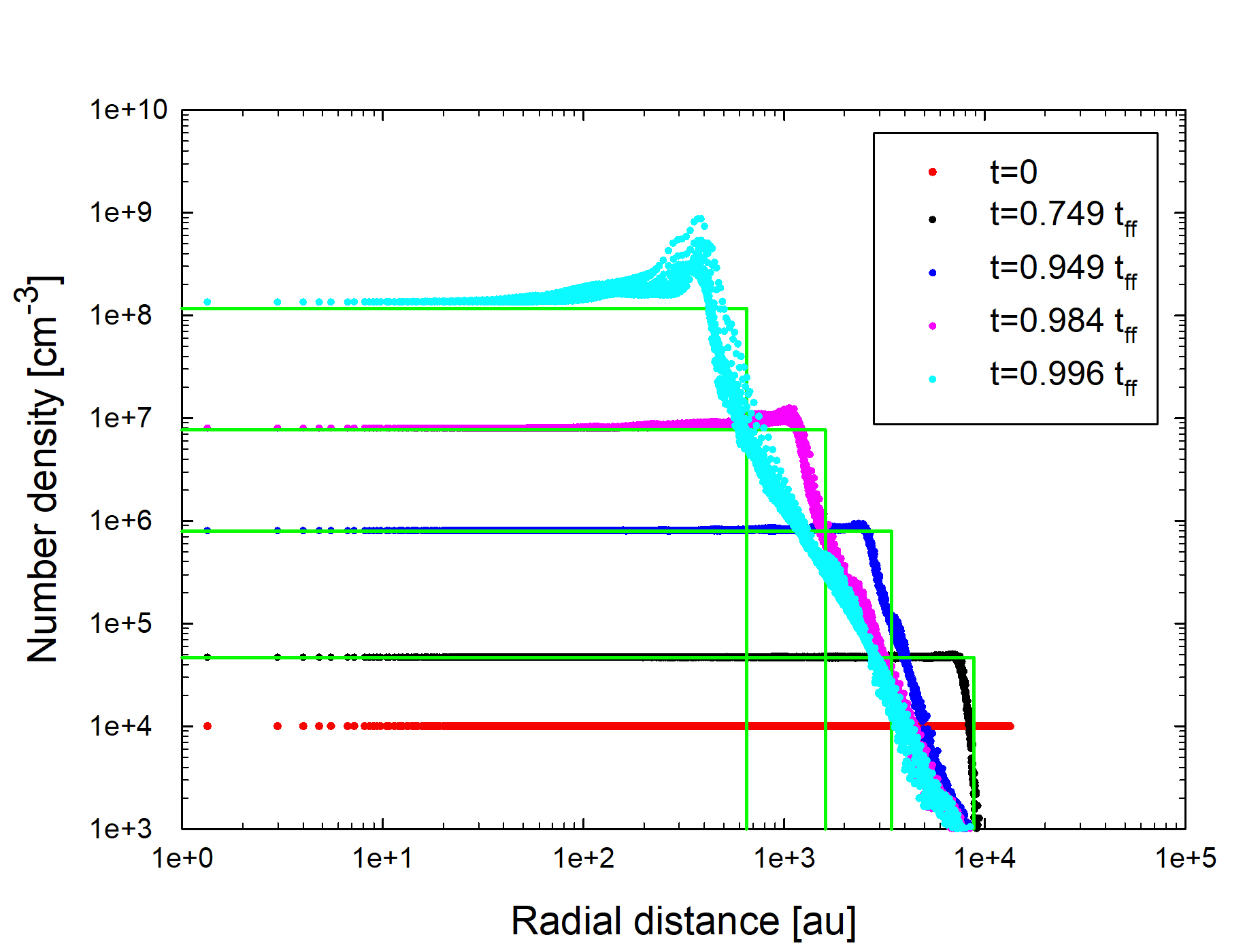} 
\par \end{centering}
\caption{Gravitational collapse of a cold gas sphere. The colored circles correspond to different times in terms of the free-fall time. The green lines present the corresponding analytic solutions.} 
\label{fig:colgas}
\end{figure}

As a first test, we consider the collapse of a cold gaseous sphere. We set a computational domain with a size of 0.15 pc in each direction and introduce a homogeneous gaseous sphere with radius 0.0652~pc. Initially, the number density of the sphere is set equal to $10^4$~cm$^{-3}$. The rest of the computational domain is filled with rarefied gas of negligible density. The gas temperature is set to 0.1~K to imitate the cold sphere collapse. The mass of the cloud is 0.52~$M_\odot$. We have chosen $m=9$ nested grids with the number of grid cells in each direction $N=64$, which corresponds to a effective minimal resolution of 1.83~au. The results are shown in Fig.~\ref{fig:colgas}. The code reproduces well the analytic solution in the inner regions, only slightly overestimating the analytic value at 99.6\% of the free fall time. In the outer regions, an anomalous peak develops, which is likely caused by degrading spatial resolution at large distances from the coordinate center on nested grids. 

%Figure~\ref{???} shows a similar test problem but for the dust sphere. We note that dust is considered as as pressureless fluid in our model.  

\subsubsection{Collapse of a rotating sphere}
\label{Sect:angmom}
In the absence of any mechanisms for angular momentum
redistribution, such as during the gravitational collapse of an axisymmetric cloud, the mass with specific angular momentum less
than or equal to $J=\Omega r^2$ should be conserved and equal to \citep{1980Norman}
\begin{equation}
    M(J) = M_0 \left[1 - \left( 1 - {J \over \Omega_0 R_0^2}  \right)  \right].
\end{equation}
Here,  $\Omega$ is the angular velocity of the collapsing cloud at a distance $r$ from the $z$-axis, and $M_0$, $\Omega_0$, and $R_0$ are the initial mass, angular velocity, and radius of the cloud. We use this property to assess the code’s ability to conserve angular momentum during the gravitational collapse. 

The setup consists of a homogeneous gaseous sphere with the initial mass $1.4~M_\odot$ and the ratio of rotational to gravitational energy set equal to 1.0\%, which is typical for prestellar clouds \citep{2002Caselli}. The initial radius of the cloud, the number density, and the angular velocity  are  $R_0=10^4$~au, $n_0=5\times 10^4$~cm$^{-3}$, and $\Omega_0=4\times 10^{-14}$~s$^{-1}$, respectively. Since we use the Cartesian grid, the cloud is initially submerged into a low-density medium. This causes the outer cloud layers to expand rather than collapse. This is undesirable as the analytic solution stands only for the pure collapse when the spherical layers fall in but do not mix with each other. To reduce this effect we set the cloud isothermal temperature to 0.1~K. We also found that the analytic solution holds only on regular grids. Therefore, we use one nested grid with numerical resolution $256^3$.

\begin{figure}
\begin{centering}
\includegraphics[width=\linewidth]{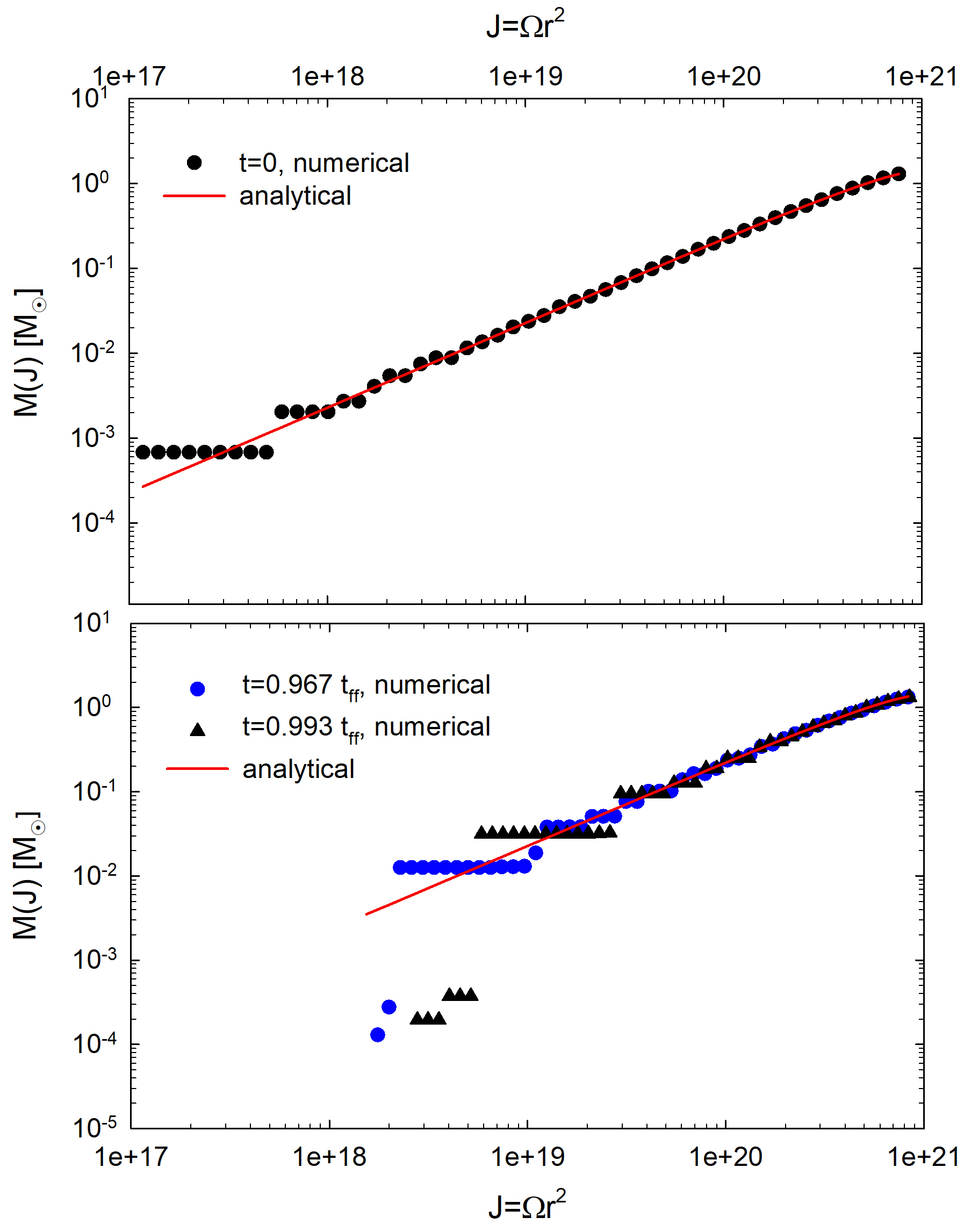} 
\par \end{centering}
\caption{Mass of the cloud with specific angular momentum less or equal to $J$. The top panel presents the $M(J)-J$ relation at the onset of gravitational collapse, while the bottom panel present the corresponding relation at different times after the onset of collapse. The red line is the analytic solution and the symbols present the numerical data.} 
\label{fig:angmom}
\end{figure}

The test results are presented in Figure~\ref{fig:angmom} at $t=96.7$\% and $t=99.3$\% that of the free-fall time, namely, when the central density has increased more than by three and four orders of magnitude, respectively. The angular momentum for the bulk of the collapsing cloud is well conserved.
The horizontal arrangement of the model data at low $J$, which can be seen even at the onset of collapse, is caused by insufficient numerical resolution and is not a signature of angular momentum nonconservation. A higher resolution would make these data points to fit better the red line.
The deviation from the analytic solution occurs only at the smallest $J$ near the rotation axis and high above and below the midplane of the collapsing cloud. These deviations are caused by small diffusion of the outer cloud layers into the external rarefied medium, which is difficult to avoid. In any case, the mass of the cloud material that shows the angular momentum nonconservation is only a minor fraction of the entire cloud mass, namely, less than 0.04\%.

\subsubsection{Protostellar accretion in spherical collapse problems }
The accretion rate of a spherically collapsing cloud on to a protostar has an analytic solution of the following form \citep{1977Shu,2010Kratter}
\begin{equation}
    \dot{M} = {c_{\rm s}^3 \over G} \times 
    \left\{  
\begin{array}{lc}
    0.975, ~& (A=2) \\   
(2A)^{3/2} /\pi, &(A \gg 2)
\end{array}
\right.
\end{equation}
for the initial gas density distribution defined as
\begin{equation}
    \rho (r) = \frac {A c_{\rm s}^2}{4 \pi G r^2}.
    \label{eq:SIS}
\end{equation}
When $A=2$, the gas density distribution turns into a marginally stable singular isothermal sphere \citep{1977Shu}. When $A>2$, the system is out of virial equilibrium and begins to collapse. 

In practice, we initialize the initial gas density distribution that is characteristic of the Bonnor-Ebert sphere with temperature 10~K, central density $n_{\rm BE,0}=4.2\times 10^5$~cm$^{-3}$ and ratio of the central to the outer densities set equal to 20. The resulting radius of the Bonnor-Ebert sphere is $\approx 8000$~au and the mass is 0.9~$M_\odot$. The initial configuration is given a positive density perturbation of 30\% to initiate the gravitational collapse.  

The accretion on the protostar is realized by setting a sink sphere in the center of the collapsing cloud with an accretion radius of 3~au. 
If the gas density in the $i$th computational cell within the sink sphere exceeds $\rho_{\rm crit}=m_{\rm H}\, \mu\, n_{\rm crit}$, where $n_{\rm crit}=10^{13}$~cm$^{-3}$, the gas density is artificially decreased and the excess mass is transferred to the star positioned in the center of the sink sphere. The following equation describes the process
\begin{equation}
  {d \rho_i \over d t} = - b \left( \rho_i - \rho_{\rm crit} \right), 
\end{equation}
which has an analytical solution of the following form
\begin{equation}
  \rho_i=\rho_{0,i} \exp^{-b \, \Delta t} + \rho_{\rm crit} \left( 1.0-\exp^{-b \, \Delta t} \right). 
\end{equation}
Here, $\rho_{0,i}$ is the gas volume density at the current time (at which the condition on the volume density is checked), $\Delta t$ is the time step, and $b$ is the rate of mass evacuation from the sink sphere set equal to the local Keplerian angular velocity $\Omega_{\rm K}$. This definition of the sink sphere is similar to that of \citet{2014Machida} but we use an analytic solution to this problem. The resulting protostellar mass accretion rate is defined as
\begin{equation}
    \dot{M} = \sum_i {\rho_{0,i} - \rho_i \over \Delta t} \times \Delta V_i,
\end{equation}
where $\Delta V_i$ is the volume of the $i$th computational cell and the summation is performed over all cells within the sink sphere for which $\rho_{0,i}>\rho_{\rm crit}$.

\begin{figure}
\begin{centering}
\includegraphics[width=\linewidth]{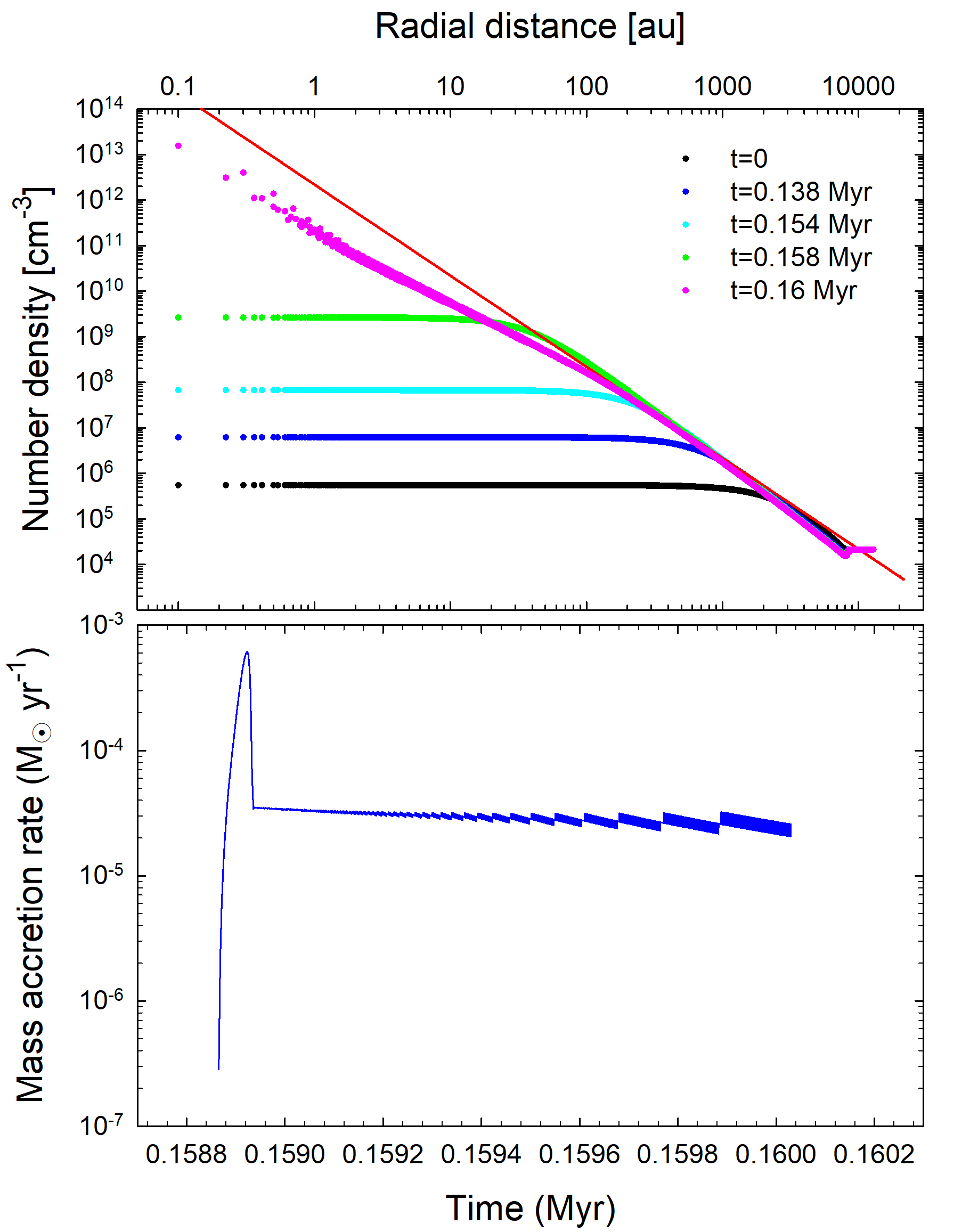} 
\par \end{centering}
\caption{Collapse of a gaseous sphere followed by star formation and protostellar accretion. The top panel shows the radial distributions of the gas number density at different time instances during the collapse and after the formation of the star. The red line is the fit to Eq.~(\ref{eq:SIS}) with $A=5$. The bottom panel displays the resulting mass accretion on to the star. The time is counted from the onset of simulations.  } 
\label{fig:arate}
\end{figure}

The top panel in Figure~\ref{fig:arate} presents the number density distributions of the collapsing cloud at different times. We used $m=12$ nested grids with N=64 grid cells in each coordinate direction.
As the collapse proceeds, the density distribution outside of the central plateau acquires the form similar to Eq.~(\ref{eq:SIS}),
as can be seen from the red line showing the density profile given by Eq.~(\ref{eq:SIS}) with $A=5.0$.  The bottom panel displays the resulting mass accretion rate on to the star vs. time. The star forms at about $t\approx0.1589$~Myr, which is manifested by a sharp increase in the mass accretion rate about this time instance. Such a peak in $\dot{M}$ is also found in other collapse simulations \citep[e.g.,][]{2005VorobyovBasu}.  Soon the mass accretion rate drops and stabilizes at  $\approx 2.5\times10^{-5} M_\odot$~yr$^{-1}$, which is only slightly higher than the analytically predicted value of $1.5\times 10^{-5} M_\odot$~yr$^{-1}$ for $A=5$. We attribute difference to deviations of the Bonnor-Ebert sphere from the form given by Eq.~(\ref{eq:SIS}).

%\begin{equation}
%    \rho (r) = \frac {A c_{\rm s}^2}{4 \pi G r^2}.
%\end{equation}

%Here, we present two (???) test problems, which test the entire code including 

\end{appendix}

\end{document}